\documentclass[12pt,oneside,openany,a4paper]{book}
\usepackage[left=2cm, right=2cm, top=3.5cm, bottom=3.5cm]{geometry}
\usepackage{pdflscape}
\usepackage{latexsym}
\usepackage{amssymb}
\usepackage{authblk}
\usepackage{graphicx}
\graphicspath{{figures/}}
\usepackage{amsmath}
\usepackage{amsthm}
\usepackage{amsfonts}
\usepackage{multirow,tabularx,booktabs}
\usepackage{threeparttable}
\usepackage{array}
\usepackage{color}
\usepackage{bm}
\usepackage[export]{adjustbox}
\usepackage[notindex,nottoc,notlof]{tocbibind}
\usepackage{subcaption}
\usepackage{cleveref}
\usepackage{appendix}
\usepackage[nosectionbib]{apacite}
\AtBeginDocument{}
\usepackage{tikz}
\usetikzlibrary{shapes.geometric, arrows}
\tikzstyle{ct1} = [rectangle, rounded corners, minimum width=3cm, text centered, minimum height=1cm, text width=5cm, draw=black]
\tikzstyle{ct2} = [rectangle, rounded corners, minimum width=3cm, text centered, minimum height=1cm, text width=3.5cm, draw=black]
\tikzstyle{ct3} = [rectangle, rounded corners, minimum width=3cm, text centered, minimum height=1cm, text width=3.5cm, draw=black]
\tikzstyle{mislabel1} = [rectangle, rounded corners, minimum width=3cm, text centered, minimum height=1cm, text width=3cm, draw=black]
\tikzstyle{mislabel2} = [rectangle, rounded corners, minimum width=3cm, text centered, minimum height=1cm, text width=3cm, draw=black]
\tikzstyle{mislabel3} = [rectangle, rounded corners, minimum width=3cm, text centered, minimum height=1cm, text width=2.5cm, draw=black]
\tikzstyle{mislabel4} = [rectangle, rounded corners, minimum width=3cm, text centered, minimum height=1cm, text width=2.5cm, draw=black]
\tikzstyle{check1a} = [rectangle, rounded corners, minimum width=3cm, minimum height=1cm, text width=12.5cm, draw=black]
\tikzstyle{check1b1} = [rectangle, rounded corners, minimum width=1cm, text centered, minimum height=1cm, text width=1cm, draw=black]
\tikzstyle{check1b2} = [rectangle, rounded corners, minimum width=3cm, minimum height=1cm, text width=6cm, draw=black]
\tikzstyle{check1c} = [rectangle, rounded corners, minimum width=3cm, minimum height=1cm, text width=7.5cm, draw=black]
\tikzstyle{check2a} = [rectangle, rounded corners, minimum width=3cm, minimum height=1cm, text width=11.5cm, draw=black]
\tikzstyle{check2b} = [rectangle, rounded corners, minimum width=3cm, text centered, minimum height=1cm, text width=13cm, draw=black]
\tikzstyle{startstop} = [rectangle, rounded corners, minimum width=3cm, minimum height=1cm,text centered, draw=black, fill=red!30]
\tikzstyle{io} = [trapezium, trapezium left angle=70, trapezium right angle=110, minimum width=3cm, minimum height=1cm, text centered, draw=black, fill=blue!30]
\tikzstyle{process} = [rectangle, minimum width=3cm, minimum height=1cm, text centered, draw=black, fill=orange!30]
\tikzstyle{decision} = [diamond, minimum width=3cm, minimum height=1cm, text centered, draw=black, fill=green!30]
\tikzstyle{arrow} = [thick,->,>=stealth]
\DeclareMathOperator*{\argmax}{arg\,max}

\newtheorem{remark}{Remark}
\newtheorem*{remark*}{Remark} 
\newtheorem{thm}{Theorem}
\newtheorem{lma}{Lemma}
\theoremstyle{definition}
\newtheorem{defn}{Definition}
\title{Statistical Inference on the Cure Time}
\author[1,2]{Yueh Wang}
\author[1]{Hung Hung}
\affil[1]{Institute of Epidemiology and Preventive Medicine, National Taiwan University}
\affil[2]{Taiwan Cancer Registry}
\date{}

\begin{document}
\maketitle
\frontmatter
\pagenumbering{roman}
\cleardoublepage

\chapter*{Abstract}
{\renewcommand\baselinestretch{1.5}\fontsize{12pt}{20pt}\selectfont
In population-based cancer survival analysis, the net survival is important for government to assess health care programs. For decades, it is observed that the net survival reaches a plateau after long-term follow-up, this is so called ``statistical cure''. Several methods were proposed to address the statistical cure. Besides, the cure time can be used to evaluate the time period of a health care program for a specific patient population, and it also can be helpful for a clinician to explain the prognosis for patients, therefore the cure time is an important health care index. However, those proposed methods assume the cure time to be infinity, thus it is inconvenient to make inference on the cure time. In this dissertation, we define a more general concept of statistical cure via conditional survival. Based on the newly defined statistical cure, the cure time is well defined. We develop cure time model methodologies and show a variety of properties through simulation. In data analysis, cure times are estimated for 22 major cancers in Taiwan, we further use colorectal cancer data as an example to conduct statistical inference via cure time model with covariate sex, age group, and stage. This dissertation provides a methodology to obtain cure time estimate, which can contribute to public health policy making.\\
\\
{\bf Key words:} net survival, conditional survival, statistical cure, cure rate, cure time
\par}

\addcontentsline{toc}{chapter}{Abstract}
\tableofcontents
\listoffigures
\mainmatter
\fontsize{12pt}{20pt} \selectfont
\chapter{Introduction}
\section{Population-based cancer survival analysis}
In the recent years, data collection are going to be faster and more effective in governments or companies. The collected data size expands rapidly in both sample size and number of variables. Therefore, using appropriate statistical methods to extract useful knowledge from the large databases becomes an important but challenging issue. From the perspective of public health, population-based cancer survival analysis focuses on analyzing national cancer databases and extracting important information, from which government is able to evaluate health care programs.

Let $D$ be the event time from the disease of interest. Researchers are often interested in the behaviour of $D$. However, in population-based cancer survival analysis, we can not observe $D$ directly, instead we observe
\begin{align*}
 T=\min(O, D),
\end{align*}
where $O$ is the event time from all causes except the disease of interest. It is said that $D$ considers a hypothetical situation where the cause of death is only due to the disease of interest \shortcite{Pohar-Perme.etal:2012}. Statistical inference of $D$ is thus often made under the independence assumption
\begin{align}
O \perp D, \label{1.1}
\end{align}
in which situation the distribution of $D$ is estimable. Based on \eqref{1.1}, the involved survival functions and the corresponding hazard functions can be expressed as
\begin{align}
S_T(t)=S_O(t)S_D(t) \Longleftrightarrow h_T(t)=h_O(t)+h_D(t), \label{1.2}
\end{align}
where $S_T(t)=P(T>t)$, $S_O(t)=P(O>t)$ is the disease-free survival, and $S_D(t)=P(D>t)$ is the net survival, and $h_T(t)$, $h_O(t)$, and $h_D(t)$ are the hazard functions of $T$, $O$, and $D$, respectively.

Under \eqref{1.1}, the idea of relative survival can be used to estimate $S_D(t)$ non-parametrically by
\begin{align*}
\mbox{RS}(t)=\frac{\widehat{S}_T(t)}{S_O(t)},
\end{align*}
where $\widehat{S}_T(t)$ is an estimate of $S_T(t)$ (e.g. the Kaplan-Meier estimator), and $S_O(t)$ can be obtained from national death certificate database. There are different kinds of relative survival estimates, depending on the method used to calculate $S_O(t)$ \shortcite{Ederer.etal:1961, Hakulinen:1982}.

\section{Statistical cure and cure rate model}
In decades, more and more complex diseases are said to be curable \shortcite{Castillo.etal:2013}. One can also observe the cure phenomenon from a diseased population after long-term follow-up, that is ``$S_D(t)$ reaches a plateau $\pi$ after long-term follow-up'', which can be formulated as
\begin{align}
\lim_{t\rightarrow\infty} S_D(t)=\pi. \label{1.8}
\end{align}
It also implies from the relation \eqref{1.2} that the excess hazard $h_D(t)$ decreases to 0 as $t$ goes to infinity. In this situation, patients will no longer die from the disease of interest, which is called ``population cure'' or ``statistical cure'' \shortcite{Dubecz.etal:2012}. The constant $\pi$ in \eqref{1.8} represents the proportion of patients that will no longer die from the disease of interest, which is called the cure rate. Notice that the concept of ``cure'' can be interpreted in individual level and population level. In the individual level, the cure can be thought of as ``medical cure'', which means asymptomatic of an individual after receiving medical treatment. In the population level, ``population cure'' or ``statistical cure'' occurs when the excess hazard decreases to zero \shortcite{Lambert.etal:2007}. In order to characterize the information of cure, we often use the cure rate model (or cure fraction model) to estimate the cure rate $\pi$.

Cure rate model has been well developed in these decades, and we can simply classify the models into mixture cure rate model \shortcite{De.etal:1999} and non-mixture cure rate model \shortcite{Andersson.etal:2011}. The mixture cure rate model considers the mixture distribution of cure and uncure for each patient. Let the cure condition $R\sim\mbox{Bernoulli}(\pi)$. If a patient will be cured, then $R=1$, otherwise $R=0$. Then, with conditions in Lemma 1, it can be shown that $S_D(t)$ is of the form
\begin{align}
S_T(t) = S_O(t)\{\pi + (1-\pi)S_u(t)\}, \label{1.5}
\end{align}
where $S_u(t)=P(D>t|R=0)$ is the survival of uncured patients, and $\pi=P(R=1)$ is the cure rate.
\begin{lma}
The following conditions implies \eqref{1.5}.
\begin{enumerate}
 \item [$(a)$] $O\perp \left(D,R\right)$,
 \item [$(b)$] $D=\infty$ if $R=1$.
\end{enumerate}
\end{lma}

Another method, the non-mixture cure rate model, derives $S_T(t)$ from a different perspective. Let $N$ be the number of metastatic-competent cancer cell number for each patient after treatment, and let $F_0(t)$ be the cdf of the event time with a metastatic-competent cancer cell. The non-mixture cure rate model assumes that $N\sim\mbox{Poisson}(\lambda)$, then it is straightforward that those patients without metastatic-competent cancer cell are considered cured, i.e. $\pi=P(N=0)=e^{-\lambda}$ is the cure rate. With conditions in Lemma 2, it can be shown that $S_D(t)$ is of the form
\begin{align}
S_T(t) = S_O(t)\pi^{F_0(t)}. \label{1.6}
\end{align}

\begin{lma}
The following conditions implies \eqref{1.6}.
\begin{enumerate}
 \item [$(a)$] $D|N=\min(D_{1}, D_{2}, \ldots, D_{N})$, 
 \item [$(b)$] $D_{1},\ldots,D_{N}\stackrel{iid}{\sim}F_0(t)$,
 \item [$(c)$] $O \perp \left(D,N\right)$,
 \item [$(d)$] $D=\infty$ if $N=0$,
\end{enumerate}
where $D_{i}$ denotes the event time from the $i$-th metastatic-competent cancer cell.
\end{lma}

Note that \eqref{1.6} can also be represented as the form of mixture cure rate model
\begin{align*}
S_O(t)\pi^{F_0(t)}=S_O(t)\left\{\pi+(1-\pi)\frac{\pi^{F_0(t)}-\pi}{1-\pi}\right\},
\end{align*}
where $\frac{\pi^{F_0(t)}-\pi}{1-\pi}$ is a proper survival function, and can be used to model $S_u(t)$.

Under appropriate modelling of $S_u(t)$ for \eqref{1.5}, or $F_0(t)$ for \eqref{1.6}, one can estimate $\pi$ via MLE inference procedure. Recently, the flexible parametric cure rate model \shortcite{Andersson.etal:2011}, which uses the restricted cubic spline function to model $F_0(t)$, is considered to be a suitable method in describing cure in a variety of cancers.

Compare with \eqref{1.2}, the above model is equivalent to model the net survival $S_D(t)$
\begin{align*}
S_D(t)=\pi+(1-\pi)S_u(t)
\end{align*}
in the mixture cure rate model, and as
\begin{align*}
S_D(t)=\pi^{F_0(t)}
\end{align*}
in the non-mixture cure rate model. Note that, in both types of cure rate models $S_D(t)$ are improper survival functions, since \eqref{1.8} tells that the long-term follow-up time of $S_D(t)$ attains $\pi$ as $t$ goes to infinity.

\section{Cure time}
Equation \eqref{1.8} indicates that the cure rate can be attained as $t$ tends to infinity. However, it is observed that the net survival may attain the cure rate after a specific time point $\tau$ within the follow-up time, instead of infinity. This specific time point $\tau$ is called ``cure time''. The government may want to know the cure time, so that the health policies can be conducted more efficiently. In Taiwan, the a cancer patient will be assigned a catastrophic illness certificate, and it should be re-evaluated after the cancer cure time. Moreover, the burden of diseases can be assessed more accurately if the government has a better estimate of cure time \shortcite{Blakely.etal:2010, Blakely.etal:2012}. An example is that the years lived with disability (YLD) needs a time point to exclude those patients who live a long period so that the disability is negligible. In pharmaceuticals, it is important to know the time when patients are identical to general population after taking new treatment. Clinicians may be also interested in cure time for precise health care suggestions to patients. There are na\"ive ways to determine the cure time $\tau$. Some non-parametric methods were suggested in practical use, one of them suggested that the estimated cure time occurred after 95\% or 99\% of the deaths had elapsed \shortcite{Woods.etal:2009, Smoll.etal:2012}. The above strategy is easy to implement but may fail to apply if the true time point occurs beyond the follow-up time, or the cure assumption (\ref{1.8}) is inappropriate. It was also suggested to use conditional relative survival to find the cure time, i.e. to find the smallest $\tau$ such that the conditional relative survival exceeds 95\% \shortcite{Janssen-Heijnen.etal:2007, Janssen-Heijnen.etal:2010, Dal.etal:2014}
\begin{align*}
\tau = \min\left\{k\;|\;\mbox{RS}(t|k) > 95\%\;\;\forall\;t\geq k\right\},
\end{align*}
where $\mbox{RS}(t|k)=\mbox{RS}(t)/\mbox{RS}(k)$ is the conditional relative survival. However, the choice of 95\% is subjective, and it is possible to see a non-negligible decreasing trend of conditional relative survival even if it exceeds 95\%. \citeA{Baade.etal:2011} proposed to use conditional relative survival to determine $\tau$ visually after which $\mbox{RS}(t|\tau)$ is nearly a constant. \citeA{Blakely.etal:2012} suggested a convenient way by visually identify the time point of non-declination in the model-based or non-parametric net survival curve. These methods still face the problem that the determination of $\tau$ is subjective. Moreover, there exists no statistical inference procedure for $\tau$ in the above mentioned methods.

The research aim of this dissertation is to give a new perspective of statistical cure, from which the information of cure time can be included. We also propose a parametric method to model $\tau$, which would be able for the researchers to make statistical inference about cure time. In application, the proposed methodology can be used not only on the population-based cancer survival analysis, but also on the clinical-based research, in which the diseased cohort with certain medical treatment can be compared to the general population.

\chapter{A New Perspective of Statistical Cure with Cure time}
\begin{figure}[htbp]
\centering
\begin{subfigure}{.5\textwidth}
  \centering
  \includegraphics[width=\textwidth]{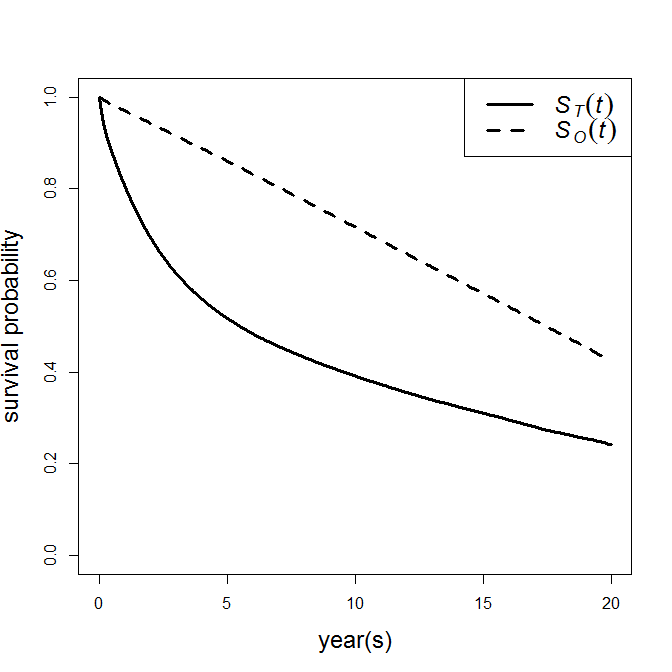}
  \caption{}
  \label{f1.5a}
\end{subfigure}%
\begin{subfigure}{.5\textwidth}
  \centering
  \includegraphics[width=\textwidth]{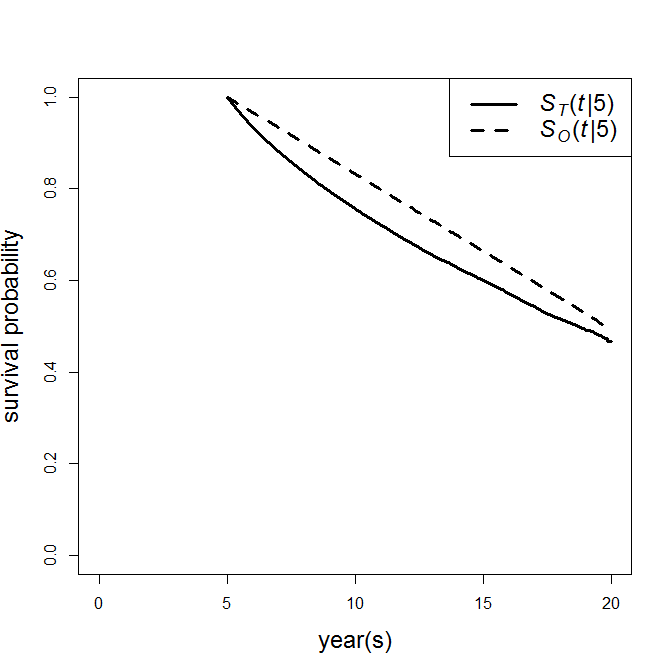}
  \caption{}
  \label{f1.5b}
\end{subfigure}
\caption{Conditional survival of colorectal cancer data and general population in Taiwan, diagnosed between 1995 and 2013 with follow-up to 2014.}
\label{f1.5}
\end{figure}

In this chapter, we propose another concept to define ``statistical cure'' from which the cure time can be directly characterized. We begin from a comparison of general population and diseased population. One can treat the general population as a pool of normal persons with negligible risk of death from the disease of interest. Therefore, it is obvious that normal persons are expected to have better survival experience than diseased population, i.e. $S_O(t)\geq S_T(t)\;\;\forall\;t >0$. Taking the colorectal cancer population and the corresponding disease-free survival in Taiwan as an example, Figure \ref{f1.5}(\subref{f1.5a}) shows that the disease-free survival $S_O(t)$ is uniformly higher than the observed survival of colorectal cancer patient.
We can also see that $S_T(t)$ decreased rapidly in the beginning, but the decreasing trend becomes similar to $S_O(t)$ when $t>5$. It implies that patients who survived at 5 years may have similar survival experience as the general population. It motivates us to find the cure time by comparing the conditional survival functions of the diseased population and general population. For any time point $k$, the conditional survival of diseased population given surviving at $k$ is defined to be
\begin{align*}
S_T(t|k)=P(T>t|T>k)=\frac{S_T(t)}{S_T(k)}\;\;\forall\;t\geq k,
\end{align*}
which can be explained as the survival probability of a person who lived upon $k$ from the beginning of follow-up (e.g.,\ diagnosis of disease). Note that $S_T(t)$ can be expressed as $S_T(t|0)$. Figure \ref{f1.5}(\subref{f1.5b}) shows two survival curves conditional on 5 years for $S_O(t|5)$ and $S_T(t|5)$. Obviously, $S_T(t|5)$ is much more close to $S_O(t|5)$, which means that those patients who has survived at 5 years would have similar survival experience to general population. This example motivates us to develop a new concept of ``statistical cure''.

\begin{defn}[$Statistical\;Cure$]
    {\it 
    The statistical cure is attained if there exists some $k>0$ such that 
    \begin{align}
    S_T(t|k)= S_O(t|k)\;\forall\;t> k. \label{2.1}
    \end{align}
    The cure time $\tau$ is defined as the minimum time point satisfying statistical cure as
    \begin{align}
    \tau = \min\left\{k\;|\;S_T(t|k)= S_O(t|k)\;\;\forall\;t> k\right\}. \label{curetime}
    \end{align}
    }
\end{defn}

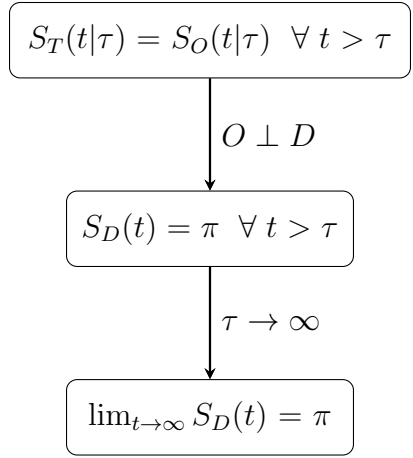
\begin{figure}[h]
	\centering
	\begin{tikzpicture}[node distance=2.5cm]
	\node (ct1) [ct1] {$S_T(t|\tau)= S_O(t|\tau)\;\;\forall\;t> \tau$};
	\node (ct2) [ct2, below of = ct1] {$S_D(t) = \pi\;\;\forall\;t> \tau$};
	\node (ct3) [ct3, below of = ct2] {$\lim_{t\rightarrow\infty} S_D(t)=\pi$};
	\draw [arrow] (ct1) -- (ct2);
	\draw [arrow] (ct2) -- (ct3);
	\draw [arrow] (ct1) -- node[anchor=west] {$O\perp D$} (ct2);
	\draw [arrow] (ct2) -- node[anchor=west] {$\tau \rightarrow \infty$} (ct3);
	\end{tikzpicture}
	\caption{Relationships between different concepts of statistical cure}
	\label{curecondition}
\end{figure}

Definition 1 means that those patients who have survived at $\tau$ can not be distinguished from the general population in the sense of conditional survival. Definition 1 also provides a connection between cure time and cure rate as summarized below.
\begin{thm}
Assume condition \eqref{1.1}. Then \eqref{2.1} is equivalent to
\begin{align}
S_D(t) = \pi\;\;\forall\;t> \tau, \label{2.2}
\end{align}
where $\pi=S_D(\tau)$ is the cure rate, and $\tau$ is the cure time.
\end{thm}
In \eqref{2.2}, $\tau$ is used to demonstrate the time that $S_D(t)$ attains $\pi$, while in \eqref{1.8}, $\tau$ is forced to be infinity. Thus \eqref{2.2} shows not only cure rate $\pi$, but also cure time $\tau$. We note that \eqref{2.1} is the most general definition for ``statistical cure'', since it does not need the assumption \eqref{1.1}, while \eqref{2.2} relies on the validity of \eqref{1.1}. Note also that \eqref{2.2} is more general than the conventional \eqref{1.8}, by noting that \eqref{2.2} becomes \eqref{1.8} if $\tau \rightarrow \infty$. Figure \ref{curecondition} summarizes the relationship between different concepts of statistical cure.
\newpage

We close this chapter by developing some properties related to Definition 1 that can help understand the meaning of statistical cure.
\begin{thm}
Assume condition \eqref{1.1}. Then, either statement (a) or (b) below is equivalent to \eqref{2.1}
\begin{enumerate}
\item [$(a)$] $T=\min(O,D)\cdot I(D\leq \tau)+O\cdot I(D>\tau)$
\item [$(b)$] $h_T(t)=h_O(t)+h_D(t)\cdot I(t\leq \tau)$
\end{enumerate}
where $I(.)$ denotes the indicator function.
\end{thm}
From Theorem 2, under the consideration of cure, the survival function of $T$ is derived to be
\begin{align}
 S_T(t) = S_O(t)U_D(t;\tau), \label{2.7}
\end{align}
where 
\begin{align}
U_D(t;\tau)=\left\{S_D(t)-S_D(\tau)\right\}\cdot I(t\leq\tau)+S_D(\tau) \label{2.9}
\end{align}
only depends on $D$ and $\tau$. Note that in Theorem 2(a), $O$ and $D$ are competing events if $D\leq\tau$, but there is only $O$ as the event time when $D>\tau$. It means that after $\tau$, patient would never die from $D$ (i.e.,\ $\tau$ is the cure time). Comparing Theorem 2(b) to hazard function in \eqref{1.2}, we know that $h_D(t)$ exists for all $t$ in \eqref{1.2} without cure. However, in the presence of cure, in some diseases the effect of $h_D(t)$ vanishes in late period of follow-up. Those who had long-term follow-up are nearly identical to normal person, and this requirement is fulfilled by assuming $h_D(t)$ disappears after a fixed time point $\tau$ as stated in Theorem 2(b). In the following chapters, we would introduce a cure time model for estimating $\tau$ based on Theorem 2.
\begin{remark}
A special case of Theorem 2(b) is $h_T(t) = h_D(t)\cdot I(t\leq \tau)$, where $O$ is assumed to be infinity. A Bayesian semi-parametric approach was proposed to estimate the cure time (in this article the cure time is called ``unknown threshold'') under this special case \cite{Nieto-Barajas:Yin:2008}, but it was not suitable for population-based studies since the semi-parametric model did not incorporate the information of $O$.
\end{remark}

\chapter{Statistical Inference Procedure of Cure Time}
\section{Data structure}
The sample is in the form of $\{Z,\delta,X\}$, where $Z=\min(T,C)$ is the last observed time, $C$ is the censoring time, $X\in\mathbb{R}^p$ is the covariate, and $\delta$ is the censoring status. Taiwan has a well-developed health care system, and considering the high-quality death certificate information should be helpful to obtain more efficient statistical inference. In this study, we propose a more general perspective to apply the cause of death information, and define a more general version of censoring status, that is
\begin{align}
\delta=\left\{
\begin{array}{ll}
0,& \mbox{if } Z=C\\
1,& \mbox{if } Z=O\\
2,& \mbox{if } Z=D\\
3,& \mbox{if } Z=T
\end{array}
\right., \label{3.5}
\end{align}
where $D$ is obtained based on the cause of death information. Note that $O$ indicates the time from all causes of death except the disease of interest, therefore the censoring should not include any other cause of death, such as car accident. 

Since covariate $X$ has been involved in estimation, \eqref{1.1} should be modified as a relaxed assumption. It is also reasonable to assume that $C$ is independent of all the last observed time, or $(O,D)$. The assumptions used in estimation can therefore be expressed below
\begin{enumerate}
 \item [(C1)] $O\perp D|X$
 \item [(C2)] $C\perp(O,D)|X.$
\end{enumerate}

In the previous population-based methodologies it was suggested using $T$ and $C$ to define $\delta$, and ignore death certificate information completely since the accuracy of death certificates are often problematic \shortcite{Howlader.etal:2010, Huang.etal:2014}. However, it is reasonable for researchers to determine whether to use the cause of death information completely or partially, according to quality of the database from health care system of their countries. In this data structure, $\delta = 3$ means that we know that the last observed time is $T$ but do not know whether $T=D$ or $T=O$, and $\delta = 3$ often occurs in the case of uncertain cause of death from death certificate database \shortcite{Naghavi.etal:2010}. Moreover, researchers can choose not to use the information of $O$ and $D$, but instead set $\delta = 3$ for an individual if his/her cause of death information is doubtful, or if the quality of the cause of death information is not reliable. Therefore, we provide a flexible way of data usage to let researchers make use of data more thoroughly.

\section{Model specification}
According to Theorem 2(b), excess hazard $h_D$ and cure time are affected by $X$. Therefore, we propose the cure time model (CTM)
\begin{align}
h_T(t|X)=h_O(t)+h_D(t|X^{(1)})\cdot I\left(t\leq \tau_{X^{(2)}}\right), \label{3.1}
\end{align}
where $X^{(1)}$ and $X^{(2)}$ are subsets of $X$, respectively. Since the cure time $\tau$ must not be negative, $\tau$ can be modelled using any link function with positive range, such as
\begin{align*}
\tau_{X^{(2)}}=\exp(\beta^TX^{(2)}),
\end{align*}
where $\beta$ is the parameter corresponding to $X^{(2)}$. $h_D(t|X^{(1)})$ is assumed to be the excess hazard function from a parametric distribution. For example, the excess hazard function can be modelled as Weibull hazard function, where the link function is set to be exponential function for both shape and scale parameters. In population-based survival analysis, we obtain the information of $h_O(t)$ through vital statistics from government.
\section{Estimation}
We use maximum likelihood estimation to obtain the estimate of $(\alpha, \beta)^T$, where $\alpha$ is the parameter of the parametric distribution to model $D$. If we model $D$ as Weibull distribution, then $\alpha = (\alpha_1, \alpha_2)^T$, where $\alpha_1$ is used to model the shape parameter in the form of $\exp(\mathbf{\alpha}_1^TX^{(1)})$, and $\alpha_2$ is used to model the scale parameter in the form of $\exp(\mathbf{\alpha}_2^TX^{(1)})$. Since $\delta$ contains four levels, we can derive the likelihood function of each level through the corresponding pdf respectively.
 \subsection*{The case of $\delta=0$}
 For a censored case with given covariate $X=x$, the observation is $(z,\delta = 0,x)$, and we have
 \begin{align*}
 P(Z>z, \delta=0|X=x)&=P(Z>z, C<T|X=x)=P(C>z, C<T|X=x)\\
 &=\int_z^\infty\int_c^\infty f_T(t|x)f_C(c)dtdc\\
 &=\int_z^\infty S_T(c|x)f_C(c)dc,
 \end{align*}
 where $f_C(z)$ is the pdf of censoring time $C$.
 The pdf of $(Z, \delta=0)$ given covariate $x$ is
 \begin{align}
 f_{Z,\delta|X}(z,0|x)&=-\frac{d}{dz}\int_z^\infty S_T(c|x)f_C(c)dc\notag\\
 &=S_T(z|x)f_C(z)\notag\\
 &=U_D(z;\tau|x)S_O(z)f_C(z), \label{3.3.1}
 \end{align}
 where $U_D(z;\tau|x)$ is similar to (\ref{2.9}), but involving covariate $x$, that is
 \begin{align}
 U_D(z;\tau|x)=\left\{S_D(z|x^{(1)})-S_D(\tau_{x^{(2)}}|x^{(1)})\right\}I\left(z\leq\tau_{x^{(2)}}\right)+S_D(\tau_{x^{(2)}}|x^{(1)}). \label{3.3.6}
 \end{align}
 Equation \eqref{3.3.6} tells that if $z\leq\tau_{x^{(2)}}$, then $U_D(z;\tau|x) = S_D(z|x^{(1)})$; if $z>\tau_{x^{(2)}}$, then $U_D(z;\tau|x)$ remains a constant $S_D(\tau_{x^{(2)}}|x^{(1)})$.
 
 In (\ref{3.3.1}), the censoring time $C$ contributed to the pdf via $f_C(z)$, and we did not observe the exact time from $O$ and $D$, but $O>z$ and $D>z$, therefore the pdf was contributed via $S_O(z)$ and $U_D(z;\tau|x)$, respectively. Since we assume that $C$, $O$, and $D$ are independent of each other, the pdf (\ref{3.3.1}) are simply the product of $U_D(z;\tau|x)$, $S_O(z)$, and $f_C(z)$.
 \subsection*{The case of $\delta=1$}
 For a case with the last observed time being $O$, the observation is $(z,\delta = 1,x)$, and we have
 \begin{align*}
 P(Z>z, \delta=1|X=x)=&P(O>z, O<D, D\leq\tau,O<C|X=x)+\\
 &P(O>z, D>\tau,O<C|X=x)\\
 =&\int_z^{\tau_{x^{(2)}}}\int_{o}^{\tau_{x^{(2)}}}\int_{o}^\infty f_D(w|x)f_O(o)f_C(c)\,dc\,dw\,do+\\
 &\int_z^\infty\int_{\tau_{x^{(2)}}}^\infty\int_{o}^\infty  f_D(w|x)f_O(o)f_C(c)\,dc\,dw\,do\\ 
 =&\int_z^{\tau_{x^{(2)}}} \{S_D(o|x^{(1)})-S_D(\tau_{x^{(2)}}|x^{(1)})\}f_O(o)S_C(o)\,do+\\
 &\int_z^\infty S_D(\tau_{x^{(2)}}|x^{(1)})f_O(o)S_C(o)\,do,
 \end{align*}
 where $S_C(.)$ is the survival function of $C$, and $f_O(.)$ is the pdf of $O$.
 The pdf of $(Z, \delta=1)$ given covariate $x$ is
 \begin{align}
 f_{Z,\delta|X}(z,1|x)=&-\frac{d}{dz}\int_z^{\tau_{x^{(2)}}} \{S_D(o|x^{(1)})-S_D(\tau_{x^{(2)}}|x^{(1)})\}f_O(o)S_C(o)\,do-\notag\\
 &\;\;\;\;\frac{d}{dz}\int_z^\infty S_D(\tau_{x^{(2)}}|x^{(1)})f_O(o)S_C(o) \,do\notag\\
 =&\{S_D(z|x^{(1)})-S_D(\tau_{x^{(2)}}|x^{(1)})\}f_O(z)S_C(z)I(z\leq\tau_{x^{(2)}})+S_D(\tau_{x^{(2)}}|x^{(1)})f_O(z)S_C(z)\notag\\
 =&\left[\{S_D(z|x^{(1)})-S_D(\tau_{x^{(2)}}|x^{(1)})\}I(z\leq\tau_{x^{(2)}})+S_D(\tau_{x^{(2)}}|x^{(1)})\right]f_O(z)S_C(z)\notag\\
 =&U_D(z;\tau|x)f_O(z)S_C(z). \label{3.3.2}
 \end{align}
 
 In (\ref{3.3.2}), $O$ contributed to the pdf via $f_O(z)$, and we did not observe the exact time from $C$ and $D$, but we know that $C>z$ and $D>z$, therefore the pdf was contributed via $S_C(z)$ and $U_D(z;\tau|x)$, respectively.
 \subsection*{The case of $\delta=2$} 
 For a case with the last observed time being $D$, we have observation $(z,\delta = 2,x)$, and
 \begin{align*}
 P(Z>z, \delta=2|X=x)&=P(D>z, D<O, D\leq\tau,D<C|X=x)\\
 &=\int_z^{\tau_{x^{(2)}}}\int_{w}^\infty\int_{w}^\infty f_D(w|x^{(1)})f_O(o)f_C(c)\,dc\,do\,dw\\
 &=\int_z^{\tau_{x^{(2)}}} f_D(w|x^{(1)})S_O(w)S_C(w)\,dw.
 \end{align*}
 The pdf of $(Z, \delta=2)$ given covariate $x$ is
 \begin{align}
 f_{Z,\delta|X}(z,2|x)&=-\frac{d}{dz}\int_z^{\tau_{x^{(2)}}} f_D(w|x^{(1)})S_O(w)S_C(w)\,dw\notag\\
 &=f_D(z|x^{(1)})S_O(z)S_C(z)I(z\leq\tau_{x^{(2)}}). \label{3.3.3}
 \end{align}
 Note that, since $(z,\delta = 2,x)$ is the event time from the disease of interest, it is an ``uncured'' case, and the last observed time $z$ is therefore smaller than the cure time $\tau_{x^{(2)}}$, that is $z\leq\tau_{x^{(2)}}$, or
 \begin{align}
 \log(z)\leq\beta^Tx^{(2)} \label{3.3.5}.
 \end{align}
 (\ref{3.3.5}) should be considered as a natural constraint for any observation with $\delta=2$ during estimation of $\beta$, which will be demonstrated later.
 
 In (\ref{3.3.3}), $D$ contributed to the pdf via $f_D(z|x^{(1)})$, and we did not observe the exact time from $O$ and $C$, but only $O>z$ and $C>z$, therefore the pdf was contributed via $S_O(z)$ and $S_C(z)$, respectively.
 \subsection*{The case of $\delta=3$} 
 For a case with the last observed time being $T$, the observation is $(z,\delta = 3,x)$, and
 \begin{align*}
 P(Z>z, \delta=3|X=x)&=P(T>z, T<C|X=x)\\
 &=\int_z^\infty\int_{t}^\infty f_T(t|x)f_C(c)\,dc\,dt\\
 &=\int_z^\infty f_T(t|x)S_C(t)\,dt.
 \end{align*}
 The pdf of $(Z, \delta=3)$ given covariate $x$ is
 \begin{align}
 f_{Z,\delta|X}(z,3|x)&=-\frac{d}{dz}\int_z^\infty f_T(t|x)S_C(t)\,dt\notag\\
 &=f_T(z|x)S_C(z)\notag\\
 &=h_T(z|x)U_D(z;\tau|x)S_O(z)S_C(z). \label{3.3.4}
 \end{align}
 In (\ref{3.3.4}), $T$ contributed to the pdf via $f_T(z|x)=h_T(z|x)U_D(z;\tau|x)S_O(z)$, and we did not observe the exact time from $C$, but only $C>z$, therefore the pdf was contributed via $S_C(z)$. Note that $h_T(z|x)=h_O(z)+h_D(z|x^{(1)})\cdot I(z\leq \tau_{x^{(2)}})$, the information of $h_O(z)$, which can be obtained through the national death certificate database, must be further included in this pdf.

By incorporating the pdfs from the corresponding levels of $\delta$, we can obtain the likelihood function $L(\alpha,\beta)$ as summarized below
\begin{thm} Given the censored data $\{z_i,\delta_i,x_i\}_{i=1}^n$, where $z_i$ and $x_i$ are the $i$-th last observed time and covariate, respectively, and $\delta_i$ is defined as (\ref{3.5}). Assume (C1) and (C2). The likelihood function $L(\alpha,\beta)$ under CTM is
\begin{align*}
L(\alpha,\beta)=
&\bigg\{\prod_{i:\delta_i=0}U_D(z_i;\tau|x_i)S_O(z_i)f_C(z_i)\bigg\}\times\bigg\{\prod_{i:\delta_i=1}U_D(z_i;\tau|x_i)f_O(z_i)S_C(z_i)\bigg\}\times\notag\\
&\bigg\{\prod_{i:\delta_i=2}f_D(z_i|x^{(1)}_i)S_O(z_i)S_C(z_i)I(z_i\leq\tau_{x^{(2)}_i})\bigg\}\times\bigg\{\prod_{i:\delta_i=3}h_T(z_i|x_i)U_D(z_i;\tau|x_i)S_O(z_i)S_C(z_i)\bigg\}.
\end{align*}
\end{thm}

Let $(\widehat{\alpha},\widehat{\beta})^T$ be the estimate of $(\alpha,\beta)^T$, we obtain $(\widehat{\alpha},\widehat{\beta})^T$ through
\begin{align*}
(\widehat{\alpha},\widehat{\beta})^T=\argmax_{\alpha,\beta}\left\{l(\alpha,\beta)-\frac{n}{2}\kappa\beta^T\beta\right\},
\end{align*}
where $l(\alpha,\beta)=\ln L(\alpha,\beta)$, and $\kappa\geq 0$ is the smoothing parameter to obtain more stable estimation. We suggest $\kappa$ to be related to the sample size $n$, such as $\kappa=1/n$ or $\kappa=1/(\sqrt{n}\log n)$, one can also set $\kappa=0$ to remove the penalty effect. Eliminating those parts independent of $\alpha,\beta$, we obtain
\begin{align}
l(\alpha,\beta)=&\sum_{i:\delta_i=0,1}\log U_D(z_i;\tau|x_i)+\sum_{i:\delta_i=2}\log\{f_D(z_i|x^{(1)}_i)I(z_i\leq\tau_{x^{(2)}_i})\}+\notag\\
&\sum_{i:\delta_i=3}\left\{\log h_T(z_i|x_i)+\log U_D(z_i;\tau|x_i)\right\}, \label{3.2}
\end{align}
Note that the censored case ($\delta = 0$), and the case dying from any cause except the disease of interest ($\delta = 1$), have equally contribution to the objective function $l(\alpha,\beta)$.

\section{Implementation}
We use the gradient descent method for $\alpha$ estimation given fixed $\beta$, which is described in Chapter 3.4.1, and use the gradient projection method to estimate $\beta$ given fixed $\alpha$ in a modified objective function, which is described in Chapter 3.4.2. The above two methods are implemented iteratively until convergence. 

\subsection{Estimation of $\alpha$ given $\beta$}
When $\beta$ is given, $\tau_{x^{(2)}_i}$ is a constant, therefore the observation can be partitioned into $\{z_i:z_i\leq\tau_{x^{(2)}_i}\}$ and $\{z_i:z_i>\tau_{x^{(2)}_i}\}$, the objective function (\ref{3.2}) can be expressed as
\begin{align*}
l_{\beta}(\alpha)=&\sum_{\{i:\delta_i=0,1;\;z_i\leq\tau_{x^{(2)}_i}\}}\log S_D(z_i|x^{(1)}_i)+\sum_{\{i:\delta_i=0,1;\;z_i>\tau_{x^{(2)}_i}\}}\log S_D(\tau_{x^{(2)}_i}|x^{(1)}_i)+\\
&\sum_{\{i:\delta_i=2;\;z_i\leq\tau_{x^{(2)}_i}\}}\log f_D(z_i|x^{(1)}_i)+\sum_{\{i:\delta_i=3;\;z_i\leq\tau_{x^{(2)}_i}\}}\left[\log S_D(z_i|x^{(1)}_i) + \log\left\{h_O(z_i)+h_D(z_i|x^{(1)}_i)\right\}\right]+\\
&\sum_{\{i:\delta_i=3;\;z_i>\tau_{x^{(2)}_i}\}}\bigg\{\log S_D(\tau_{x^{(2)}_i}|x^{(1)}_i) + \log h_O(z_i)\bigg\}
\end{align*}
with derivative
\begin{align*}
\frac{\partial}{\partial\alpha}l_{\beta}(\alpha)=&\sum_{\{i:\delta_i=0,1;\;z_i\leq\tau_{x^{(2)}_i}\}}\{S_D(z_i|x^{(1)}_i)\}^{-1}\frac{\partial}{\partial\alpha}S_D(z_i|x^{(1)}_i)+\\
&\sum_{\{i:\delta_i=0,1;\;z_i>\tau_{x^{(2)}_i}\}}\{S_D(\tau_{x^{(2)}_i}|x^{(1)}_i)\}^{-1}\frac{\partial}{\partial\alpha}S_D(\tau_{x^{(2)}_i}|x^{(1)}_i)+\\
&\sum_{\{i:\delta_i=2;\;z_i\leq\tau_{x^{(2)}_i}\}}\{f_D(z_i|x^{(1)}_i)\}^{-1}\frac{\partial}{\partial\alpha}f_D(z_i|x^{(1)}_i)+\\
&\sum_{\{i:\delta_i=3;\;z_i\leq\tau_{x^{(2)}_i}\}}\left[\{S_D(z_i|x^{(1)}_i)\}^{-1}\frac{\partial}{\partial\alpha}S_D(z_i|x^{(1)}_i)+\right.\\
&\;\;\;\;\;\;\;\;\;\;\;\;\;\;\;\;\;\;\;\;\;\left.\left\{h_O(z_i)+h_D(z_i|x^{(1)}_i)\right\}^{-1}\frac{\partial}{\partial\alpha}\left\{h_O(z_i)+h_D(z_i|x^{(1)}_i)\right\}\right]+\\
&\sum_{\{i:\delta_i=3;\;z_i>\tau_{x^{(2)}_i}\}}\{S_D(\tau_{x^{(2)}_i}|x^{(1)}_i)\}^{-1}\frac{\partial}{\partial\alpha}S_D(\tau_{x^{(2)}_i}|x^{(1)}_i).
\end{align*}
We use the gradient descent method to optimize $l(\alpha, \beta)$ given $\beta$, which is the same as optimizing $l_{\beta}(\alpha)$
\begin{align*}
\widehat{\alpha}(\beta)=\argmax_{\alpha}l_{\beta}(\alpha).
\end{align*}

\subsection{Estimation of $\beta$ given $\alpha$}
As $\alpha=\widehat{\alpha}(\beta)$ is given, $\sum_{\delta_i=2}\log\big\{f_D(z_i|x^{(1)}_i)I\big(z_i\leq\tau_{x^{(2)}_i}\big)\big\}$ in $l(\alpha, \beta)$ implies that for $i$-th observation from $D_i$ ($\delta_i = 2$ if $Z_i=D_i$), the cure time $\tau_{x^{(2)}_i}$ must be larger than $D_i$, therefore $z_i\leq\tau_{x^{(2)}_i}$ is naturally a linear constraint, that should be considered in the optimization of $l(\alpha, \beta)$. Let $l_{\alpha}(\beta)$ be $l(\alpha, \beta)$ given $\alpha$, and without the information of those $\delta_i=2$
\begin{align}
l_{\alpha}(\beta)=&\sum_{i:\delta_i=0,1}\log U_D(z_i;\tau|\alpha, x_i) +\sum_{i:\delta_i=3}\{\log h_T(z_i|\alpha, x_i)+\log U_D(z_i|\alpha, x_i)\}. \label{3.6}
\end{align}
The optimization of $l_{\alpha}(\beta)$ is equivalent to the following optimization problem
\begin{align*}
\begin{array}{lll}
\mbox{maximize }&   l_{\alpha}(\beta)-\frac{n}{2}\kappa\beta^T\beta &\\
\mbox{subject to }& \log(z_i) \leq \beta^Tx^{(2)}_i, & \forall\;i\; \mbox{such that}\;\delta_i = 2.
\end{array}
\end{align*}
Note that, in $l_{\alpha}(\beta)$ the indicator function $I(z_i\leq\tau_{x^{(2)}_i})$ leads to non-differentiation at $\beta$. To deal with this problem, \citeA{Ma:Huang:2007} suggested using the sigmoid function
\begin{align*}
R(u;\sigma_n) = \left\{1+\exp\left(-\frac{u}{\sigma_n}\right)\right\}^{-1}
\end{align*}
to approximate $I\big(u\geq 0\big)$, where the tuning parameter $\sigma_n$ is a sequence of positive numbers satisfying $\lim_{n\rightarrow\infty}\sigma_n=0$. Note that $\lim_{\sigma_n\rightarrow 0}R(u;\sigma_n)=I\big(u\geq 0\big)$.

Denote the modified objective function of (\ref{3.6}), $l_{\alpha, \sigma_n}(\beta)$ as
\begin{align*}
l_{\alpha, \sigma_n}(\beta)=&\sum_{i:\delta_i=0,1}\log U_{\alpha,\sigma_n}(z_i;\tau|x_i)+\sum_{i:\delta_i=3}\big\{\log h_{\alpha,\sigma_n}(z_i|x_i)+\log U_{\alpha,\sigma_n}(z_i|x_i)\big\},
\end{align*}
where
\begin{align*}
\begin{array}{l}
h_{\alpha,\sigma_n}(z_i|x_i)=h_O(z_i)+h_D(z_i|\alpha, x^{(1)}_i)R(\tau_{x^{(2)}_i}-z_i;\sigma_n),\\
U_{\alpha,\sigma_n}(z_i;\tau|x_i)=\{S_D(z_i|\alpha, x^{(1)}_i)-S_D(\tau_{x^{(2)}_i}|\alpha, x^{(1)}_i)\}R(\tau_{x^{(2)}_i}-z_i;\sigma_n)+S_D(\tau_{x^{(2)}_i}|\alpha, x^{(1)}_i).
\end{array}
\end{align*}
All the indicator functions $I(z_i\leq\tau_{x^{(2)}_i})\;\;i\in\{i|\delta_i=0,1,3\}$ in $l(\alpha,\beta)$ are replaced by sigmoid function $R(\tau_{x^{(2)}_i}-z_i;\sigma_n)$, thus the modified objective function $l_{\alpha, \sigma_n}(\beta)$ is differentiable at $\beta$. In $l(\alpha,\beta)$, those observations with $\delta_i=2$ do not contribute to $l_{\alpha, \sigma_n}(\beta)$, but become natural linear constraints for $(\alpha,\beta)^T$.

For a fixed $\alpha$, the gradient projection method \shortcite{Luenberger:Ye:2008} is used to solve the optimization problem
\begin{align*}
\begin{array}{lll}
\mbox{maximize }&   l_{\alpha, \sigma_n}(\beta)-\frac{n}{2}\kappa\beta^T\beta &\\
\mbox{subject to }& \log(z_i) \leq \beta^Tx^{(2)}_i, & \forall\;i\; \mbox{such that}\;\delta_i = 2.
\end{array}
\end{align*}
The derivative of $l_{\alpha, \sigma_n}(\beta)$ is
\begin{align*}
\frac{\partial}{\partial\beta}l_{\alpha, \sigma_n}(\beta)=&\frac{\partial}{\partial\beta}\sum_{i:\delta_i=0,1}\log\big[U_{\alpha, \sigma_n}(z_i;\tau|x_i)\big]+\\
&\frac{\partial}{\partial\beta}\sum_{i:\delta_i=3}\big\{\log\big[U_{\alpha, \sigma_n}(z_i;\tau|x_i)\big]+\log\big[h_{\alpha, \sigma_n}(z_i|x_i)\big]\big\}\\
=&\sum_{i:\delta_i=0,1}\{U_{\alpha, \sigma_n}(z_i;\tau|x_i)\}^{-1}\frac{\partial}{\partial\beta}U_{\alpha, \sigma_n}(z_i;\tau|x_i)+\\
&\;\sum_{i:\delta_i=3}\left[\{U_{\alpha, \sigma_n}(z_i;\tau|x_i)\}^{-1}\frac{\partial}{\partial\beta}U_{\alpha, \sigma_n}(z_i;\tau|x_i)+\{h_s(z_i|x_i)\}^{-1}\frac{\partial}{\partial\beta}h_{\alpha, \sigma_n}(z_i|x_i)\right],
\end{align*}
where
\begin{align*}
\frac{\partial}{\partial\beta}h_{\alpha, \sigma_n}(z_i|x_i)&=h_D(z_i|\alpha, x^{(1)}_i)\frac{\partial}{\partial\beta}R(\tau_{x^{(2)}_i}-z_i;\sigma_n),\\
\frac{\partial}{\partial\beta}U_{\alpha, \sigma_n}(z_i;\tau|x_i)&=-\frac{\partial}{\partial\beta}S_D(\tau_{x^{(2)}_i}|\alpha, x^{(1)}_i)R(\tau_{x^{(2)}_i}-z_i;\sigma_n)+\\
&\;\;\;\;\left\{S_D(z_i|\alpha, x^{(1)}_i)-S_D(\tau_{x^{(2)}_i}|\alpha, x^{(1)}_i)\right\}\frac{\partial}{\partial\beta}R(\tau_{x^{(2)}_i}-z_i;\sigma_n)+\frac{\partial}{\partial\beta}S_D(\tau_{x^{(2)}_i}|\alpha, x^{(1)}_i).
\end{align*}
We use the gradient projection method \shortcite{Luenberger:Ye:2008} to obtain the estimate of $\beta$ given $\alpha$.

\subsection{Estimation of $(\alpha,\beta)^T$}
We use an iterative algorithm below to obtain $(\widehat{\alpha},\widehat{\beta})^T$:
\\
\\
\fbox{\begin{minipage}{40em}
{\bf Estimation algorithm}
\begin{enumerate}
 \item Given an initial vector $\widehat{\beta}^{(0)}$, obtain the first iterative estimate of $\alpha$, $\widehat{\alpha}^{(1)}$, by optimizing $l_{\widehat{\beta}^{(0)}}(\alpha)$ using gradient descent.
 \item Obtain the first iterative estimate of $\beta$, $\widehat{\beta}^{(1)}$, by optimizing $l_{\widehat{\alpha}^{(1)}, \sigma_n}(\beta)$ using gradient projection method.
 \item Repeat step 1. and step 2. until obtaining $(\widehat{\alpha},\widehat{\beta})^T=(\widehat{\alpha}^{(\infty)},\widehat{\beta}^{(\infty)})^T$.
\end{enumerate}
\end{minipage}}
\\
\section{Standard error}
The parametric bootstrap method is used to generate the null distribution and estimate the standard error of $(\widehat{\alpha},\widehat{\beta})^T$.\\
\fbox{\begin{minipage}{40em}
{\bf Parametric bootstrap algorithm for CTM}
\begin{enumerate}
	\item For $i=1,\ldots,n$, let $\delta^{(c)}_i=1$ if $\delta_i = 0$, $\delta^{(c)}_i=0$ if $\delta_i \neq 0$. Fit Weibull model 
	to $\{Z_i, \delta^{(c)}_i\}_{i=1}^n$ to obtain the censoring distribution estimate $\widehat{\alpha}_{c}
	 $.
	\item For $b=1,\ldots,B$, for $i=1,\ldots,n$,
	\begin{enumerate}
		\item Generate $C^{(b)}_i$ from Weibull model with parameter $\widehat{\alpha}_{c}$.
		\item Generate $O^{(b)}_i$ by using vital statistics from government (in Taiwan we have the hazard of each calendar year 1985-2017, each 0-100 years old, and each sex category). Calculate the conditional survival $S_i(t|y_i,a_i,s_i)$ with corresponding diagnosed year $y_i$, age $a_i$, and sex $s_i$. Then draw a random sample $u^{(b)}_i$ from uniform($0,1$), and $S_i^{-1}(u^{(b)}_i|y,a,s)$ is $O^{(b)}_i$.
		\item Generate $D^{(b)}_i$ from $S_D(t|\widehat{\alpha},X_i)$.
		\item Calculate $T^{(b)}_i = \min(O^{(b)}_i, D^{(b)}_i)\cdot I(D^{(b)}_i\leq \widehat{\tau}_i)+O^{(b)}_i\cdot I(D^{(b)}_i> \widehat{\tau}_i)$, where $\widehat{\tau}_i=\exp(\widehat{\beta}^TX_i)$.
		\item If $C^{(b)}_i < T^{(b)}_i$ then $\delta^{(b)}_i = 0$ and $Z^{(b)}_i=C^{(b)}_i$. Otherwise generate $v^{(b)}_i$ from Bernoulli$\left(\frac{\widehat{P}(Z=T)}{\widehat{P}(Z=O)+\widehat{P}(Z=D)+\widehat{P}(Z=T)}\right)$.
		If $v^{(b)}_i=1$ then $\delta^{(b)}_i = 3$ and $Z^{(b)}_i=T^{(b)}_i$. If $v^{(b)}_i=0$ and $T^{(b)}_i = D^{(b)}_i$, then $\delta^{(b)}_i = 2$ and $Z^{(b)}_i=D^{(b)}_i$. If $v^{(b)}_i=0$ and $T^{(b)}_i = O^{(b)}_i$, then $\delta^{(b)}_i = 1$ and $Z^{(b)}_i=O^{(b)}_i$.
	\end{enumerate}
	\item Obtain the bootstrapped estimate $(\widehat{\alpha}^{(b)},\widehat{\beta}^{(b)})^T$ using $\{Z^{(b)}_i,\delta^{(b)}_i,X_i\}_{i=1}^{n}$ and algorithm stated in section 3.4.3.
	\item Obtain standard error by $\{\widehat{\alpha}^{(b)},\widehat{\beta}^{(b)}\}_{b=1}^B$.
\end{enumerate}
\end{minipage}
}
\\
\begin{remark}[$\sigma_n$ selection]
It is convenient to choose a suitable $\sigma_n$ before optimizing $l_{\alpha, \sigma_n}(\beta)$. Let $\sigma_n=n^{-\frac{1}{w}}$, where $w\in\mathbb{R}^+$ can be several candidates. A small $w$ makes $R(u;\sigma_n)$ a better approximation to $I(u\geq 0)$ as $u\rightarrow 0$, but maybe more unstable in differentiation, thus there is a trade-off in selecting an appropriate $\sigma_n$. One can use cross-validation to select $\sigma_n$. However, for convenience one can just subjectively choose one of the candidates mentioned above, since different $\sigma_n$'s give almost the same results in estimation. Here we use $\sigma_n=n^{-\frac{1}{2}}$ in the following simulation chapter and data analysis chapter.
\end{remark}
\begin{remark}
If $\beta_0$ is the only parameter to be estimated in cure time, i.e.,\ $\tau=\beta_0$, then we suggest to obtain the estimate and standard error of $\tau$ using grid search directly. Specifically, let $l_{\alpha}(\beta)$ be
\begin{align*}
l_{\alpha}(\beta)=\sum_{\delta_i=0,1}\log U_D(z_i;\tau|\alpha)+\sum_{i:\delta_i=3}\big\{\log U_D(z_i;\tau|\alpha)+\log h_T(z_i|\alpha)\big\}
\end{align*}
where
\begin{align*}
\begin{array}{l}
h_T(z_i|\alpha)=h_O(z_i)+h_D(z_i|\alpha)I(z_i\leq\tau)\\
U_D(z_i;\tau|\alpha)=\big\{S_D(z_i|\alpha)-S_D(\tau|\alpha)\big\}I(z_i\leq\tau)+S_D(\tau|\alpha).
\end{array}
\end{align*}
We suggest to optimize $l_{\alpha}(\beta)$ using grid search instead of optimizing $l_{\alpha, \sigma_n}(\beta)$. Note that in this case the estimation does not involve the sigmoid function approximation.
\end{remark}

\chapter{Simulation Studies}
In this chapter, we conduct simulation to evaluate our proposed method under three simulation studies. In (S1), four datasets with different distributions are used to validate the methodology from different aspects. (S2) focuses on demonstrating an application issue in the sense of mislabelling within $\delta$ ($O$ misbabel to $D$, or vice versa), which makes the simulated dataset much closer to the real world. (S3) shows the robustness of the cure time estimation against the misspecification of the model for $S_D(t)$. Although one can use different covariates in modelling $S_D(t)$ and $\tau$, respectively, it is natural for a practitioner to use the same covariate $X$ to describe the behavior of $D$ and the cure time. Therefore, we use the same covariate $X$ to model all parameters (i.e.,\ $X^{(1)} = X^{(2)}=X$) in our simulation studies. For each setting, we generate 200 datasets, each with sample size $n=500$. The covariate is $X=(X_0, X_1, X_2)^T$, where $X_0$ is set to be 1 for the intercept, and $(X_1, X_2)^T$ is generated from the normal distribution with mean vector $\mathbf{0}$ and covariance matrix
\begin{align*}
\Sigma = \left(
\begin{array}{cc}
1 & 0.5 \\
0.5 & 1 \\
\end{array}
\right).
\end{align*}
Conditional on $X$, $D$ is generated from the Weibull distribution with the shape parameter $\exp(\alpha_{1}^TX)$, where $\alpha_{1} = (\alpha_{10},\alpha_{11},\alpha_{12})^T$, and the scale parameter $\exp(\alpha_{2}^TX)$, where $\alpha_{2} = (\alpha_{20} ,\alpha_{21} ,\alpha_{22})^T$. The cure time parameter $\tau$ is modelled as $\tau = \exp(\beta^TX)$, where $\beta = (\beta_0, \beta_1, \beta_2)^T$. The life table of the general population in Taiwan is used to generate $O$. $C$ is generated from the Weibull distribution to achieve different censoring rates. For each setting, we calculate mean and standard deviation (SD) of the estimates, and obtain the standard error (SE) and the square root mean squared error (SMSE), from 200 bootstrapped samples. For convenience we subjectively choose $\sigma_n = n^{-\frac{1}{2}}$, since different $\sigma_n$'s give similar results.

\section{Simulation results under (S1)}
In (S1), we evaluate the behaviors of the proposed method under different combinations of $q_C=P(Z=C)$, $q_O=P(Z=O)$, $q_D=P(Z=D)$, and $q_T=P(Z=T)$. (S1)-1 represents the ideal situation in which we know exactly if $Z=D$ or $Z=O$ (i.e.\ $q_T=0$) for each individual without censoring. Note that in real world it is a rare case that a dataset contains almost no censored sample, and it is also rare that we know exactly if $Z=D$ or $Z=O$ for each patient, due to the difficulty in identifying the underlying cause of death for all patients. Therefore the information of cause of death in death certificate may contain the garbage codes, which motivates the usage of $\delta=3$. (S1)-2 is the same as (S1)-1 except that $q_C$ increases. (S1)-3 is the same as (S1)-1, except that $\alpha_{20} = 3.912$, such that $q_D$ becomes smaller than (S1)-1. Unlike those settings with $q_T=0$, in (S1)-4 we set the $q_T$ to be higher. Table \ref{table:II} reports the true parameter values (True), mean of estimates (Mean), SD, and SE.

{\scriptsize
\begin{center}
\begin{threeparttable}[h]
\caption{Simulation results of (S1) under different settings for $\delta$.}
\label{table:II}
\begin{tabular}{@{}ccrrrrrrrrr@{}}
\toprule
\multirow{7}{.8cm}{(S1)-1} & \multicolumn{10}{c}{$(q_C, q_O, q_D, q_T)=(1\%, 41\%, 58\%, 0\%)$} \\																				
\cmidrule(l){2-11}																				
&	Parameter     	&	$\alpha_{10}$	&	$\alpha_{11}$	&	$\alpha_{12}$	&	$\alpha_{20}$	&	$\alpha_{21}$	&	$\alpha_{22}$	&	$\beta_{0}$	&	$\beta_{1}$	&	$\beta_{2}$	\\
\cmidrule(l){2-11}																				
&	True	&	-0.693	&	-0.110	&	0.040	&	1.946	&	0.130	&	0.070	&	1.792	&	0.250	&	-0.300	\\
&	Mean	&	-0.679	&	-0.105	&	0.042	&	1.926	&	0.130	&	0.060	&	1.755	&	0.250	&	-0.305	\\
&	SD	&	0.051	&	0.065	&	0.065	&	0.127	&	0.146	&	0.141	&	0.024	&	0.040	&	0.036	\\
&	SE	&	0.063	&	0.074	&	0.074	&	0.145	&	0.166	&	0.168	&	0.021	&	0.039	&	0.040	\\
\toprule																				
\multirow{7}{.8cm}{(S1)-2} & \multicolumn{10}{c}{$(q_C, q_O, q_D, q_T)=(34\%, 11\%, 55\%, 0\%)$} \\																				
\cmidrule(l){2-11}																				
&	Parameter     	&	$\alpha_{10}$	&	$\alpha_{11}$	&	$\alpha_{12}$	&	$\alpha_{20}$	&	$\alpha_{21}$	&	$\alpha_{22}$	&	$\beta_{0}$	&	$\beta_{1}$	&	$\beta_{2}$	\\
\cmidrule(l){2-11}																				
&	True	&	-0.693	&	-0.110	&	0.040	&	1.946	&	0.130	&	0.070	&	1.792	&	0.250	&	-0.300	\\
&	Mean	&	-0.677	&	-0.112	&	0.040	&	1.935	&	0.155	&	0.065	&	1.743	&	0.346	&	-0.299	\\
&	SD	&	0.054	&	0.068	&	0.070	&	0.136	&	0.159	&	0.151	&	0.029	&	0.060	&	0.055	\\
&	SE	&	0.065	&	0.075	&	0.076	&	0.154	&	0.174	&	0.175	&	0.032	&	0.057	&	0.057	\\
\toprule																				
\multirow{7}{.8cm}{(S1)-3} & \multicolumn{10}{c}{$(q_C, q_O, q_D, q_T)=(3\%, 69\%, 28\%, 0\%)$} \\																				
\cmidrule(l){2-11}																				
&	Parameter     	&	$\alpha_{10}$	&	$\alpha_{11}$	&	$\alpha_{12}$	&	$\alpha_{20}$	&	$\alpha_{21}$	&	$\alpha_{22}$	&	$\beta_{0}$	&	$\beta_{1}$	&	$\beta_{2}$	\\
\cmidrule(l){2-11}																				
&	True	&	-0.693	&	-0.110	&	0.040	&	3.912	&	0.130	&	0.070	&	1.792	&	0.250	&	-0.300	\\
&	Mean	&	-0.668	&	-0.105	&	0.032	&	3.861	&	0.116	&	0.108	&	1.741	&	0.243	&	-0.292	\\
&	SD	&	0.082	&	0.090	&	0.097	&	0.255	&	0.292	&	0.303	&	0.033	&	0.071	&	0.065	\\
&	SE	&	0.098	&	0.110	&	0.110	&	0.309	&	0.333	&	0.336	&	0.033	&	0.060	&	0.060	\\
\toprule																				
\multirow{7}{.8cm}{(S1)-4} & \multicolumn{10}{c}{$(q_C, q_O, q_D, q_T)=(2\%, 12\%, 17\%, 69\%)$} \\																				
\cmidrule(l){2-11}																				
&	Parameter     	&	$\alpha_{10}$	&	$\alpha_{11}$	&	$\alpha_{12}$	&	$\alpha_{20}$	&	$\alpha_{21}$	&	$\alpha_{22}$	&	$\beta_{0}$	&	$\beta_{1}$	&	$\beta_{2}$	\\
\cmidrule(l){2-11}																				
&	True	&	-0.693	&	-0.110	&	0.040	&	1.946	&	0.130	&	0.070	&	1.792	&	0.250	&	-0.300	\\
&	Mean	&	-0.660	&	-0.106	&	0.037	&	1.904	&	0.108	&	0.068	&	1.762	&	0.251	&	-0.298	\\
&	SD	&	0.054	&	0.073	&	0.067	&	0.126	&	0.147	&	0.144	&	0.030	&	0.036	&	0.038	\\
&	SE	&	0.065	&	0.076	&	0.076	&	0.146	&	0.165	&	0.166	&	0.029	&	0.039	&	0.040	\\

\bottomrule
\end{tabular}
\end{threeparttable}
\end{center}
}
Since (S1)-1 is the ideal situation, we can obtain correct estimates, and the bootstrapped SE is similar to the SD for all parameters. The main difference between (S1)-1 and (S1)-2 is $q_C$ and $q_O$, In (S1)-1, $q_C = 1\%$ and $q_O = 41\%$; in (S1)-2, $q_C = 34\%$ and $q_O = 11\%$. The SD and the SE in (S1)-2 are all slightly larger than that of the corresponding estimates in (S1)-1. It makes sense that the higher SD exists in a data containing more censored cases.

The main difference between (S1)-1 and (S1)-3 is $q_O$ and $q_D$, In (S1)-1, $q_O = 41\%$ and $q_D = 58\%$; in (S1)-3, $q_O = 69\%$ and $q_D = 28\%$. The smaller $q_D$ in (S1)-3 means fewer information of constraints contribute to the estimation process, thus less efficient estimates are obtained. Therefore, the SD and the SE in (S1)-3 are all larger than that of the corresponding estimates in (S1)-1.

(S1)-4 aims to mimic the situation that most of the exact statuses for $(O,D)$ are not available, where one can only observe $Z=T$ for most of the uncensored subjects. The SD and the SE in (S1)-4 are slightly larger than that of the corresponding estimates in (S1)-1. Note that the information used in estimation is quite different between (S1)-1 and (S1)-4. The estimation even works in a $(O,D)$ unclear dataset, with similar SD between (S1)-1 and (S1)-4. Also note that the similar estimation between (S1)-1 and (S1)-4 may be resulted from that (S1)-1 naturally contains higher percentage of $D$. The comparison with lower percentage of $D$ will be further demonstrated in study-(S2).

\section{Simulation results under (S2)}
In this simulation, we show a situation in which the status of $O$ is partly mislabelling to $D$ and vice versa, which implies a poor cause of death quality. To avoid using wrong information of $(O,D)$, we arbitrarily set a portion of $(O,D)$ to be $T$. We also arbitrarily set all $(O,D)$ to be $T$ to see the robustness of our method.

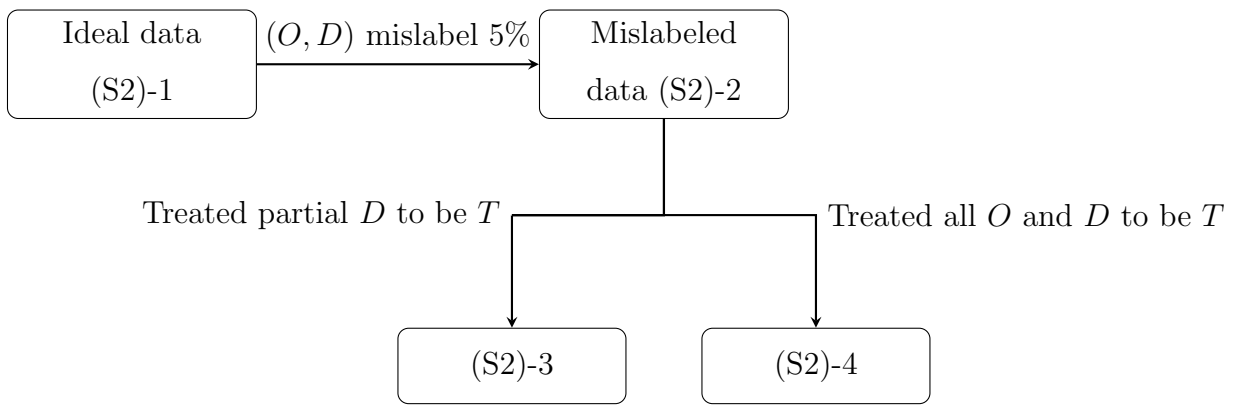
\begin{figure}[h]
\centering
\begin{tikzpicture}[node distance=4cm]
\node (mislabel1) [mislabel1] {Ideal data (S2)-1};
\node (mislabel2) [mislabel2, right of = mislabel1, xshift=3cm] {Mislabeled data (S2)-2};
\node (mislabel3) [mislabel3, below of = mislabel2, xshift=-2cm] {(S2)-3};
\node (mislabel4) [mislabel4, below of = mislabel2, xshift=2cm] {(S2)-4};
\draw [arrow] (mislabel1) -- node[anchor=south] {$(O,D)$ mislabel 5\%} (mislabel2);
\draw [arrow] (mislabel2) |- (5,-2) -| node[anchor=east] {Treated partial $D$ to be $T$} (mislabel3);
\draw [arrow] (mislabel2) |- (5,-2) -| node[anchor=west] {Treated all $O$ and $D$ to be $T$} (mislabel4);
\end{tikzpicture}
\caption{$(O,D)$ mislabelling simulation design}
\label{f4.1}
\end{figure}

(S2)-1 represents the ideal case where the exact status for $(O,D)$ are available. (S2)-2 mimics the situation where some status of $(O,D)$ are wrongly identified, where the proportions that $\delta=2$ is mislabelled as $\delta=1$ and $\delta=1$ is mislabelled as $\delta=2$ are in total 5\%. (S2)-3 uses the same data with (S2)-2, except that 50\% of subjects with $\delta=2$ are treated as $\delta=3$ to enter the analysis. (S2)-4 adopts the same strategy with (S2)-3 but all uncensored subjects are coded as $\delta=3$. Since coding $\delta=3$ will not be affected by the mislabelling between $(O,D)$, we would expect a better analysis result of (S2)-3 and (S2)-4 than (S2)-2 in the presence of mislabelling. Figure \ref{f4.1} summarizes the relationships between different settings. Simulation results are reported in Table \ref{table:III}.
\\

{\scriptsize
\begin{center}
\begin{threeparttable}[h]
\caption{Mislabelling $(O,D)$ simulation result.}
\label{table:III}
\begin{tabular}{@{}ccrrrrrrrrr@{}}
\toprule
\multirow{7}{.8cm}{(S2)-1} & \multicolumn{10}{c}{$(q_C, q_O, q_D, q_T)=(2\%, 57\%, 41\%, 0\%)$,  mislabelling rate = 0\%} \\																				
\cmidrule(l){2-11}																				
&	Parameter     	&	$\alpha_{10}$	&	$\alpha_{11}$	&	$\alpha_{12}$	&	$\alpha_{20}$	&	$\alpha_{21}$	&	$\alpha_{22}$	&	$\beta_{0}$	&	$\beta_{1}$	&	$\beta_{2}$	\\
\cmidrule(l){2-11}																				
&	True	&	-0.693	&	-0.110	&	0.040	&	3.401	&	0.300	&	-0.400	&	2.303	&	0.250	&	-0.300	\\
&	Mean	&	-0.663	&	-0.112	&	0.046	&	3.338	&	0.293	&	-0.408	&	2.256	&	0.245	&	-0.293	\\
&	SD	&	0.066	&	0.078	&	0.079	&	0.172	&	0.199	&	0.202	&	0.031	&	0.051	&	0.048	\\
&	SMSE	&	0.073	&	0.078	&	0.079	&	0.183	&	0.199	&	0.202	&	0.056	&	0.051	&	0.049	\\
\toprule																				
\multirow{7}{.8cm}{(S2)-2} & \multicolumn{10}{c}{$(q_C, q_O, q_D, q_T)=(2\%, 57\%, 41\%, 0\%)$,  mislabelling rate = 5\%} \\																				
\cmidrule(l){2-11}																				
&	Parameter     	&	$\alpha_{10}$	&	$\alpha_{11}$	&	$\alpha_{12}$	&	$\alpha_{20}$	&	$\alpha_{21}$	&	$\alpha_{22}$	&	$\beta_{0}$	&	$\beta_{1}$	&	$\beta_{2}$	\\
\cmidrule(l){2-11}																				
&	True	&	-0.693	&	-0.110	&	0.040	&	3.401	&	0.300	&	-0.400	&	2.303	&	0.250	&	-0.300	\\
&	Mean	&	-0.967	&	-0.020	&	-0.037	&	4.589	&	0.000	&	-0.120	&	4.099	&	0.042	&	-0.013	\\
&	SD	&	0.058	&	0.073	&	0.081	&	0.253	&	0.351	&	0.368	&	0.250	&	0.539	&	0.515	\\
&	SMSE	&	0.280	&	0.116	&	0.111	&	1.214	&	0.461	&	0.462	&	1.814	&	0.576	&	0.589	\\
\toprule																				
\multirow{7}{.8cm}{(S2)-3} & \multicolumn{10}{c}{$(q_C, q_O, q_D, q_T)=(2\%, 57\%, 20.5\%, 20.5\%)$,  mislabelling rate = 5\%} \\																				
\cmidrule(l){2-11}																				
&	Parameter     	&	$\alpha_{10}$	&	$\alpha_{11}$	&	$\alpha_{12}$	&	$\alpha_{20}$	&	$\alpha_{21}$	&	$\alpha_{22}$	&	$\beta_{0}$	&	$\beta_{1}$	&	$\beta_{2}$	\\
\cmidrule(l){2-11}																				
&	True	&	-0.693	&	-0.110	&	0.040	&	3.401	&	0.300	&	-0.400	&	2.303	&	0.250	&	-0.300	\\
&	Mean	&	-0.755	&	-0.093	&	0.023	&	3.854	&	0.258	&	-0.353	&	2.412	&	0.131	&	-0.121	\\
&	SD	&	0.134	&	0.089	&	0.093	&	0.537	&	0.316	&	0.320	&	0.735	&	0.283	&	0.311	\\
&	SMSE	&	0.147	&	0.090	&	0.094	&	0.701	&	0.318	&	0.323	&	0.742	&	0.307	&	0.359	\\
\toprule																				
\multirow{7}{.8cm}{(S2)-4} & \multicolumn{10}{c}{$(q_C, q_O, q_D, q_T)=(2\%, 0\%, 0\%, 98\%)$,  mislabelling rate = 5\%} \\																				
\cmidrule(l){2-11}																				
&	Parameter     	&	$\alpha_{10}$	&	$\alpha_{11}$	&	$\alpha_{12}$	&	$\alpha_{20}$	&	$\alpha_{21}$	&	$\alpha_{22}$	&	$\beta_{0}$	&	$\beta_{1}$	&	$\beta_{2}$	\\
\cmidrule(l){2-11}																				
&	True	&	-0.693	&	-0.110	&	0.040	&	3.401	&	0.300	&	-0.400	&	2.303	&	0.250	&	-0.300	\\
&	Mean	&	-0.586	&	-0.093	&	0.013	&	3.168	&	0.202	&	-0.284	&	2.505	&	0.212	&	-0.320	\\
&	SD	&	0.078	&	0.089	&	0.087	&	0.165	&	0.210	&	0.201	&	0.636	&	0.230	&	0.261	\\
&	SMSE	&	0.133	&	0.091	&	0.091	&	0.286	&	0.231	&	0.232	&	0.666	&	0.232	&	0.261	\\

\bottomrule
\end{tabular}
\end{threeparttable}
\end{center}
}

Obviously, (S2)-2 shows worse estimation result than (S2)-1 because of the mislabelling. Moreover, mislabelling affects a lot even when the mislabelling rate is small. In the estimation of (S2)-2, the mislabelling rate 5\% causes much more bias and much higher SD of all estimates than (S2)-1. Both (S2)-3 and (S2)-4 give more accurate estimates than that of (S2)-2 even if there exists mislabelling information. (S2)-4 performs better than (S2)-3 from the perspective of SMSE, since (S2)-3 can still affected by mislabelling, while (S2)-4 does not. It implies that in real applications, one is suggested to set the status of an unclear cause of death to be $\delta=3$ ($Z=T$) to avoid poor estimation.

\section{Simulation results under (S3)}
In this simulation study, we show the robustness of the cure time estimation when $S_D(t)$ was misspecified. In order to see how bias affect the cure time estimate, we estimate $\tau$ by using both the correct distribution (Weibull distribution), and incorrect distribution (log-normal distribution) to model the distribution of $D$. Note that in this simulation, the shape parameter of Weibull distribution corresponds to increasing hazards. However, the log-normal distribution has the limitation to model an increasing hazard, therefore it is obvious that the misspecifying distribution of $D$ would lead to a poor estimation.

$D$ is generated from Weibull distribution such that $q_O > q_D$, which means that this simulated patient population is more likely to death from $O$ (general cause) than $D$ (disease). That is, we simulate mild disease patient population. In (S3)-1 we use the almost uncensored data; in (S3)-2 we use the same data as (S3)-1, but $(O,D)$ is converted to $T$ (denoted by $(O,D)\rightarrow T$); in (S3)-3 we use the censored data; in (S3)-4 weuse the same data as (S3)-3, but $(O,D)$ is all converted to $T$. Simulation results are reported in Table \ref{table:IV}.

{\scriptsize
\begin{center}
\begin{threeparttable}[h]
\caption{Influence of distribution misspecification on mild disease patient population.}
\label{table:IV}
\begin{tabular}{@{}lcrrrrrrr@{}}
\toprule
\multicolumn{2}{c}{$S_D(t)$}		&	\multicolumn{3}{c}{Weibull (correct)}					&		&	\multicolumn{3}{c}{Log-normal (incorrect)}\\
\midrule
\multicolumn{2}{c}{$\tau = \exp(\beta^TX)$}   	&	$\beta_{0}$	&	$\beta_{1}$	&	$\beta_{2}$	&		&	$\beta_{0}$	&	$\beta_{1}$	&	$\beta_{2}$	\\
\midrule
\multirow{5}{.8cm}{(S3)-1} & \multicolumn{8}{c}{$(q_C, q_O, q_D, q_T)=(1\%, 41\%, 58\%, 0\%)$}\\
\cmidrule(l){2-9}
&	True	&	2.303	&	0.250	&	-0.300	&		&	2.303	&	0.250	&	-0.300	\\
&	Mean	&	2.289	&	0.249	&	-0.298	&		&	2.290	&	0.253	&	-0.302	\\
&	SD	&	0.008	&	0.015	&	0.016	&		&	0.009	&	0.017	&	0.016	\\
&	SMSE	&	0.016	&	0.015	&	0.016	&		&	0.015	&	0.017	&	0.017	\\
\toprule
\multirow{5}{.8cm}{(S3)-2} & \multicolumn{8}{c}{$(q_C, q_O, q_D, q_T)=(1\%, 0\%, 0\%, 99\%)$}\\
\cmidrule(l){2-9}
&	True	&	2.303	&	0.250	&	-0.300	&		&	2.303	&	0.250	&	-0.300	\\
&	Mean	&	2.296	&	0.249	&	-0.299	&		&	2.304	&	0.255	&	-0.306	\\
&	SD	&	0.015	&	0.021	&	0.022	&		&	0.023	&	0.028	&	0.026	\\
&	SMSE	&	0.016	&	0.021	&	0.022	&		&	0.023	&	0.028	&	0.027	\\
\toprule
\multirow{5}{.8cm}{(S3)-3} & \multicolumn{8}{c}{$(q_C, q_O, q_D, q_T)=(41\%, 16\%, 43\%, 0\%)$}\\
\cmidrule(l){2-9}
&	True	&	2.303	&	0.250	&	-0.300	&		&	2.303	&	0.250	&	-0.300	\\
&	Mean	&	2.271	&	0.249	&	-0.298	&		&	2.274	&	0.254	&	-0.304	\\
&	SD	&	0.027	&	0.043	&	0.044	&		&	0.030	&	0.043	&	0.047	\\
&	SMSE	&	0.042	&	0.043	&	0.044	&		&	0.041	&	0.043	&	0.047	\\
\toprule
\multirow{5}{.8cm}{(S3)-4} & \multicolumn{8}{c}{$(q_C, q_O, q_D, q_T)=(41\%, 0\%, 0\%, 59\%)$}\\
\cmidrule(l){2-9}
&	True	&	2.303	&	0.250	&	-0.300	&		&	2.303	&	0.250	&	-0.300	\\
&	Mean	&	2.287	&	0.250	&	-0.301	&		&	2.307	&	0.248	&	-0.305	\\
&	SD	&	0.048	&	0.051	&	0.057	&		&	0.067	&	0.063	&	0.064	\\
&	SMSE	&	0.050	&	0.051	&	0.057	&		&	0.067	&	0.063	&	0.064	\\
\bottomrule
\end{tabular}
\end{threeparttable}
\end{center}
}

In Table \ref{table:IV}, we can see that the correct (Weibull) and incorrect (log-normal) distribution specification have similar cure time SMSE in all settings. The estimation of completed data ((S3)-1 and (S3)-2) have smaller SMSE than that of censored data ((S3)-3 and (S3)-4), respectively. The estimation of data without $(O,D)\rightarrow T$ ((S3)-1 and (S3)-3) have smaller SMSE than that with $(O,D)\rightarrow T$ ((S3)-2 and (S3)-4), respectively. (S3) shows the robustness on the estimation of cure time when distribution is misspecified. This property ensures the correctness of cure time estimation even if a wrong parametric distribution is specified for $S_D(t)$.
\newpage

\chapter{Data Analysis}
\section{Cure time estimation of 22 major cancers in Taiwan}
We analyze the cancer data from the National Cancer Registry Database of the Taiwan Cancer Registry, in which all patients were diagnosed between 2004 and 2015, and follow-up to 2016. Taiwan Cancer Registry database is good for population-based cancer survival analysis because of both high qualities of cancer registry database and death certificate database.

In order to conduct a reliable estimation of the cure time, we suggest the following steps, where we use colorectal cancer as an example to illustrate the analysis procedure.
\begin{enumerate}
 \item Draw $S_T(t|c)$ and $S_O(t|c),\forall\;t>c$ in some subjectively chosen time points such as $c\in \{1,3,5,\ldots\}$ to check if statistical cure exists. According to Figure \ref{f7.1}, we can see that Figure \ref{f7.1}(\subref{f7.1f}) satisfies statistical cure \eqref{2.1}, that is, $S_T(t|10) = S_O(t|10)\;\;\forall\;t> 10$. It is suggested that the cure time exists and is between 7 years and 10 years. Therefore we can use CTM to estimate the cure time.
\begin{figure}[h]
\centering
\begin{subfigure}{.3\textwidth}
  \centering
  \includegraphics[width=\textwidth]{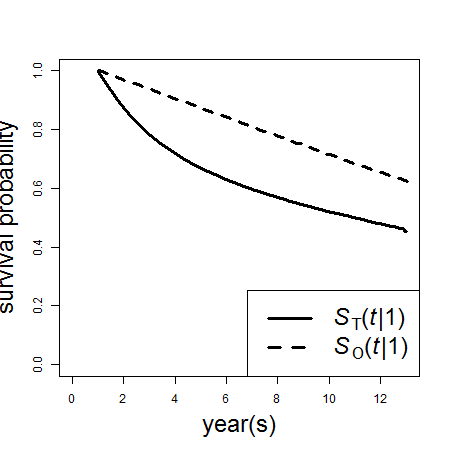}
  \caption{$k=1$}
  \label{f7.1b}
\end{subfigure}%
\begin{subfigure}{.3\textwidth}
  \centering
  \includegraphics[width=\textwidth]{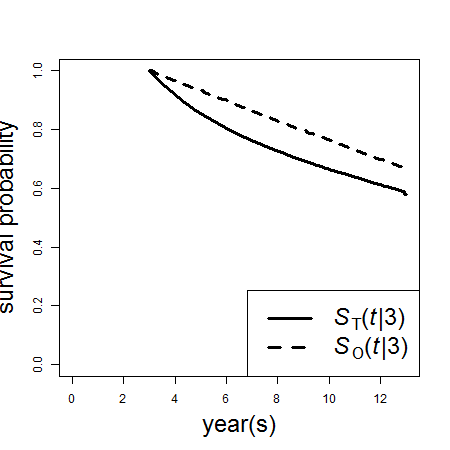}
  \caption{$k=3$}
  \label{f7.1c}
\end{subfigure}%
\begin{subfigure}{.3\textwidth}
  \centering
  \includegraphics[width=\textwidth]{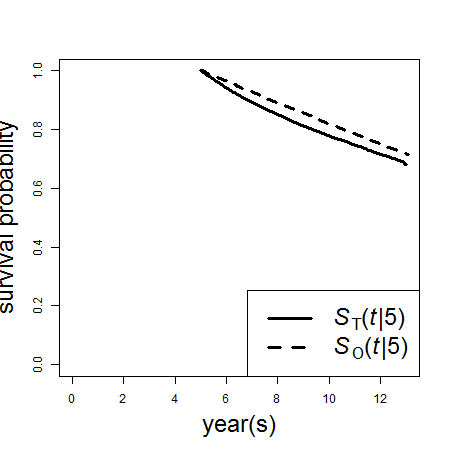}
  \caption{$k=5$}
  \label{f7.1d}
\end{subfigure}\\
\begin{subfigure}{.3\textwidth}
  \centering
  \includegraphics[width=\textwidth]{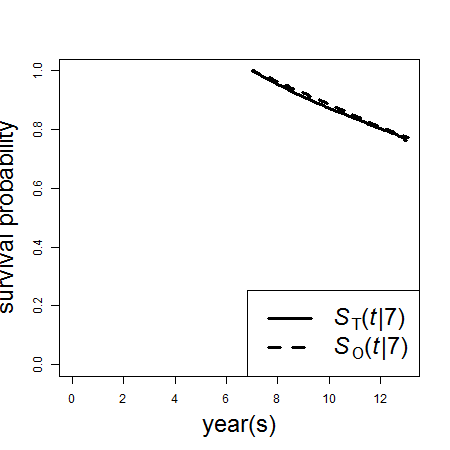}
  \caption{$k=7$}
  \label{f7.1e}
\end{subfigure}%
\begin{subfigure}{.3\textwidth}
  \centering
  \includegraphics[width=\textwidth]{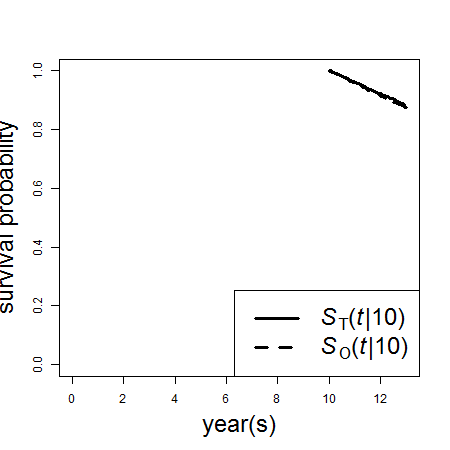}
  \caption{$k=10$}
  \label{f7.1f}
\end{subfigure}%
\begin{subfigure}{.3\textwidth}
  \centering
  \includegraphics[width=\textwidth]{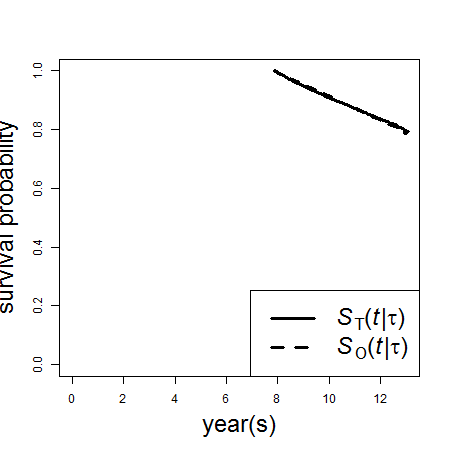}
  \caption{$k=\widehat{\tau}$}
  \label{f8}
\end{subfigure}
\caption{$S_T(t|k)$ and $S_O(t|k)$ of colorectal cancer data and general population in Taiwan, for $k=1$, $3$, $5$, $7$, $10$, and $\widehat{\tau}$ years.}
\label{f7.1}
\end{figure}
 \item Estimate cure time $\tau$ through the cure time model. We use Weibull model, log-normal model, and log-logistic model, to model $D$. After obtaining the estimated cure time of the corresponding three parametric distributions, we choose the estimated cure time with the largest log-likelihood value among the three parametric models. The estimated cure time of, for example, colorectal cancer, is $\widehat{\tau}=7.85$ years from fitting a log-normal model.
 \item Draw $S_T(t|\widehat{\tau})$ and $S_O(t|\widehat{\tau})$ to check if $\widehat{\tau}$ satisfies the definition of statistical cure: $S_T(t|\tau)=S_O(t|\tau), \forall\;t>\tau$. From Figure \ref{f7.1}(\subref{f8}) we can see that 7.85 years seems to be a reasonable estimate, since it satisfies the definition of statistical cure, that is, $S_T(t|7.85) = S_O(t|7.85)\;\;\forall\;t> 7.85$.
\end{enumerate}

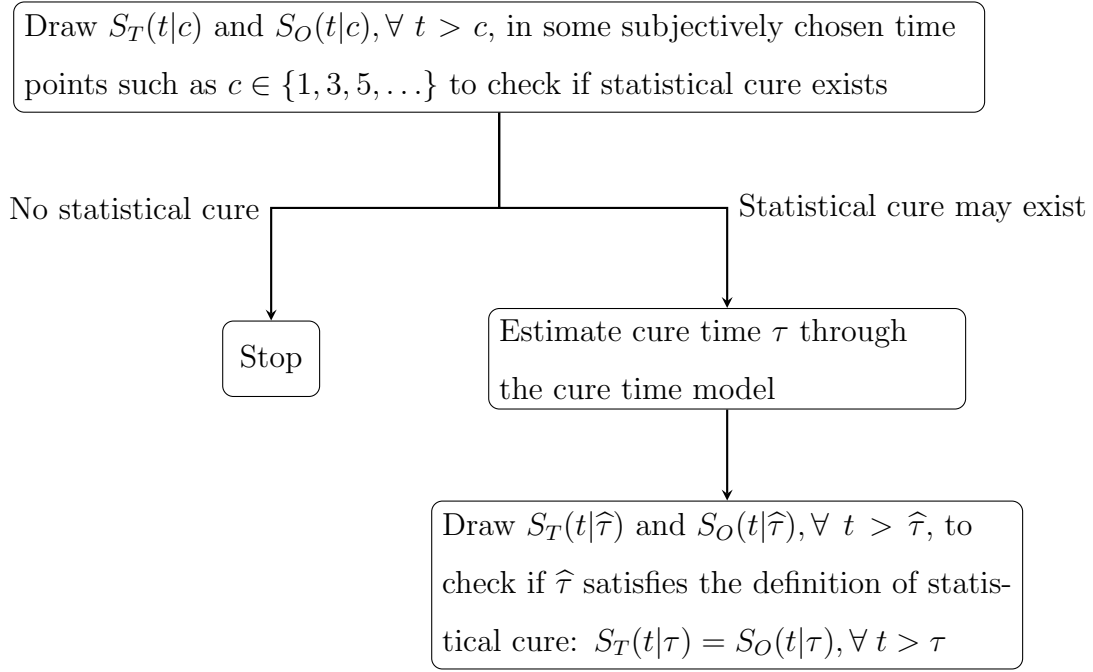
\begin{figure}[h]
\centering
\begin{tikzpicture}[node distance=4cm]
\node (check1a) [check1a] {Draw $S_T(t|c)$ and $S_O(t|c),\forall\;t>c$, in some subjectively chosen time points such as $c\in \{1,3,5,\ldots\}$ to check if statistical cure exists};
\node (check1b1) [check1b1, below of = check1a, xshift=-3cm] {Stop};
\node (check1b2) [check1b2, right of = check1b1, xshift=2cm] {Estimate cure time $\tau$ through the cure time model};
\node (check1c) [check1c, below of = check1b2, yshift=1cm] {Draw $S_T(t|\widehat{\tau})$ and $S_O(t|\widehat{\tau}), \forall\;t>\widehat{\tau}$, to check if $\widehat{\tau}$ satisfies the definition of statistical cure: $S_T(t|\tau)=S_O(t|\tau), \forall\;t>\tau$};
\draw [arrow] (check1a) |- (-3,-2) -| node[anchor=east] {No statistical cure} (check1b1);
\draw [arrow] (check1a) |- (3,-2) -| node[anchor=west] {Statistical cure may exist} (check1b2);
\draw [arrow] (check1b2) -- (check1c);
\end{tikzpicture}
\caption{Cure time estimation flow chart}
\label{f6}
\end{figure}

The above steps are summarized in Figure \ref{f6}.
\\

In Global Burden of Disease \shortcite{GBD:2015}, the cure time is roughly defined as 10 years for all diseases. In The Burden of Cancer New Zealand 2006 \shortcite{Blakely.etal:2010}, the cure time is defined based on visually identification through the relative survival curves. Using our proposed method, we calculated cure time on 22 cancer sites and compare with the result from New Zealand. We also calculate cure time using a convenient method from \citeA{Janssen-Heijnen.etal:2007, Janssen-Heijnen.etal:2010, Dal.etal:2014}, which is the smallest time point such that the conditional relative survival exceeds 95\% (denoted by CRS95) or 99\% (denoted by CRS99). The results are shown in Table \ref{table:VI}.

\begin{figure}
\centering
\begin{subfigure}{.3\textwidth}
  \centering
  \includegraphics[width=\textwidth]{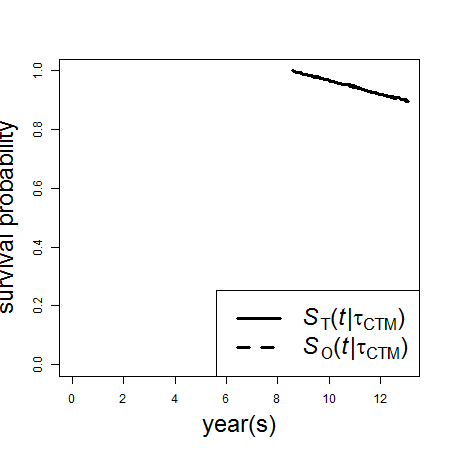}
  \caption{$\widehat{\tau}=8.58$ from CTM}
  \label{f10a}
\end{subfigure}%
\begin{subfigure}{.3\textwidth}
  \centering
  \includegraphics[width=\textwidth]{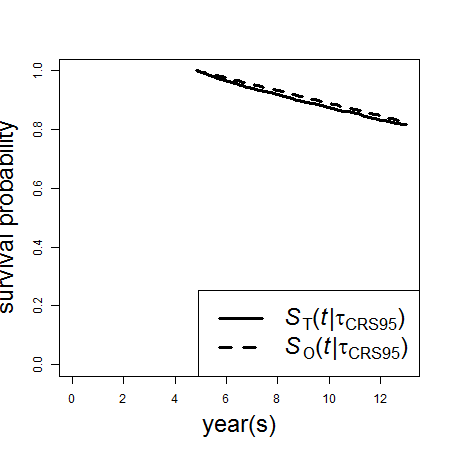}
  \caption{$\widehat{\tau}=4.84$ from CRS95}
  \label{f10b}
\end{subfigure}\\
\begin{subfigure}{.3\textwidth}
  \centering
  \includegraphics[width=\textwidth]{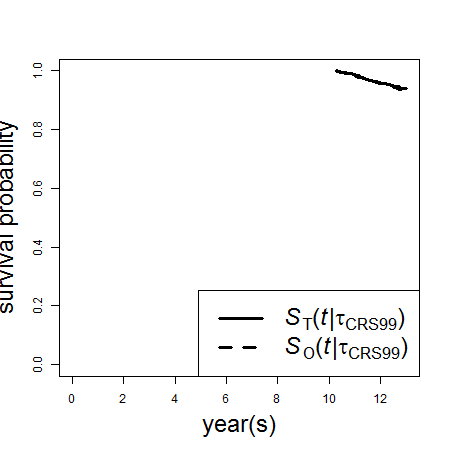}
  \caption{$\widehat{\tau}=10.28$ from CRS99}
  \label{f10c}
\end{subfigure}%
\begin{subfigure}{.3\textwidth}
  \centering
  \includegraphics[width=\textwidth]{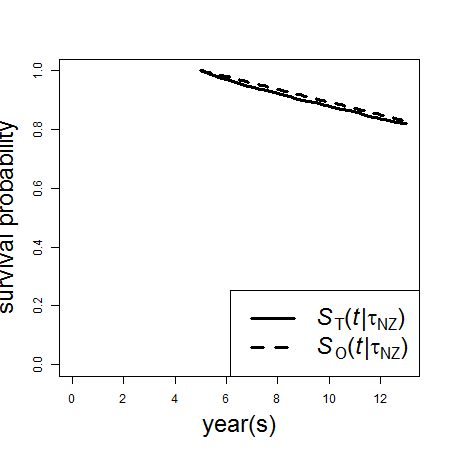}
  \caption{$\widehat{\tau}=5$ from New Zealand}
  \label{f10d}
\end{subfigure}\\
	\begin{subfigure}{.3\textwidth}
  \centering
  \includegraphics[width=\textwidth]{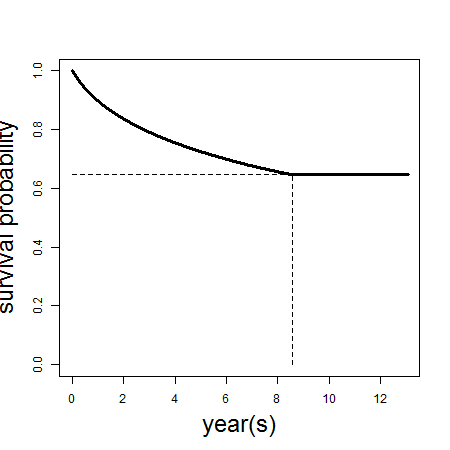}
  \caption{CTM-based net survival}
  \label{f10e}
\end{subfigure}%
\begin{subfigure}{.3\textwidth}
  \centering
  \includegraphics[width=\textwidth]{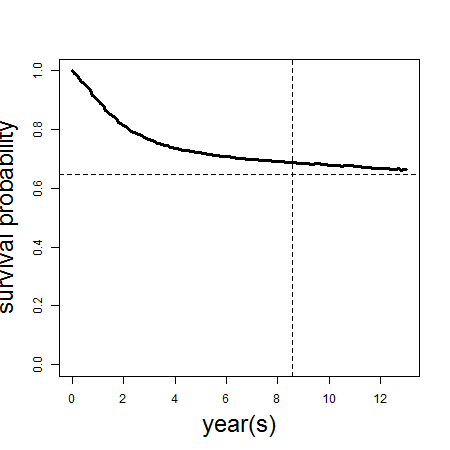}
  \caption{Relative survival}
  \label{f10f}
\end{subfigure}
\caption{$S_T(t|k)$ and $S_O(t|k)$ of cervical cancer data and general population in Taiwan, where $k$ is cure time estimated from CTM, CRS95, CRS99, and New Zealand (2006). (\ref{f10a}-\ref{f10d}). Model-based net survival and relative survival (\ref{f10e}, \ref{f10f}), horizontal and vertical dashed lines represent locations of CTM-estimated cure time and cure rate, respectively.}
\label{f10}
\end{figure}

From Table \ref{table:VI}, the cancer sites gallbladder, kidney and other urinary, larynx, leukaemia, oesophagus, ovary, and testis have the estimated cure time similar to both the CRS method CRS95 and CRS99. The rest 15 cancer sites, however, have large differences comparing to the results of CRS95 and CRS99. Among these sites, the objectively estimated cure time from data may be more reasonable than either the CRS method, or visually identification using relative survival curve. Although \citeA{Janssen-Heijnen.etal:2007, Janssen-Heijnen.etal:2010} suggested to use 95\% and 99\% as the threshold, the CRS method is not reliable since the threshold is determined subjectively. Figure \ref{f10} shows the conditional survival from cervical cancer with estimated cure time from different methods. We can see that the estimated cure time from CRS95 (4.84 years) (Figure \ref{f10}(\subref{f10b})) and from \citeA{Blakely.etal:2010} (5 years) (Figure \ref{f10}(\subref{f10d})) obviously do not attain statistical cure. The estimated cure time from CTM (8.58 years) (Figure \ref{f10}(\subref{f10a})) and from CRS99 (10.28 years) (Figure \ref{f10}(\subref{f10c})) both attain the statistical cure, but CTM gives a more reasonable and smaller estimate than CRS99, which is, by Definition 1, the smallest time point satisfying \eqref{2.1}. We show the same graphical judgement from Figure \ref{fA1_Bladder} to Figure \ref{fA1_Uterus} for the rest cancer sites. 

We further observe that the differences between the CTM estimated cure time and the last observed time are all less than 1 year among kidney and other urinary, liver, oesophagus, and ovary, which implies that we may not obtain stable cure time estimate until those corresponding follow-up time are long enough. One can still calculate these cure time, but we do not recommend using these results in application, since all we know about the statistical cure information of these cancer sites is that the cure time are larger than the last observed time.

{\scriptsize
\begin{center}
\begin{threeparttable}[h]
\caption{Statistical cure time of 22 cancer sites in Taiwan, diagnosed between 2004 and 2015, follow-up to 2016.}
\label{table:VI}
\begin{tabular}{@{}lrrrrrrr@{}}
\toprule
\multirow{4}{*}{Cancer type} & \multirow{4}{*}{Case Number} & \multirow{4}{1.3cm}{Follow-up (years)} & \multicolumn{4}{c}{Cure time (years)} & \multirow{4}{*}{Cure rate (se)}\\
\cmidrule(l){4-7}
& & & \multicolumn{3}{c}{Taiwan} & \multirow{2}{1.7cm}{New Zealand (years)} & \\
\cmidrule(l){4-6}
& & & CTM (se) & CRS95\tnote{a} & CRS99\tnote{b} & & \\
\midrule
Bladder	&	14,210 	&	9.65	&	6.56	(0.029)	&	6.68	&	8.88	&	10	&	0.56	(0.005)	\\
Bone and connective	&	471 	&	11.78	&	5.36	(0.881)	&	7.65	&	7.65	&	10	&	0.66	(0.047)	\\
Breast (female)\tnote{1}	&	100,845 	&	13.00	&	12.15\tnote{*}	(0.020)	&	7.67	&	11.58	&	20	&	0.75	(0.002)	\\
Cervix	&	18,625 	&	13.00	&	8.58	(0.018)	&	4.84	&	10.28	&	5	&	0.65	(0.005)	\\
Colorectum\tnote{2}	&	126,309 	&	13.00	&	7.85	(0.099)	&	5.07	&	7.97	&	8	&	0.54	(0.003)	\\
Gallbladder	&	1,101 	&	6.83	&	5.74	(0.479)	&	5.74	&	5.74	&	7	&	0.16	(0.023)	\\
Hodgkin	&	1,298 	&	12.12	&	7.16	(1.250)	&	3.02	&	7.05	&	10	&	0.79	(0.024)	\\
Kidney and other urinary	&	3,352 	&	6.96	&	7.00\tnote{*}	(0.535)	&	6.48	&	6.48	&	10	&	0.52	(0.025)	\\
Larynx	&	2,549 	&	7.98	&	4.27	(0.521)	&	3.64	&	4.06	&	10	&	0.63	(0.023)	\\
Leukaemia	&	13,109 	&	12.50	&	8.01	(1.064)	&	8.01	&	8.01	&	10	&	0.37	(0.018)	\\
Lip mouth pharynx\tnote{6}	&	80,256 	&	13.00	&	7.66	(0.005)	&	9.50	&	12.18	&	10	&	0.45	(0.002)	\\
Liver\tnote{5}	&	116,843 	&	13.00	&	12.01\tnote{*}	(0.061)	&	10.68	&	12.27	&	7	&	0.18	(0.001)	\\
Lung trachea bronchus\tnote{3}	&	112,862 	&	13.00	&	6.90	(0.003)	&	9.08	&	11.38	&	6	&	0.14	(0.001)	\\
Non-Hodgkin lymphoma	&	14,052 	&	12.96	&	6.00	(0.592)	&	12.86	&	12.86	&	20	&	0.58	(0.012)	\\
Oesophagus	&	17,379 	&	9.00	&	7.33	(0.060)	&	8.02	&	8.45	&	6	&	0.09	(0.002)	\\
Ovary	&	8,263 	&	7.99	&	7.51\tnote{*}	(0.249)	&	6.34	&	7.50	&	10	&	0.57	(0.011)	\\
Pancreas	&	2,674 	&	7.71	&	3.52	(0.315)	&	6.39	&	6.39	&	5	&	0.11	(0.010)	\\
Prostate\tnote{4}	&	32,779 	&	9.00	&	7.31	(0.022)	&	2.87	&	5.38	&	20	&	0.72	(0.004)	\\
Stomach\tnote{9}	&	26,954 	&	9.46	&	5.36	(0.007)	&	5.79	&	8.33	&	6	&	0.33	(0.003)	\\
Testis	&	294 	&	7.32	&	4.02	(0.537)	&	4.02	&	4.02	&	3	&	0.88	(0.034)	\\
Thyroid\tnote{8}	&	4,032 	&	7.52	&	6.31	(0.671)	&	1.46	&	6.31	&	5	&	0.94	(0.006)	\\
Uterus\tnote{7}	&	13,005 	&	12.15	&	5.76	(0.116)	&	2.77	&	5.37	&	6	&	0.80	(0.005)	\\
\bottomrule
\end{tabular}
\begin{tablenotes}
\item[a,b] The smallest time point such that the conditional relative survival exceeds 95\% (CRS95) or 99\% (CRS99) \shortcite{Janssen-Heijnen.etal:2007, Janssen-Heijnen.etal:2010, Dal.etal:2014}
\item[1-9] 1st to 9th major cancers in Taiwan Cancer Registry Annual Report 2016.
\item[*]No statistical cure. The estimated cure times, which are mostly close to the last observed time, are still shown in the table.
\end{tablenotes}
\end{threeparttable}
\end{center}
}
\section{Taiwan colorectal cancer data analysis}
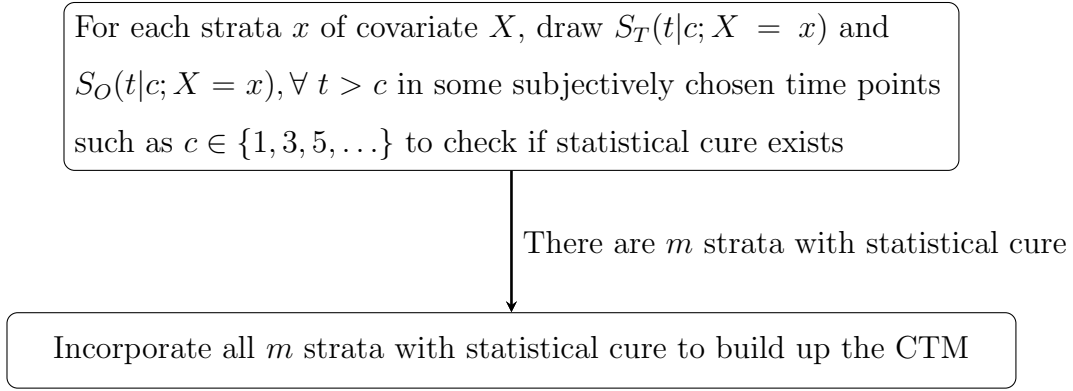
\begin{figure}[h]
\centering
\begin{tikzpicture}[node distance=3.5cm]
\node (check2a) [check2a] {For each strata $x$ of covariate $X$, draw $S_T(t|c;X=x)$ and $S_O(t|c;X=x),\forall\;t>c$ in some subjectively chosen time points such as $c\in \{1,3,5,\ldots\}$ to check if statistical cure exists};
\node (check2b) [check2b, below of = check2a] {Incorporate all $m$ strata with statistical cure to build up the CTM};
\draw [arrow] (check2a) -- (check2b);
\draw [arrow] (check2a) -- node[anchor=west] {There are $m$ strata with statistical cure} (check2b);
\end{tikzpicture}
\caption{CTM estimation flow chart}
\label{f9}
\end{figure}

In population-based studies, although it is enough for one to apply the method described in Section 5.1 to obtain cure time estimate, some drawbacks should be noted. First, one may obtain not reliable cure time estimate if the strata has few amount of patients. Second, the statistical inference related to the cure time comparison may be hard to conduct. In this section, we aim to reveal usage of CTM by incorporating covariates. In order to make sure that the CTM can be used with covariates, we suggest the following steps, and using colorectal cancer as an example:
\begin{enumerate}
 \item For each strata $x$ of covariate $X$, draw $S_T(t|c;X=x)$ and $S_O(t|c;X=x),\forall\;t>c$ in some subjectively chosen time points such as $c\in \{1,3,5,\ldots\}$ to check if statistical cure exists. In this example, $X = (\mbox{Age, Sex, Stage})$ and the strata is $(\mbox{Age, Sex, Stage}) = (\mbox{60-69, Male, II})$. According to Figure \ref{f9.1}, we can see that Figure \ref{f9.1}(\subref{f9.1e}) and Figure \ref{f9.1}(\subref{f9.1f}) satisfies statistical cure \eqref{2.1} visually.
 \newpage
\begin{figure}[h]
\centering
\begin{subfigure}{.3\textwidth}
  \centering
  \includegraphics[width=\textwidth]{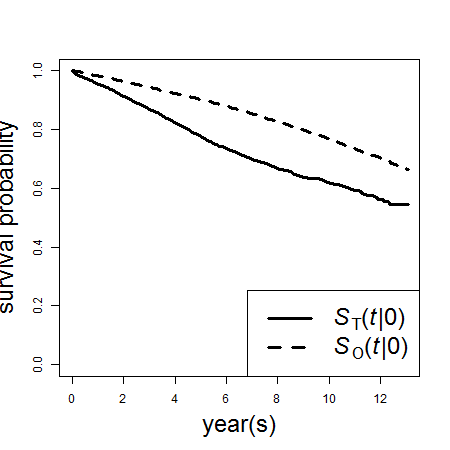}
  \caption{$k=0$}
  \label{f9.1a}
\end{subfigure}%
\begin{subfigure}{.3\textwidth}
  \centering
  \includegraphics[width=\textwidth]{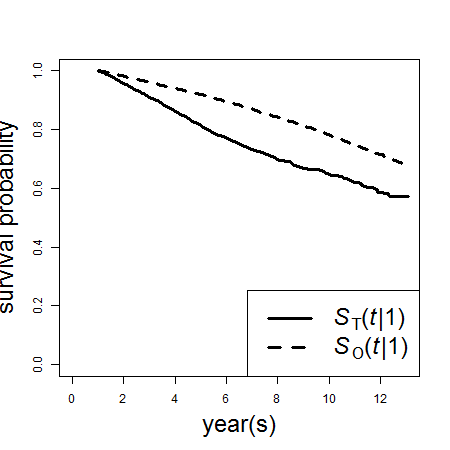}
  \caption{$k=1$}
  \label{f9.1b}
\end{subfigure}%
\begin{subfigure}{.3\textwidth}
  \centering
  \includegraphics[width=\textwidth]{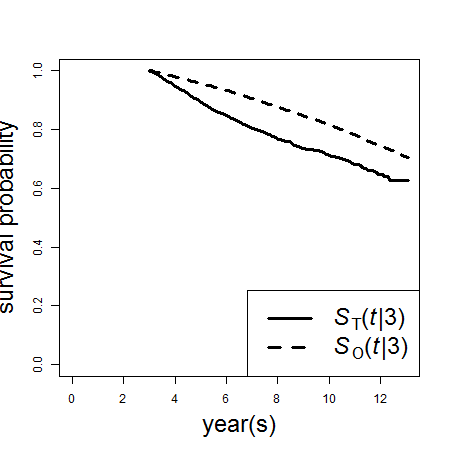}
  \caption{$k=3$}
  \label{f9.1c}
\end{subfigure}\\
\begin{subfigure}{.3\textwidth}
  \centering
  \includegraphics[width=\textwidth]{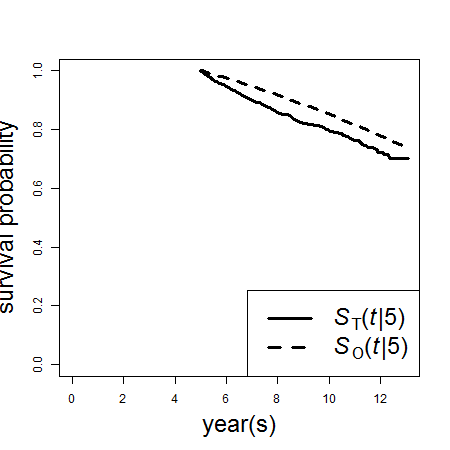}
  \caption{$k=5$}
  \label{f9.1d}
\end{subfigure}%
\begin{subfigure}{.3\textwidth}
  \centering
  \includegraphics[width=\textwidth]{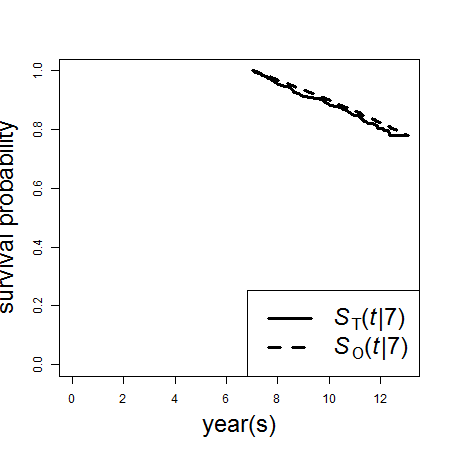}
  \caption{$k=7$}
  \label{f9.1e}
\end{subfigure}%
\begin{subfigure}{.3\textwidth}
  \centering
  \includegraphics[width=\textwidth]{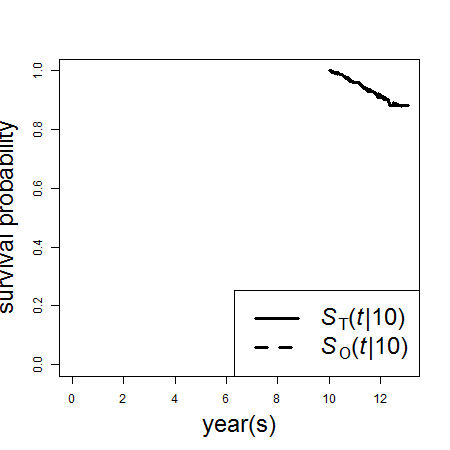}
  \caption{$k=10$}
  \label{f9.1f}
\end{subfigure}
\caption{$S_T(t|k)$ and $S_O(t|k)$ of 60-69 years old, male, stage II colorectal cancer data and corresponding general population in Taiwan, for $k=0$, 1, 3, 5, 7, and 10 years.}
\label{f9.1}
\end{figure}
 \item Suppose there are $m$ strata with statistical cure, we then incorporate these $m$ strata to build up the cure time model.
\end{enumerate}
The steps above are summarized in Figure \ref{f9}.
\newpage

We use the process stated on Figure \ref{f9}. There is no statistical cure being observed on all strata containing age group 80+ and stage IV, therefore we exclude all strata containing age group 80+ and stage IV. One can imagine that there is no statistical cure in older or late stage patient population.

{\scriptsize
\begin{center}
\begin{threeparttable}[h]
\caption{Cure time model on colorectal cancer of Taiwan Cancer Registry.}
\label{table:VII}
\begin{tabular}{@{}crrrrrrr@{}}
\toprule
\multicolumn{8}{c}{$(q_C, q_O, q_D, q_T)=(73\%, 4\%, 16\%, 7\%)$} \\																	
\midrule																	
\multirow{2.6}{*}{} & \multirow{2.6}{*}{Intercept}		&	 Sex	&	 \multicolumn{3}{c}{Age}	&	 \multicolumn{2}{c}{Stage}	\\
\cmidrule(l){3-3} \cmidrule(l){4-6} \cmidrule(l){7-8}
		&		&	 Male	&	 50-59	&	 60-69	&	 70-79	&	  II	&	 III	\\
\midrule																	
$\mu = \alpha_1^TX$&$\alpha_{10}$	&	$\alpha_{11}$ &	$\alpha_{12}$	&	$\alpha_{13}$	&	$\alpha_{14}$	&	$\alpha_{15}$	&	$\alpha_{16}$ \\
\midrule
	Estimate	&	3.9190	&	-0.3123	&	0.1252	&	-0.1120	&	-0.9219	&	-0.3702	&	-1.1455	\\
	SE	&	0.0405	&	0.0286	&	0.0470	&	0.0433	&	0.0379	&	0.0396	&	0.0360	\\
	p-value	&	$<0.0001$	&	$<0.0001$	&	0.0040	&	$<0.0001$	&	$<0.0001$	&	$<0.0001$	&	$<0.0001$	\\
\toprule																	
$\sigma = \alpha_2^TX$	&	$\alpha_{20}$	&	$\alpha_{21}$ &	$\alpha_{22}$	&	$\alpha_{23}$	&	$\alpha_{24}$	&	$\alpha_{25}$	&	$\alpha_{26}$ \\
\midrule
	Estimate	&	1.6166	&	-0.0090	&	0.1443	&	0.2115	&	0.2304	&	0.0800	&	-0.1152	\\
	SE	&	0.0312	&	0.0232	&	0.0348	&	0.0344	&	0.0310	&	0.0310	&	0.0280	\\
	p-value	&	$<0.0001$	&	0.2560	&	$<0.0001$	&	$<0.0001$	&	$<0.0001$	&	$<0.0001$	&	0.0120	\\
\toprule																	
$\tau = \exp(\beta^TX)$	&	$\beta_{0}$	&	$\beta_{1}$ &	$\beta_{2}$	&	$\beta_{3}$	&	$\beta_{4}$	&	$\beta_{5}$	&	$\beta_{6}$ \\
\midrule
	Estimate	&	2.2404	&	-0.0012	&	0.0060	&	0.0021	&	0.0024	&	-0.0278	&	-0.0371	\\
	SE	&	0.0054	&	0.0027	&	0.0046	&	0.0040	&	0.0040	&	0.0051	&	0.0048	\\
	p-value	&	$<0.0001$	&	0.7040	&	0.1560	&	0.4720	&	0.3880	&	$<0.0001$	&	$<0.0001$	\\
\bottomrule
\end{tabular}
\end{threeparttable}
\end{center}
}

After excluding those patients with age group 80+ or stage IV, 75,944 colorectal cancer patients in Taiwan are included in the analysis. The covariate 
\begin{align*}
X = \{\mbox{Sex (female is reference), Age group (50- is reference), Stage (stage I is reference)}\}
\end{align*}
is used to build up the CTM. Log-normal distribution is used to model $D$ with parameter $\mu = \alpha_1^TX$ and $\sigma = \alpha_2^TX$, where $\alpha_1 = \{\alpha_{10}, \alpha_{11}, \ldots, \alpha_{16}\}$, and $\alpha_2 = \{\alpha_{20}, \alpha_{21}, \ldots, \alpha_{26}\}$. The standard error (SE) and two-sided p-value are obtained from 500 bootstrapping.

Table \ref{table:VII} shows the estimation result. Note that both sex and age group show non-significant effect relative to cure time, and the cure time of both stage II and stage III are significant lower than the cure time of stage I. With fixed sex and age group, a stage I colorectal cancer patient has the estimated cure time $\exp(2.2404) = 9.3971$ years, the estimated cure time is $\exp(2.2404-0.0278) = 9.1394$ years for stage II patient, and the estimated cure time is $\exp(2.2404-0.0371) = 9.0548$ years for stage III patient. Those statistical cure stage III patients may be distinguishable from uncure patients earlier than that of stage II patients, and those statistical cure stage II patients may be distinguishable from uncure patients earlier than that of stage I patients. This example provides a typical application guide in both National Burden of Disease studies and population-based studies.

\chapter{Discussion}
For decades, cure rate can be obtained by the appropriate cure rate model to help public health policy making. Cure time, however, was obtained by visually identifying the time point of non-declination in the non-parametric net survival curve. This study enables us to obtain cure time estimate with solid statistical properties. In this study, we propose a new definition of statistical cure in (\ref{2.1}), and develop a parametric method CTM to estimate the cure time.

The cure time model has several good properties, and we investigate these properties through simulations. (S1) shows that the CTM estimation works in highly censored data; (S2) shows the robustness of CTM by estimating the cure time from mislabelled data sets; (S3) shows the robustness of CTM against the misspecification of $S_D(.)$

In application, practitioners can obtain the estimate of cure time through the proposed CTM. Covariates can also be involved in the cure time model with a well-prepared data in which each strata has a cure time. 

To estimate cure time, we propose a random variable representation in Theorem 2, that is
\begin{align*}
 T=\min(O,D)\cdot I(D\leq \tau)+O\cdot I(D>\tau)
\end{align*}
with cure time $\tau$ embedded in $T$. Note that the conventional mixture cure rate model has a similar form
\begin{align*}
 T=\min(O,D)\cdot I(R=1)+O\cdot I(R=0),
\end{align*}
and so does non-mixture cure rate model
\begin{align*}
 T=\min(O,D)\cdot I(N>0)+O\cdot I(N=0),
\end{align*}
where $R\sim\mbox{Bernoulli}(\pi)$ and $N\sim\mbox{Poisson}(\lambda)$, and $\pi=P(N=0)=e^{-\lambda}$. The advantage of the proposed random variable representation (\ref{2.2}) is that one can obtain both cure time $\tau$ and cure rate $P(D>\tau)$, while the conventional cure rate model can only obtain the cure rate.

The cause of death information usage issue is important. In the previous population-based methodologies it was suggested using $T$ and ignore death certificate information ($O,D$) completely \shortcite{Howlader.etal:2010, Huang.etal:2014}. In this study we derive the likelihood (Theorem 3) that allows the usage of partial or full death certificate information, to help obtain more efficient estimate. In Taiwan, we have high quality of death certificate system, the ignorance of this information does not make sense. The conventional approaches that ignore all death certificate information are just a special case in our perspective of likelihood function derivation. Using this concept, researchers may improve the conventional cure rate model with $(O,D)$ involving in likelihood function.

Although the proposed methodology enable us to obtain the cure time point estimate, we should keep in mind that there are assumptions and limitations. Since the cure time model can be used if the cure time is assumed to be in the model, we have to use graphical check as a diagnostics tool stated in Chapter 5. Although we show consistency and robustness in the simulation setting, so far we do not derive the asymptotics and robustness properties of cure time estimate. Moreover, it is time consuming in obtaining bootstrapped standard error, the computational cost is non-negligible, especially in large population size data analysis. The above limitations are all important to be dealt with in future direction.

\bibliographystyle{apacite}
\bibliography{paper}
\appendix
\chapter{Proof of Lemma 1}
\begin{proof}
Derive $S(t)$ directly, we have
\begin{align*}
S(t)&=P(T>t|R=1)P(R=1)+P(T>t|R=0)P(R=0)\\
&=P(\min(O, D)>t|R=1)\pi+P(\min(O, D)>t|R=0)(1-\pi)\\
&=P(O>t|R=1)\pi+P(\min(O, D)>t)(1-\pi)\\
&=P(O>t|R=1)\pi+P(O>t|R=0)P(D>t|R=0)(1-\pi)\\
&=P(O>t)\pi+P(O>t)P(D>t|R=0)(1-\pi)\\
&=P(O>t)\left[\pi+(1-\pi)P(D>t|R=0)\right]\\
&=S_O(t)\left[\pi+(1-\pi)S_u(t)\right].
\end{align*}
\end{proof}
\chapter{Proof of Lemma 2}
\begin{proof}
Deriving directly, we have
{\normalsize
\begin{align*}
S(t)&=P(T>t|N=0)P(N=0)+\sum^\infty_{n=1}P(T>t|N=n)P(N=n)\\
&=P(\min(O, D)>t|N=0)P(N=0)+\sum^\infty_{n=1}P(\min(O, D)>t|N=n)P(N=n)\\
&=P(O>t|N=0)P(N=0)+\sum^\infty_{n=1}P(O>t|N=n)P(D>t|N=n)P(N=n)\\
&=P(O>t)P(N=0)+\sum^\infty_{n=1}P(O>t)P(D>t|N=n)P(N=n)\\
&=P(O>t)\left[P(N=0)+\sum^\infty_{n=1}P(\min{(D_{1}, \ldots, D_{n})}>t)P(N=n)\right]\\
&=P(O>t)\left[P(N=0)+\sum^\infty_{n=1}[1-F_0(t)]^nP(N=n)\right]\\
&=P(O>t)\sum^\infty_{n=0}[1-F_0(t)]^n\frac{\lambda^ne^{-\lambda}}{n!}\\
&=P(O>t)\frac{e^{-\lambda}}{e^{-[\lambda (1-F_0(t))]}}\sum^\infty_{n=0}\frac{[(1-F_0(t))\lambda]^ne^{-[(1-F_0(t))\lambda]}}{n!}\\
&=P(O>t)\frac{e^{-\lambda}}{e^{-[\lambda (1-F_0(t))]}}=P(O>t)e^{-\lambda F_0(t)}=S_O(t)\pi^{F_0(t)}.
\end{align*}
}
\end{proof}
\chapter{Proof of Theorem 1}
\begin{proof}
By \eqref{1.1}, the left hand side of \eqref{2.1} can be expressed as
\begin{align*}
S(t|\tau)=\frac{S_O(t)S_D(t)}{S_O(\tau)S_D(\tau)}\;\;\forall\;t\geq\tau.
\end{align*}
And the right hand side of \eqref{2.1} is
\begin{align*}
S_O(t|\tau)=\frac{S_O(t)}{S_O(\tau)}\;\;\forall\;t\geq\tau.
\end{align*}
After some simplification, we have
\begin{align*}
S_D(t)=S_D(\tau)\;\;\forall\;t\geq\tau.
\end{align*}
\end{proof}
\chapter{Proof of Theorem 2}
\begin{proof}
We first show that \eqref{2.1} implies Theorem 2(b). Since $S(t|\tau) = \exp\left\{-\left[H(t)-H(\tau)\right]\right\}$, \eqref{2.1} can also be expressed as 
\begin{align}
H(t)-H(\tau)= H_O(t)-H_O(\tau). \label{proof:1}
\end{align}
Considering the assumption of independence between $O$ and $D$ \eqref{1.1}, the left hand side of \eqref{proof:1} can be expressed as
\begin{align*}
H_O(t)+H_D(t)-H_O(\tau)-H_D(\tau)= H_O(t)-H_O(\tau),
\end{align*}
which implies that
\begin{align*}
H_D(t)-H_D(\tau)=\int_{\tau}^t h_D(u)du= 0.
\end{align*}
Since $h_D(t)\geq 0\;\;\forall\;t$, we must have $h_D(t) = 0\;\;\forall\;t > \tau$.
Thus Theorem 2(b) follows.

We next show that Theorem 2(b) implies \eqref{2.1}. Integrate both sides of Theorem 2(b) gives
\begin{align*}
H(t) = H_O(t)+H_D(\tau)\;\;\forall\;t > \tau,
\end{align*}
and we have
\begin{align}
S(t) = S_O(t)S_D(\tau)\;\;\forall\;t > \tau. \label{proof:2}
\end{align}
Divide left sides of \eqref{proof:2} by $S(\tau)$, and right side of \eqref{proof:2} by $S_O(\tau)S_D(\tau)$, we get \eqref{2.1}.

Then we show that Theorem 2(a) implies \eqref{2.1}.
Under \eqref{1.1}, the survival function of $T$ is
\begin{align*}
S(t)=P(T>t) &= P(\min(O,D)\cdot I(D\leq \tau)+O\cdot I(D>\tau) > t)\\
&=P(\min(O,D)>t, D\leq \tau) + P(O>t, D>\tau)\\
&=P(O>t)\left[P(t<D\leq\tau)+P(D>\tau)\right]\\
&=S_O(t)\left\{\left[S_D(t)-S_D(\tau)\right]\cdot I(t\leq\tau)+S_D(\tau)\right\}.
\end{align*}
Then we have
\begin{align*}
S(t|\tau)=\frac{S(t)}{S(\tau)}=\frac{S_O(t)\left\{\left[S_D(t)-S_D(\tau)\right]\cdot I(t\leq\tau)+S_D(\tau)\right\}}{S_O(t)S_D(\tau)}=S_O(t|\tau)\;\;\forall t>\tau.
\end{align*}
\eqref{2.1} follows.

Finally we show that Theorem 2(b) implies Theorem 2(a). Under \eqref{1.1}, Theorem 2(b) can be expressed as
\begin{align*}
S(t)&=S_O(t)\left[S_D(t)I(t\leq\tau)+S_D(\tau)I(t>\tau)\right]\\
&=S_O(t)\left\{\left[S_D(t)-S_D(\tau)\right]\cdot I(t\leq\tau)+S_D(\tau)\right\}\\
&=P(O>t)P(t<D<\tau)+P(O>t)P(D>\tau)\\
&=P(O>t, t<D<\tau)+P(O>t, D>\tau)\\
&=P(\min(O,D)\cdot I(D\leq \tau)+O\cdot I(D>\tau) > t).
\end{align*}
Thus Theorem 2(a) follows.
\end{proof}
\chapter{Figures of $S_T(t|k)$ and $S_O(t|k)$ of 21 major cancers in Taiwan}

\begin{figure}
\centering
\begin{subfigure}{.3\textwidth}
  \centering
  \includegraphics[width=\textwidth]{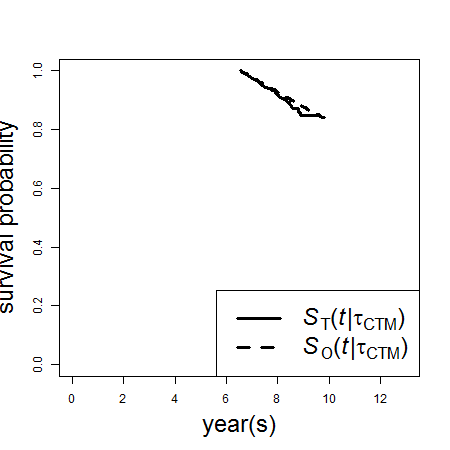}
  \caption{$\widehat{\tau}=6.56$ from CTM}
  \label{fA1_Bladder_CTM}
\end{subfigure}%
\begin{subfigure}{.3\textwidth}
  \centering
  \includegraphics[width=\textwidth]{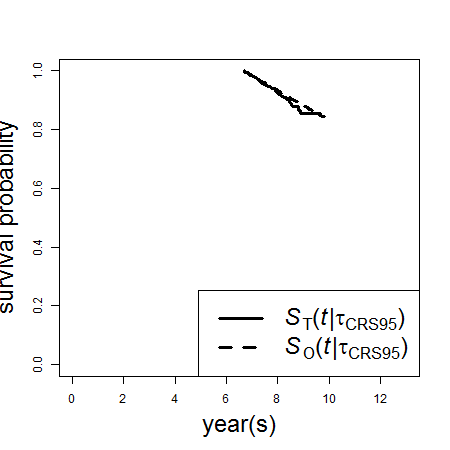}
  \caption{$\widehat{\tau}=6.68$ from CRS95}
  \label{fA1_Bladder_CRS95}
\end{subfigure}\\
\begin{subfigure}{.3\textwidth}
  \centering
  \includegraphics[width=\textwidth]{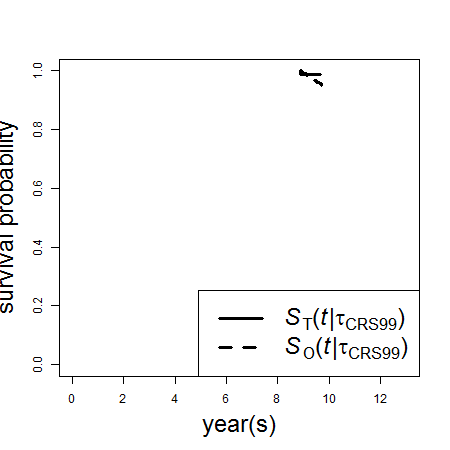}
  \caption{$\widehat{\tau}=8.88$ from CRS99}
  \label{fA1_Bladder_CRS99}
\end{subfigure}%
\begin{subfigure}{.3\textwidth}
  \centering
  \includegraphics[width=\textwidth]{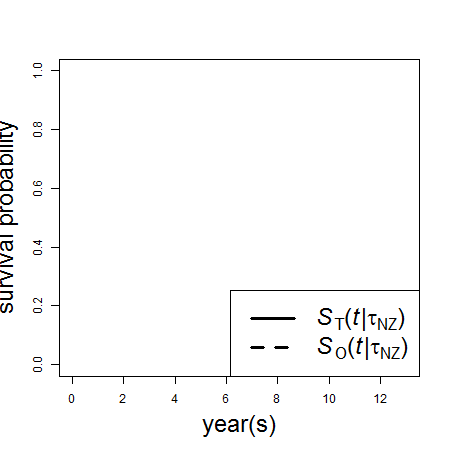}
  \caption{$\widehat{\tau}=10$ from New Zealand}
  \label{fA1_Bladder_NZ2006}
\end{subfigure}\\
\begin{subfigure}{.3\textwidth}
  \centering
  \includegraphics[width=\textwidth]{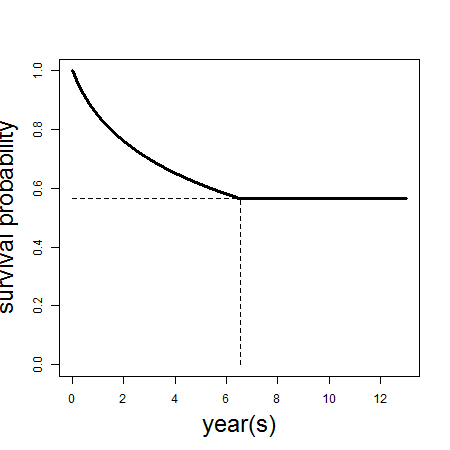}
  \caption{CTM-based net survival}
  \label{fA2_Bladder_MRS}
\end{subfigure}%
\begin{subfigure}{.3\textwidth}
  \centering
  \includegraphics[width=\textwidth]{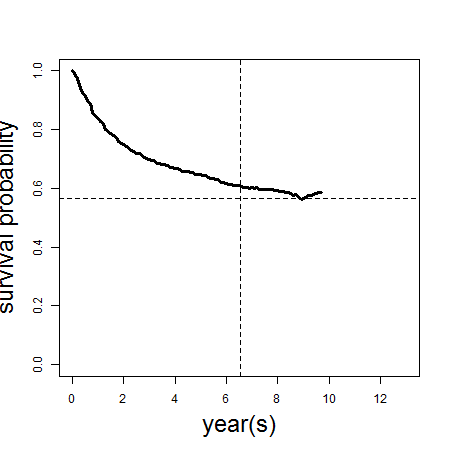}
  \caption{Relative survival}
  \label{fA2_Bladder_RS}
\end{subfigure}
\caption{$S_T(t|k)$ and $S_O(t|k)$ of bladder cancer data and general population in Taiwan, where $k$ is cure time estimated from CTM, CRS95, CRS99, and New Zealand (2006) (\ref{fA1_Bladder_CTM}-\ref{fA1_Bladder_NZ2006}). Model-based net survival and relative survival (\ref{fA2_Bladder_MRS}, \ref{fA2_Bladder_RS}), horizontal and vertical dashed lines represent locations of CTM-estimated cure time and cure rate, respectively. Note that in \ref{fA1_Bladder_NZ2006} there is no $S_T(t|k)$ and $S_O(t|k)$ since the follow-up time (9.65 years) is smaller than the cure time (10 years).}
\label{fA1_Bladder}
\end{figure}

\begin{figure}
	\centering
	\begin{subfigure}{.3\textwidth}
		\centering
		\includegraphics[width=\textwidth]{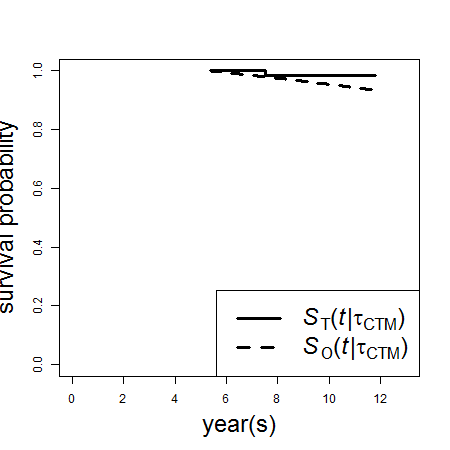}
		\caption{$\widehat{\tau}=5.36$ from CTM}
		\label{fA1_Bone_and_connective_CTM}
	\end{subfigure}%
	\begin{subfigure}{.3\textwidth}
		\centering
		\includegraphics[width=\textwidth]{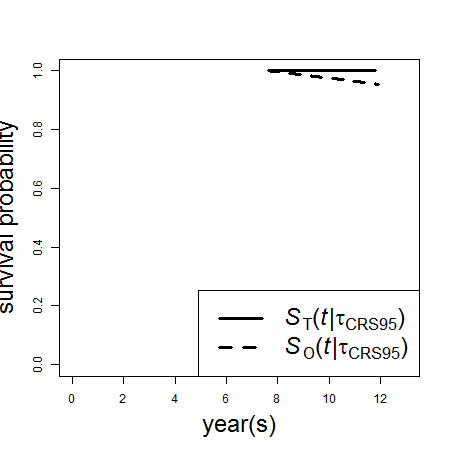}
		\caption{$\widehat{\tau}=7.65$ from CRS95}
		\label{fA1_Bone_and_connective_CRS95}
	\end{subfigure}\\
	\begin{subfigure}{.3\textwidth}
		\centering
		\includegraphics[width=\textwidth]{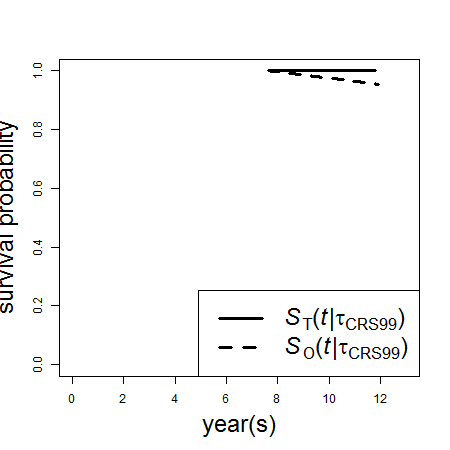}
		\caption{$\widehat{\tau}=7.65$ from CRS99}
		\label{fA1_Bone_and_connective_CRS99}
	\end{subfigure}%
	\begin{subfigure}{.3\textwidth}
		\centering
		\includegraphics[width=\textwidth]{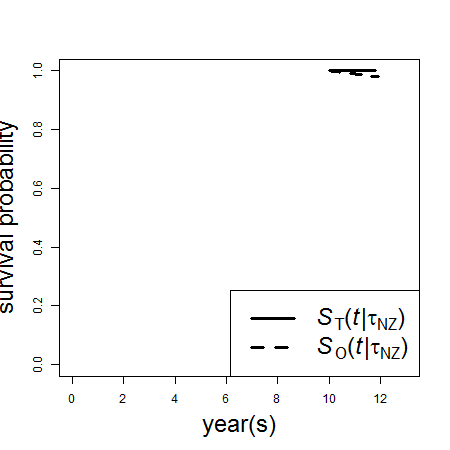}
		\caption{$\widehat{\tau}=10$ from New Zealand}
		\label{fA1_Bone_and_connective_NZ2006}
	\end{subfigure}\\
	\begin{subfigure}{.3\textwidth}
  \centering
  \includegraphics[width=\textwidth]{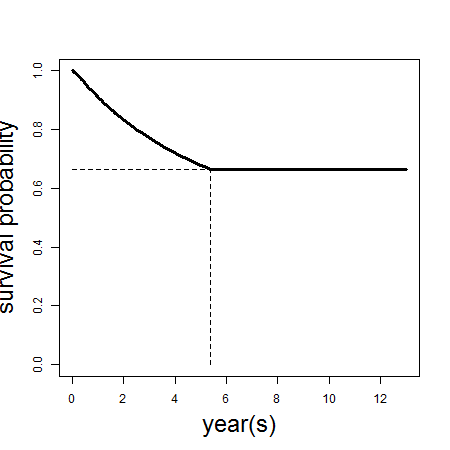}
  \caption{CTM-based net survival}
  \label{fA2_Bone_and_connective_MRS}
\end{subfigure}%
\begin{subfigure}{.3\textwidth}
  \centering
  \includegraphics[width=\textwidth]{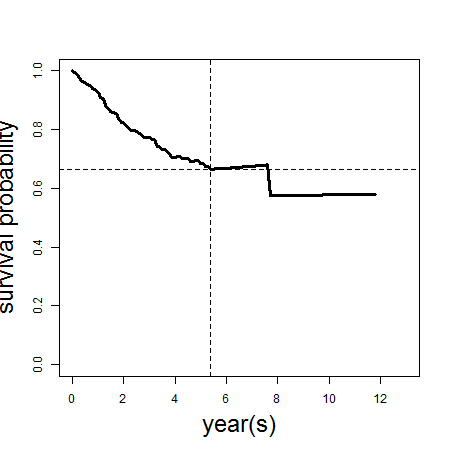}
  \caption{Relative survival}
  \label{fA2_Bone_and_connective_RS}
\end{subfigure}
	\caption{$S_T(t|k)$ and $S_O(t|k)$ of Bone and connective cancer data and general population in Taiwan, where $k$ is cure time estimated from CTM, CRS95, CRS99, and New Zealand (2006) (\ref{fA1_Bone_and_connective_CTM}-\ref{fA1_Bone_and_connective_NZ2006}). Model-based net survival and relative survival (\ref{fA2_Bone_and_connective_MRS}, \ref{fA2_Bone_and_connective_RS}), horizontal and vertical dashed lines represent locations of CTM-estimated cure time and cure rate, respectively.}
	\label{fA1_Bone_and_connective}
\end{figure}

\begin{figure}
	\centering
	\begin{subfigure}{.3\textwidth}
		\centering
		\includegraphics[width=\textwidth]{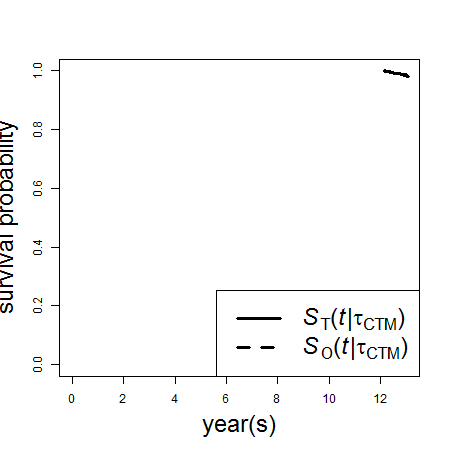}
		\caption{$\widehat{\tau}=12.15$ from CTM}
		\label{fA1_Breast_female_CTM}
	\end{subfigure}%
	\begin{subfigure}{.3\textwidth}
		\centering
		\includegraphics[width=\textwidth]{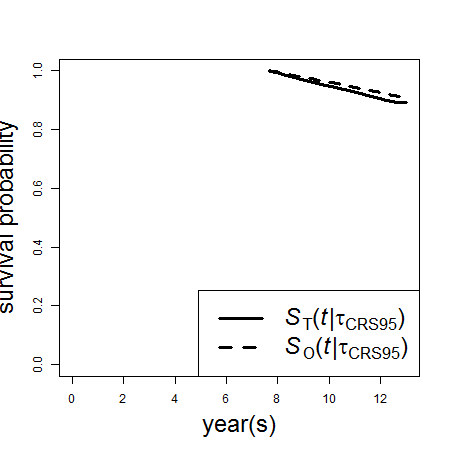}
		\caption{$\widehat{\tau}=7.67$ from CRS95}
		\label{fA1_Breast_female_CRS95}
	\end{subfigure}\\
	\begin{subfigure}{.3\textwidth}
		\centering
		\includegraphics[width=\textwidth]{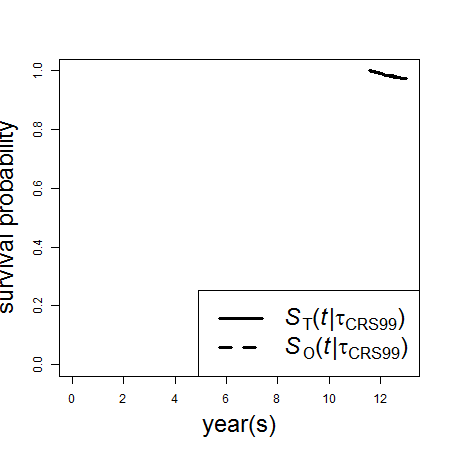}
		\caption{$\widehat{\tau}=11.58$ from CRS99}
		\label{fA1_Breast_female_CRS99}
	\end{subfigure}%
	\begin{subfigure}{.3\textwidth}
		\centering
		\includegraphics[width=\textwidth]{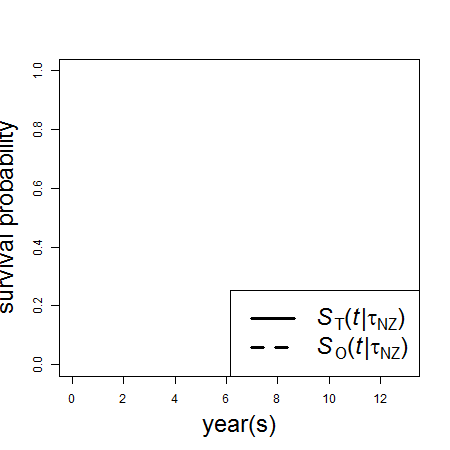}
		\caption{$\widehat{\tau}=20$ from New Zealand}
		\label{fA1_Breast_female_NZ2006}
	\end{subfigure}\\
	\begin{subfigure}{.3\textwidth}
  \centering
  \includegraphics[width=\textwidth]{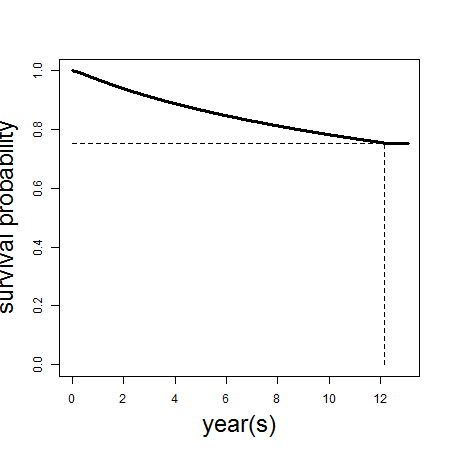}
  \caption{CTM-based net survival}
  \label{fA2_Breast_female_MRS}
\end{subfigure}%
\begin{subfigure}{.3\textwidth}
  \centering
  \includegraphics[width=\textwidth]{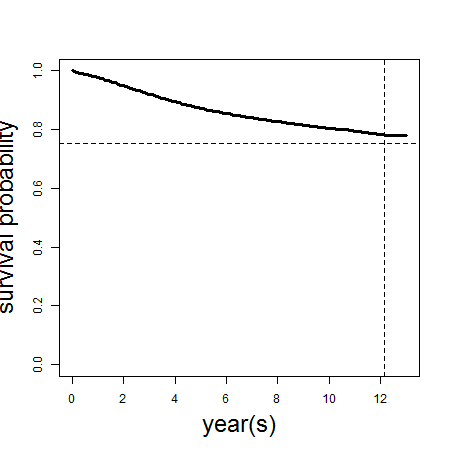}
  \caption{Relative survival}
  \label{fA2_Breast_female_RS}
\end{subfigure}
	\caption{$S_T(t|k)$ and $S_O(t|k)$ of female breast cancer data and general population in Taiwan, where $k$ is cure time estimated from CTM, CRS95, CRS99, and New Zealand (2006) (\ref{fA1_Breast_female_CTM}-\ref{fA1_Breast_female_NZ2006}). Model-based net survival and relative survival (\ref{fA2_Breast_female_MRS}, \ref{fA2_Breast_female_RS}), horizontal and vertical dashed lines represent locations of CTM-estimated cure time and cure rate, respectively. Note that in \ref{fA1_Breast_female_NZ2006} there is no $S_T(t|k)$ and $S_O(t|k)$ since the follow-up time (13 years) is smaller than the cure time (20 years).}
	\label{fA1_Breast_female}
\end{figure}

\begin{figure}
	\centering
	\begin{subfigure}{.3\textwidth}
		\centering
		\includegraphics[width=\textwidth]{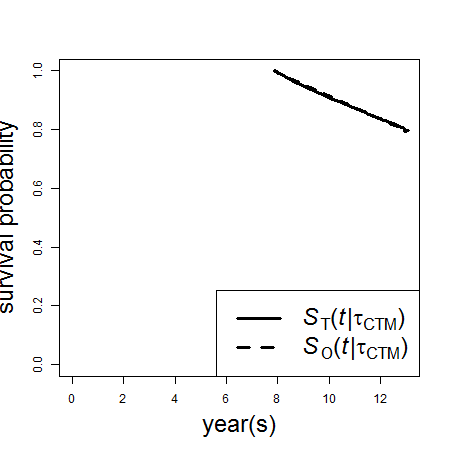}
		\caption{$\widehat{\tau}=7.85$ from CTM}
		\label{fA1_Colorectum_CTM}
	\end{subfigure}%
	\begin{subfigure}{.3\textwidth}
		\centering
		\includegraphics[width=\textwidth]{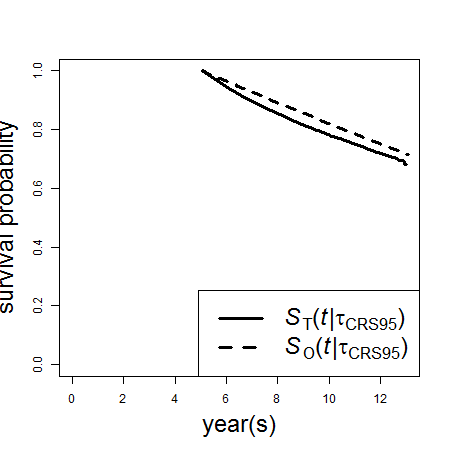}
		\caption{$\widehat{\tau}=5.07$ from CRS95}
		\label{fA1_Colorectum_CRS95}
	\end{subfigure}\\
	\begin{subfigure}{.3\textwidth}
		\centering
		\includegraphics[width=\textwidth]{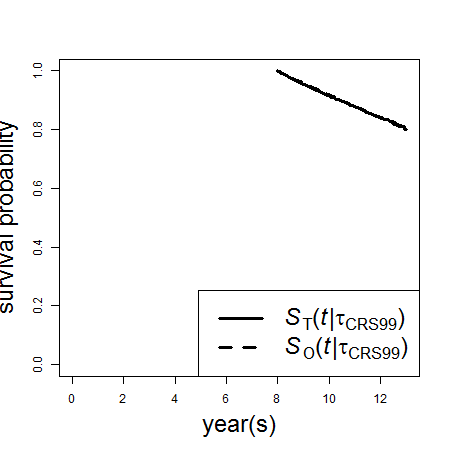}
		\caption{$\widehat{\tau}=7.97$ from CRS99}
		\label{fA1_Colorectum_CRS99}
	\end{subfigure}%
	\begin{subfigure}{.3\textwidth}
		\centering
		\includegraphics[width=\textwidth]{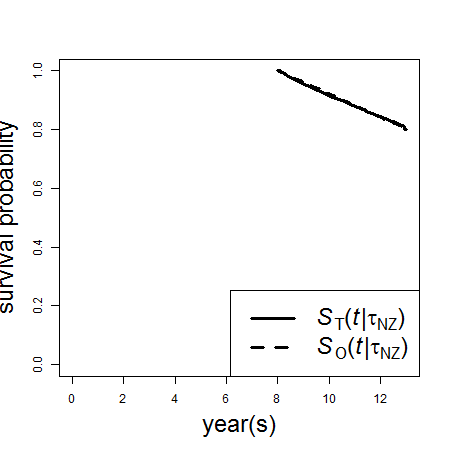}
		\caption{$\widehat{\tau}=8$ from New Zealand}
		\label{fA1_Colorectum_NZ2006}
	\end{subfigure}\\
	\begin{subfigure}{.3\textwidth}
  \centering
  \includegraphics[width=\textwidth]{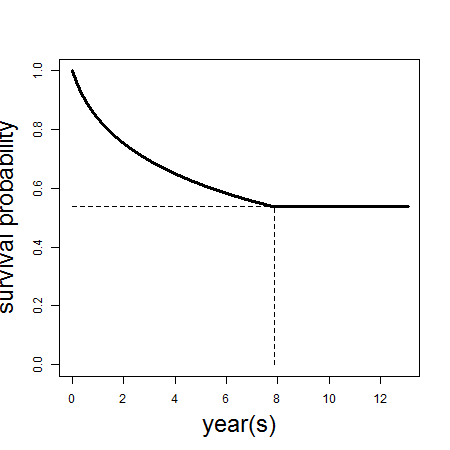}
  \caption{CTM-based net survival}
  \label{fA2_Colorectum_MRS}
\end{subfigure}%
\begin{subfigure}{.3\textwidth}
  \centering
  \includegraphics[width=\textwidth]{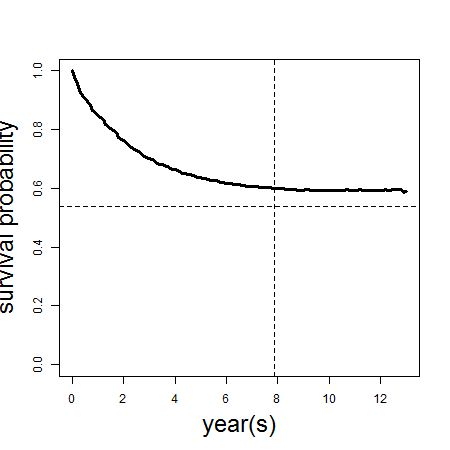}
  \caption{Relative survival}
  \label{fA2_Colorectum_RS}
\end{subfigure}
	\caption{$S_T(t|k)$ and $S_O(t|k)$ of colorectal cancer data and general population in Taiwan, where $k$ is cure time estimated from CTM, CRS95, CRS99, and New Zealand (2006) (\ref{fA1_Colorectum_CTM}-\ref{fA1_Colorectum_NZ2006}). Model-based net survival and relative survival (\ref{fA2_Colorectum_MRS}, \ref{fA2_Colorectum_RS}), horizontal and vertical dashed lines represent locations of CTM-estimated cure time and cure rate, respectively.}
	\label{fA1_Colorectum}
\end{figure}

\begin{figure}
	\centering
	\begin{subfigure}{.3\textwidth}
		\centering
		\includegraphics[width=\textwidth]{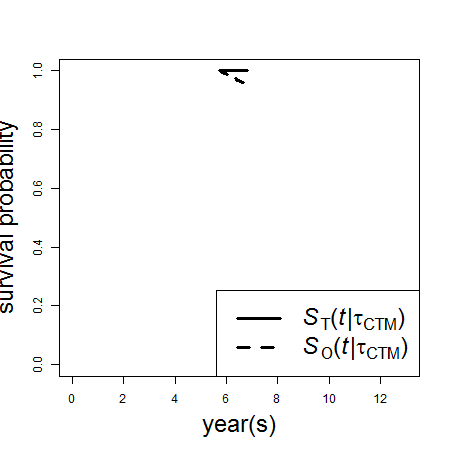}
		\caption{$\widehat{\tau}=5.74$ from CTM}
		\label{fA1_Gallbladder_CTM}
	\end{subfigure}%
	\begin{subfigure}{.3\textwidth}
		\centering
		\includegraphics[width=\textwidth]{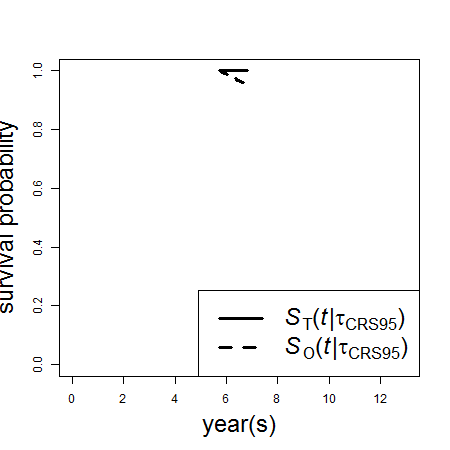}
		\caption{$\widehat{\tau}=5.74$ from CRS95}
		\label{fA1_Gallbladder_CRS95}
	\end{subfigure}\\
	\begin{subfigure}{.3\textwidth}
		\centering
		\includegraphics[width=\textwidth]{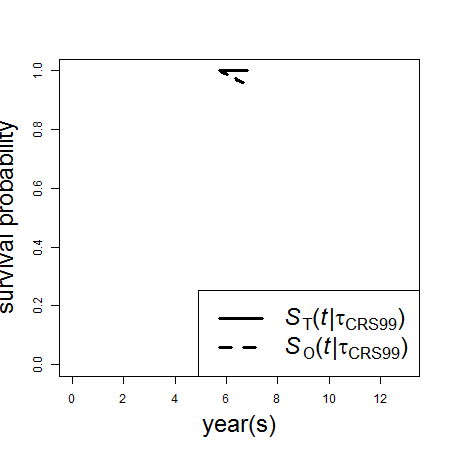}
		\caption{$\widehat{\tau}=5.74$ from CRS99}
		\label{fA1_Gallbladder_CRS99}
	\end{subfigure}%
	\begin{subfigure}{.3\textwidth}
		\centering
		\includegraphics[width=\textwidth]{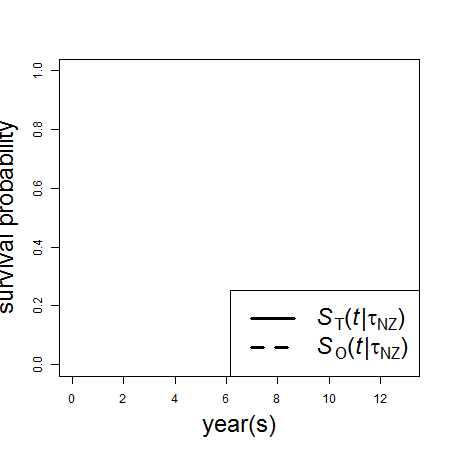}
		\caption{$\widehat{\tau}=7$ from New Zealand}
		\label{fA1_Gallbladder_NZ2006}
	\end{subfigure}\\
	\begin{subfigure}{.3\textwidth}
  \centering
  \includegraphics[width=\textwidth]{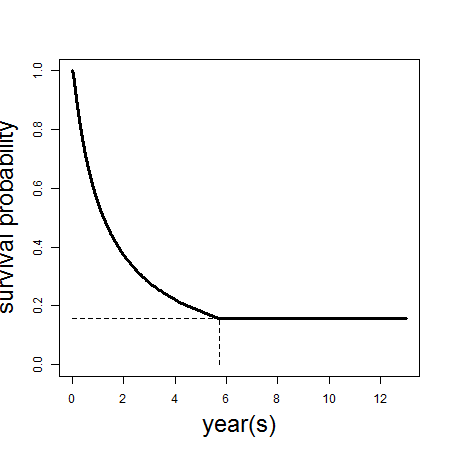}
  \caption{CTM-based net survival}
  \label{fA2_Gallbladder_MRS}
\end{subfigure}%
\begin{subfigure}{.3\textwidth}
  \centering
  \includegraphics[width=\textwidth]{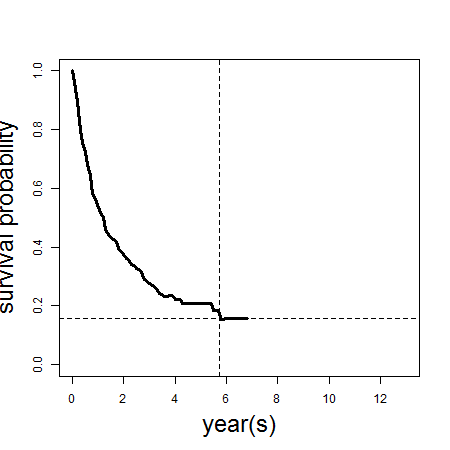}
  \caption{Relative survival}
  \label{fA2_Gallbladder_RS}
\end{subfigure}
	\caption{$S_T(t|k)$ and $S_O(t|k)$ of gallbladder cancer data and general population in Taiwan, where $k$ is cure time estimated from CTM, CRS95, CRS99, and New Zealand (2006). (\ref{fA1_Gallbladder_CTM}-\ref{fA1_Gallbladder_NZ2006}). Model-based net survival and relative survival (\ref{fA2_Gallbladder_MRS}, \ref{fA2_Gallbladder_RS}), horizontal and vertical dashed lines represent locations of CTM-estimated cure time and cure rate, respectively. Note that in \ref{fA1_Gallbladder_NZ2006} there is no $S_T(t|k)$ and $S_O(t|k)$ since the follow-up time (6.83 years) is smaller than the cure time (7 years).}
	\label{fA1_Gallbladder}
\end{figure}

\begin{figure}
	\centering
	\begin{subfigure}{.3\textwidth}
		\centering
		\includegraphics[width=\textwidth]{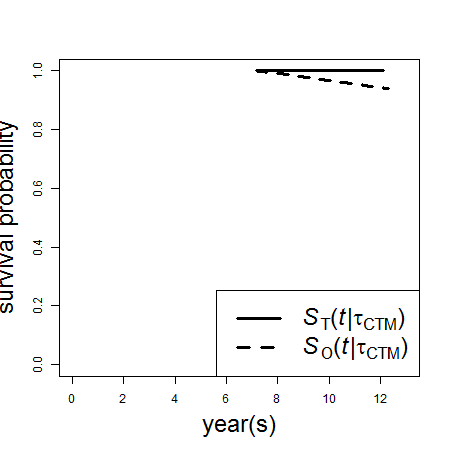}
		\caption{$\widehat{\tau}=7.16$ from CTM}
		\label{fA1_Hodgkin_CTM}
	\end{subfigure}%
	\begin{subfigure}{.3\textwidth}
		\centering
		\includegraphics[width=\textwidth]{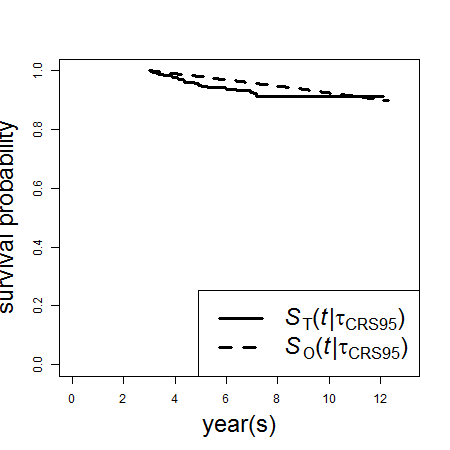}
		\caption{$\widehat{\tau}=3.02$ from CRS95}
		\label{fA1_Hodgkin_CRS95}
	\end{subfigure}\\
	\begin{subfigure}{.3\textwidth}
		\centering
		\includegraphics[width=\textwidth]{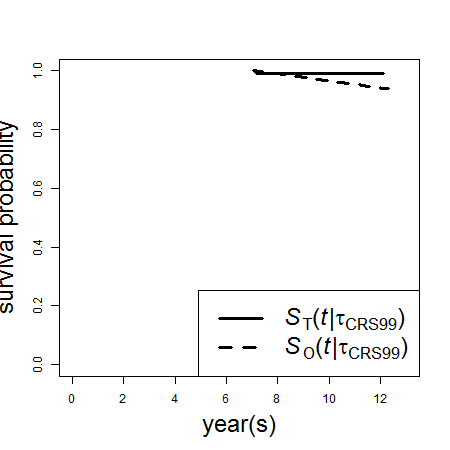}
		\caption{$\widehat{\tau}7.05$ from CRS99}
		\label{fA1_Hodgkin_CRS99}
	\end{subfigure}%
	\begin{subfigure}{.3\textwidth}
		\centering
		\includegraphics[width=\textwidth]{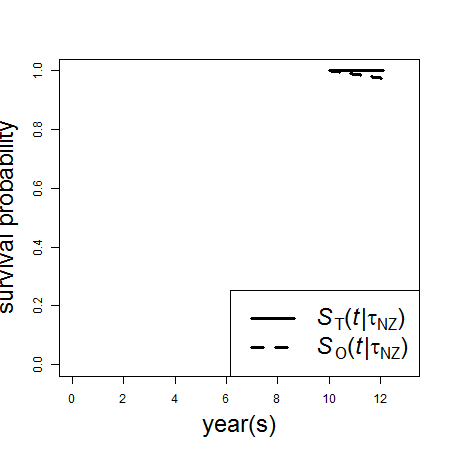}
		\caption{$\widehat{\tau}=10$ from New Zealand}
		\label{fA1_Hodgkin_NZ2006}
	\end{subfigure}\\
	\begin{subfigure}{.3\textwidth}
  \centering
  \includegraphics[width=\textwidth]{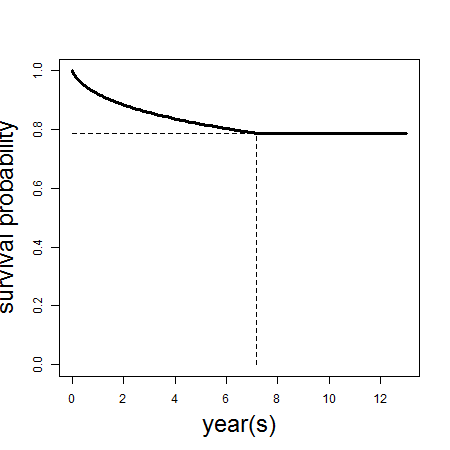}
  \caption{CTM-based net survival}
  \label{fA2_Hodgkin_MRS}
\end{subfigure}%
\begin{subfigure}{.3\textwidth}
  \centering
  \includegraphics[width=\textwidth]{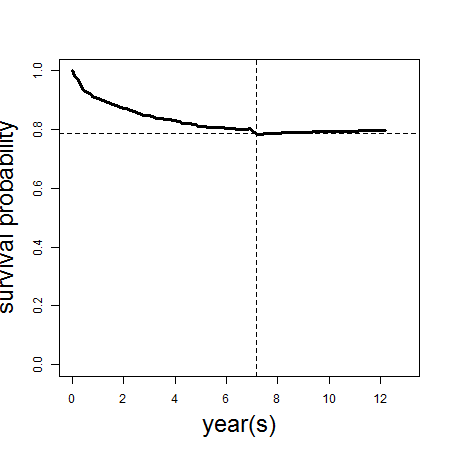}
  \caption{Relative survival}
  \label{fA2_Hodgkin_RS}
\end{subfigure}
	\caption{$S_T(t|k)$ and $S_O(t|k)$ of Hodgkin lymphoma data and general population in Taiwan, where $k$ is cure time estimated from CTM, CRS95, CRS99, and New Zealand (2006). (\ref{fA1_Hodgkin_CTM}-\ref{fA1_Hodgkin_NZ2006}). Model-based net survival and relative survival (\ref{fA2_Hodgkin_MRS}, \ref{fA2_Hodgkin_RS}), horizontal and vertical dashed lines represent locations of CTM-estimated cure time and cure rate, respectively.}
	\label{fA1_Hodgkin}
\end{figure}

\begin{figure}
	\centering
	\begin{subfigure}{.3\textwidth}
		\centering
		\includegraphics[width=\textwidth]{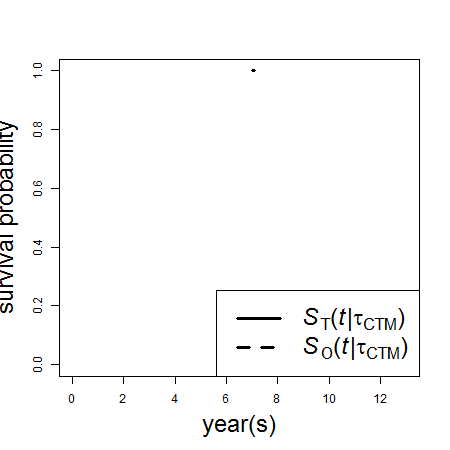}
		\caption{$\widehat{\tau}=7$ from CTM}
		\label{fA1_Kidney_and_other_urinary_CTM}
	\end{subfigure}%
	\begin{subfigure}{.3\textwidth}
		\centering
		\includegraphics[width=\textwidth]{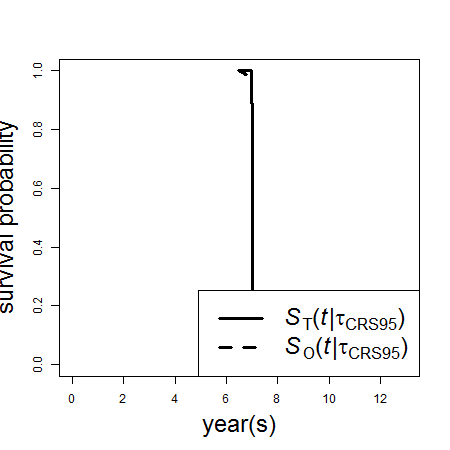}
		\caption{$\widehat{\tau}=6.48$ from CRS95}
		\label{fA1_Kidney_and_other_urinary_CRS95}
	\end{subfigure}\\
	\begin{subfigure}{.3\textwidth}
		\centering
		\includegraphics[width=\textwidth]{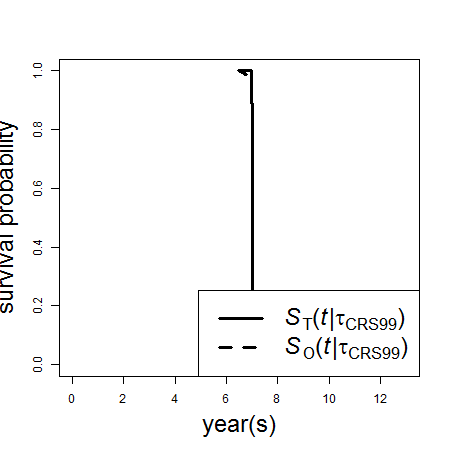}
		\caption{$\widehat{\tau}=6.48$ from CRS99}
		\label{fA1_Kidney_and_other_urinary_CRS99}
	\end{subfigure}%
	\begin{subfigure}{.3\textwidth}
		\centering
		\includegraphics[width=\textwidth]{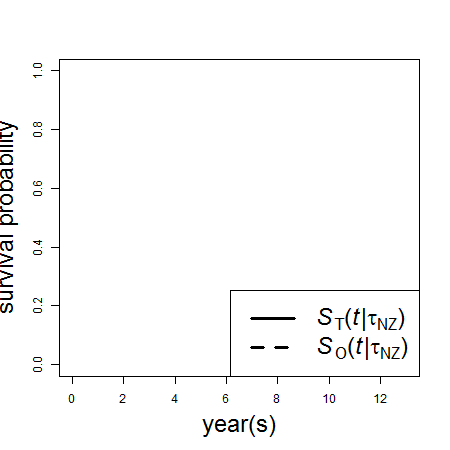}
		\caption{$\widehat{\tau}=10$ from New Zealand}
		\label{fA1_Kidney_and_other_urinary_NZ2006}
	\end{subfigure}\\
	\begin{subfigure}{.3\textwidth}
  \centering
  \includegraphics[width=\textwidth]{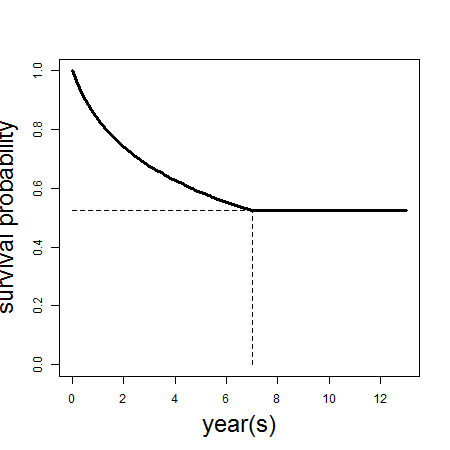}
  \caption{CTM-based net survival}
  \label{fA2_Kidney_and_other_urinary_MRS}
\end{subfigure}%
\begin{subfigure}{.3\textwidth}
  \centering
  \includegraphics[width=\textwidth]{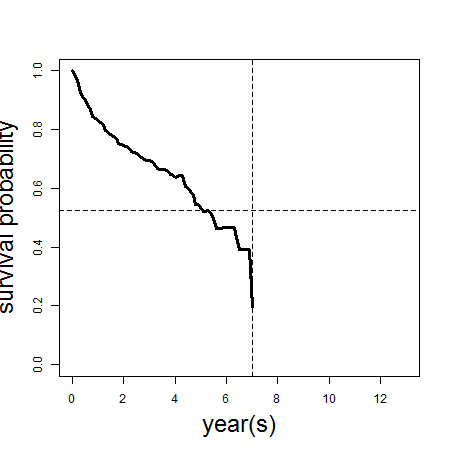}
  \caption{Relative survival}
  \label{fA2_Kidney_and_other_urinary_RS}
\end{subfigure}
	\caption{$S_T(t|k)$ and $S_O(t|k)$ of kidney and other urinary cancer data and general population in Taiwan, where $k$ is cure time estimated from CTM, CRS95, CRS99, and New Zealand (2006). (\ref{fA1_Kidney_and_other_urinary_CTM}-\ref{fA1_Kidney_and_other_urinary_NZ2006}). Model-based net survival and relative survival (\ref{fA2_Kidney_and_other_urinary_MRS}, \ref{fA2_Kidney_and_other_urinary_RS}), horizontal and vertical dashed lines represent locations of CTM-estimated cure time and cure rate, respectively. Note that in \ref{fA1_Kidney_and_other_urinary_NZ2006} there is no $S_T(t|k)$ and $S_O(t|k)$ since the follow-up time (6.96 years) is smaller than the cure time (10 years).}
	\label{fA1_Kidney_and_other_urinary}
\end{figure}

\begin{figure}
	\centering
	\begin{subfigure}{.3\textwidth}
		\centering
		\includegraphics[width=\textwidth]{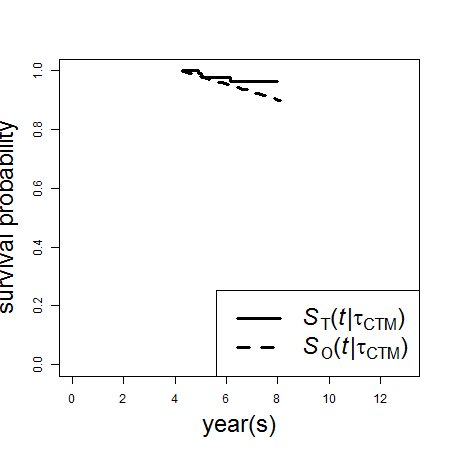}
		\caption{$\widehat{\tau}=4.27$ from CTM}
		\label{fA1_Larynx_CTM}
	\end{subfigure}%
	\begin{subfigure}{.3\textwidth}
		\centering
		\includegraphics[width=\textwidth]{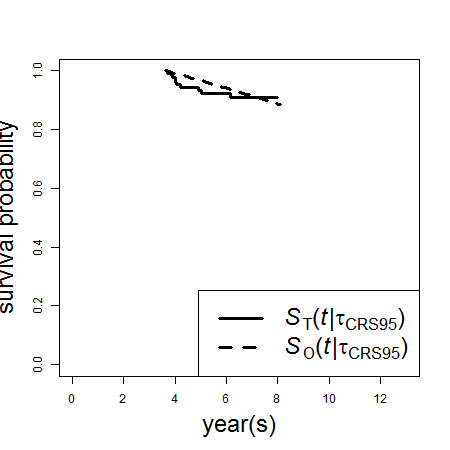}
		\caption{$\widehat{\tau}=3.64$ from CRS95}
		\label{fA1_Larynx_CRS95}
	\end{subfigure}\\
	\begin{subfigure}{.3\textwidth}
		\centering
		\includegraphics[width=\textwidth]{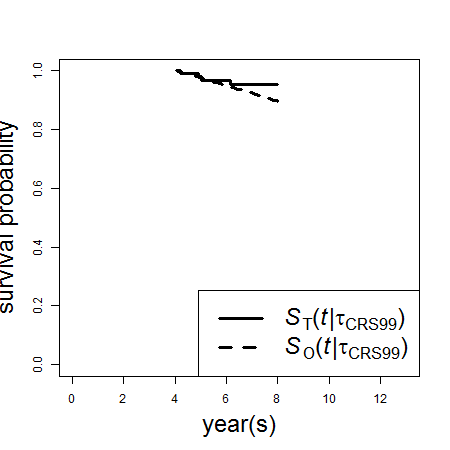}
		\caption{$\widehat{\tau}=4.06$ from CRS99}
		\label{fA1_Larynx_CRS99}
	\end{subfigure}%
	\begin{subfigure}{.3\textwidth}
		\centering
		\includegraphics[width=\textwidth]{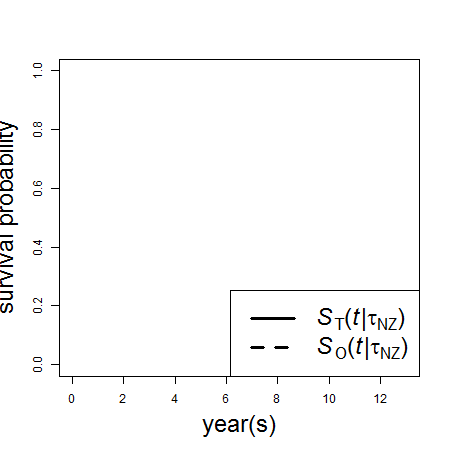}
		\caption{$\widehat{\tau}=10$ from New Zealand}
		\label{fA1_Larynx_NZ2006}
	\end{subfigure}\\
	\begin{subfigure}{.3\textwidth}
  \centering
  \includegraphics[width=\textwidth]{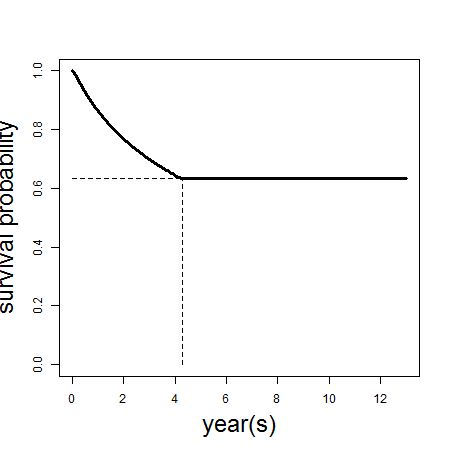}
  \caption{CTM-based net survival}
  \label{fA2_Larynx_MRS}
\end{subfigure}%
\begin{subfigure}{.3\textwidth}
  \centering
  \includegraphics[width=\textwidth]{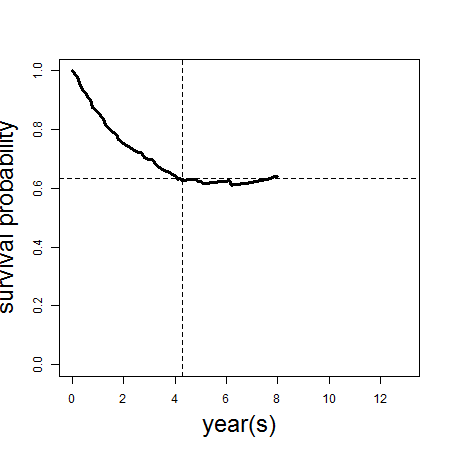}
  \caption{Relative survival}
  \label{fA2_Larynx_RS}
\end{subfigure}
	\caption{$S_T(t|k)$ and $S_O(t|k)$ of larynx cancer data and general population in Taiwan, where $k$ is cure time estimated from CTM, CRS95, CRS99, and New Zealand (2006). (\ref{fA1_Larynx_CTM}-\ref{fA1_Larynx_NZ2006}). Model-based net survival and relative survival (\ref{fA2_Larynx_MRS}, \ref{fA2_Larynx_RS}), horizontal and vertical dashed lines represent locations of CTM-estimated cure time and cure rate, respectively. Note that in \ref{fA1_Larynx_NZ2006} there is no $S_T(t|k)$ and $S_O(t|k)$ since the follow-up time (7.98 years) is smaller than the cure time (10 years).}
	\label{fA1_Larynx}
\end{figure}

\begin{figure}
	\centering
	\begin{subfigure}{.3\textwidth}
		\centering
		\includegraphics[width=\textwidth]{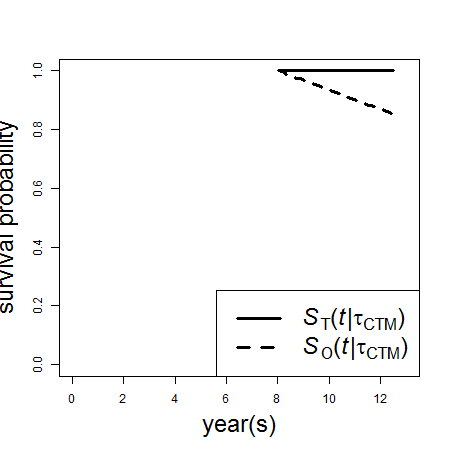}
		\caption{$\widehat{\tau}=8.01$ from CTM}
		\label{fA1_Leukaemia_CTM}
	\end{subfigure}%
	\begin{subfigure}{.3\textwidth}
		\centering
		\includegraphics[width=\textwidth]{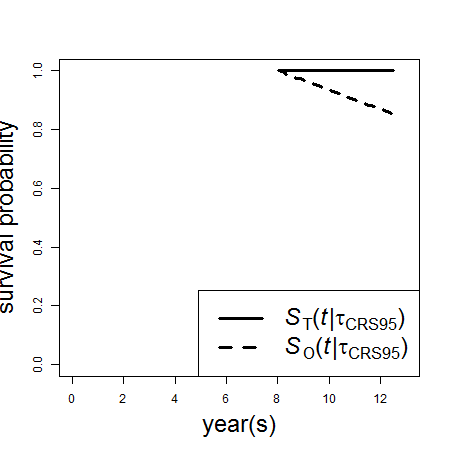}
		\caption{$\widehat{\tau}=8.01$ from CRS95}
		\label{fA1_Leukaemia_CRS95}
	\end{subfigure}\\
	\begin{subfigure}{.3\textwidth}
		\centering
		\includegraphics[width=\textwidth]{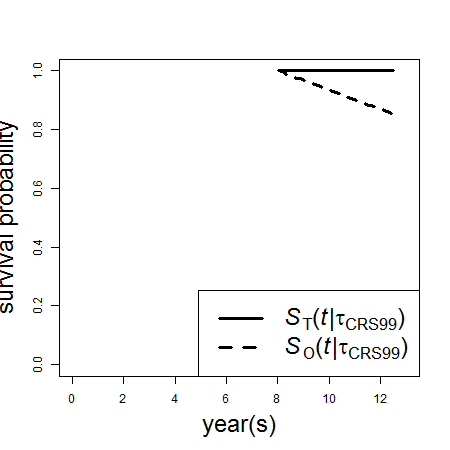}
		\caption{$\widehat{\tau}=8.01$ from CRS99}
		\label{fA1_Leukaemia_CRS99}
	\end{subfigure}%
	\begin{subfigure}{.3\textwidth}
		\centering
		\includegraphics[width=\textwidth]{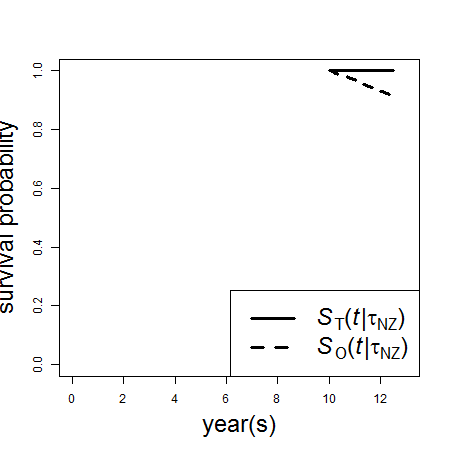}
		\caption{$\widehat{\tau}=10$ from New Zealand}
		\label{fA1_Leukaemia_NZ2006}
	\end{subfigure}\\
	\begin{subfigure}{.3\textwidth}
  \centering
  \includegraphics[width=\textwidth]{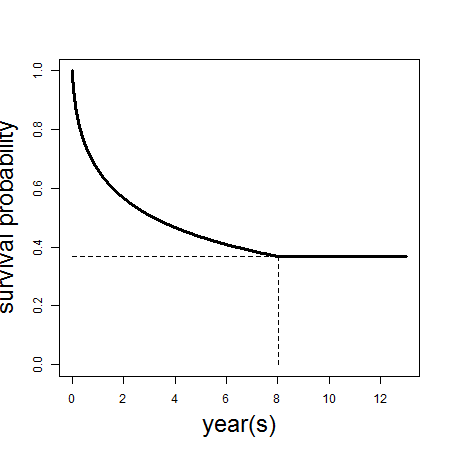}
  \caption{CTM-based net survival}
  \label{fA2_Leukaemia_MRS}
\end{subfigure}%
\begin{subfigure}{.3\textwidth}
  \centering
  \includegraphics[width=\textwidth]{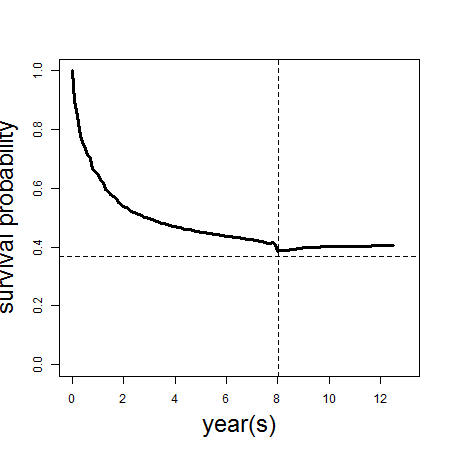}
  \caption{Relative survival}
  \label{fA2_Leukaemia_RS}
\end{subfigure}
	\caption{$S_T(t|k)$ and $S_O(t|k)$ of leukaemia data and general population in Taiwan, where $k$ is cure time estimated from CTM, CRS95, CRS99, and New Zealand (2006). (\ref{fA1_Leukaemia_CTM}-\ref{fA1_Leukaemia_NZ2006}). Model-based net survival and relative survival (\ref{fA2_Leukaemia_MRS}, \ref{fA2_Leukaemia_RS}), horizontal and vertical dashed lines represent locations of CTM-estimated cure time and cure rate, respectively.}
	\label{fA1_Leukaemia}
\end{figure}

\begin{figure}
	\centering
	\begin{subfigure}{.3\textwidth}
		\centering
		\includegraphics[width=\textwidth]{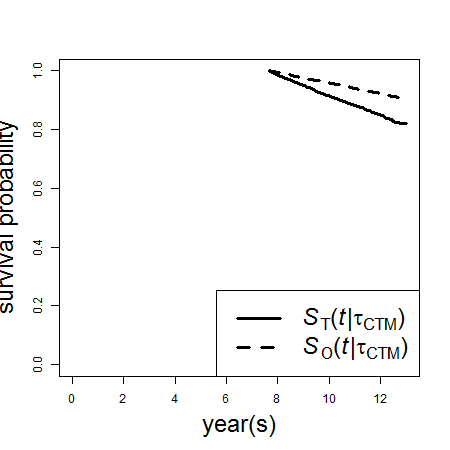}
		\caption{$\widehat{\tau}=7.66$ from CTM}
		\label{fA1_Lip_mouth_pharynx_CTM}
	\end{subfigure}%
	\begin{subfigure}{.3\textwidth}
		\centering
		\includegraphics[width=\textwidth]{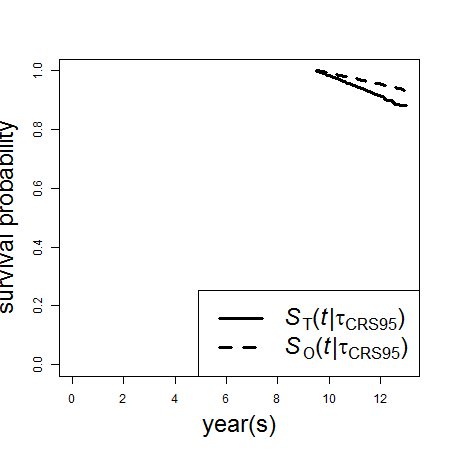}
		\caption{$\widehat{\tau}=9.50$ from CRS95}
		\label{fA1_Lip_mouth_pharynx_CRS95}
	\end{subfigure}\\
	\begin{subfigure}{.3\textwidth}
		\centering
		\includegraphics[width=\textwidth]{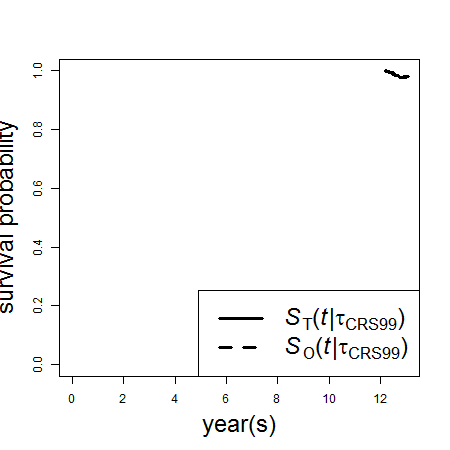}
		\caption{$\widehat{\tau}=12.18$ from CRS99}
		\label{fA1_Lip_mouth_pharynx_CRS99}
	\end{subfigure}%
	\begin{subfigure}{.3\textwidth}
		\centering
		\includegraphics[width=\textwidth]{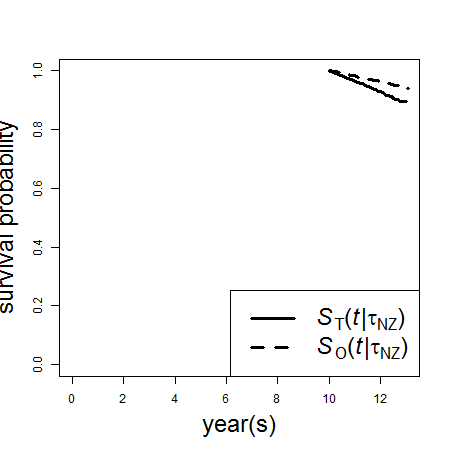}
		\caption{$\widehat{\tau}=10$ from New Zealand}
		\label{fA1_Lip_mouth_pharynx_NZ2006}
	\end{subfigure}\\
	\begin{subfigure}{.3\textwidth}
  \centering
  \includegraphics[width=\textwidth]{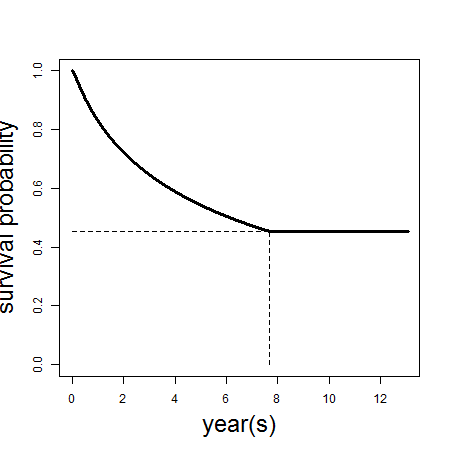}
  \caption{CTM-based net survival}
  \label{fA2_Lip_mouth_pharynx_MRS}
\end{subfigure}%
\begin{subfigure}{.3\textwidth}
  \centering
  \includegraphics[width=\textwidth]{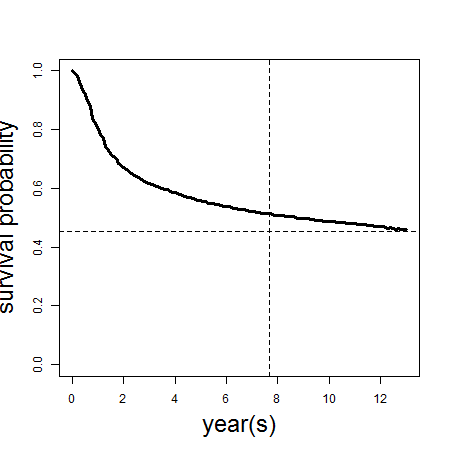}
  \caption{Relative survival}
  \label{fA2_Lip_mouth_pharynx_RS}
\end{subfigure}
	\caption{$S_T(t|k)$ and $S_O(t|k)$ of lip, mouth, and pharynx cancer data and general population in Taiwan, where $k$ is cure time estimated from CTM, CRS95, CRS99, and New Zealand (2006). (\ref{fA1_Lip_mouth_pharynx_CTM}-\ref{fA1_Lip_mouth_pharynx_NZ2006}). Model-based net survival and relative survival (\ref{fA2_Lip_mouth_pharynx_MRS}, \ref{fA2_Lip_mouth_pharynx_RS}), horizontal and vertical dashed lines represent locations of CTM-estimated cure time and cure rate, respectively.}
	\label{fA1_Lip_mouth_pharynx}
\end{figure}

\begin{figure}
	\centering
	\begin{subfigure}{.3\textwidth}
		\centering
		\includegraphics[width=\textwidth]{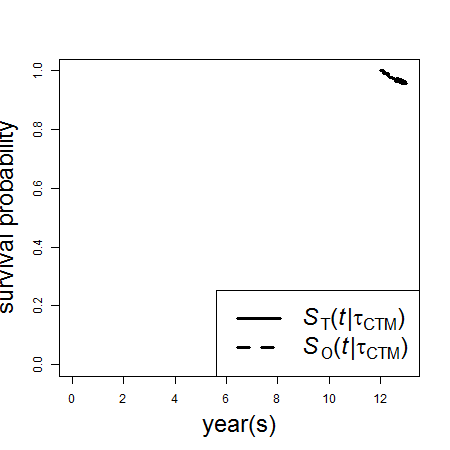}
		\caption{$\widehat{\tau}=12.01$ from CTM}
		\label{fA1_Liver_CTM}
	\end{subfigure}%
	\begin{subfigure}{.3\textwidth}
		\centering
		\includegraphics[width=\textwidth]{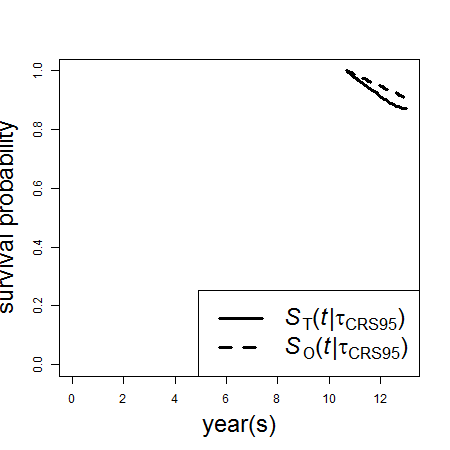}
		\caption{$\widehat{\tau}=10.68$ from CRS95}
		\label{fA1_Liver_CRS95}
	\end{subfigure}\\
	\begin{subfigure}{.3\textwidth}
		\centering
		\includegraphics[width=\textwidth]{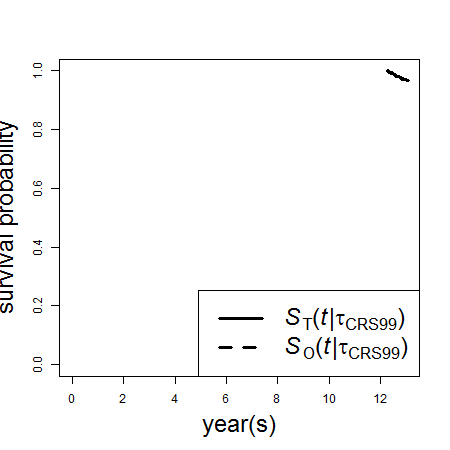}
		\caption{$\widehat{\tau}=12.27$ from CRS99}
		\label{fA1_Liver_CRS99}
	\end{subfigure}%
	\begin{subfigure}{.3\textwidth}
		\centering
		\includegraphics[width=\textwidth]{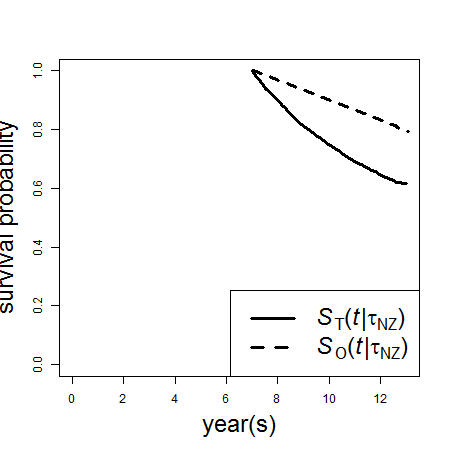}
		\caption{$\widehat{\tau}=7$ from New Zealand}
		\label{fA1_Liver_NZ2006}
	\end{subfigure}\\
	\begin{subfigure}{.3\textwidth}
  \centering
  \includegraphics[width=\textwidth]{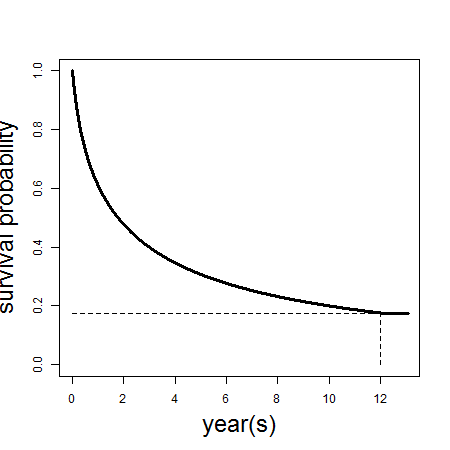}
  \caption{CTM-based net survival}
  \label{fA2_Liver_MRS}
\end{subfigure}%
\begin{subfigure}{.3\textwidth}
  \centering
  \includegraphics[width=\textwidth]{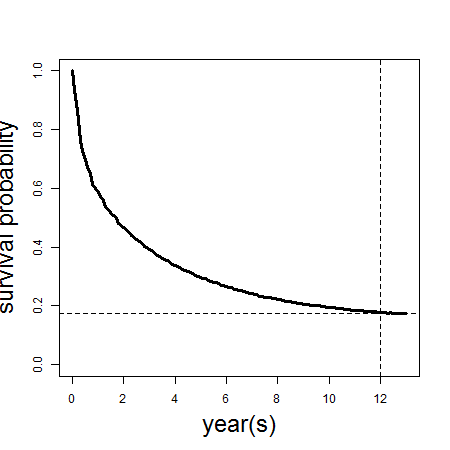}
  \caption{Relative survival}
  \label{fA2_Liver_RS}
\end{subfigure}
	\caption{$S_T(t|k)$ and $S_O(t|k)$ of liver cancer data and general population in Taiwan, where $k$ is cure time estimated from CTM, CRS95, CRS99, and New Zealand (2006). (\ref{fA1_Liver_CTM}-\ref{fA1_Liver_NZ2006}). Model-based net survival and relative survival (\ref{fA2_Liver_MRS}, \ref{fA2_Liver_RS}), horizontal and vertical dashed lines represent locations of CTM-estimated cure time and cure rate, respectively.}
	\label{fA1_Liver}
\end{figure}

\begin{figure}
	\centering
	\begin{subfigure}{.3\textwidth}
		\centering
		\includegraphics[width=\textwidth]{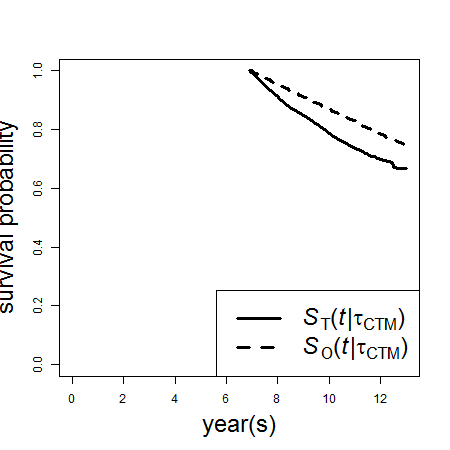}
		\caption{$\widehat{\tau}=6.90$ from CTM}
		\label{fA1_Lung_trachea_bronchus_CTM}
	\end{subfigure}%
	\begin{subfigure}{.3\textwidth}
		\centering
		\includegraphics[width=\textwidth]{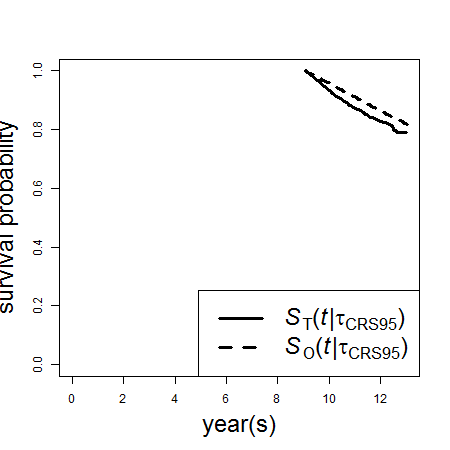}
		\caption{$\widehat{\tau}=9.08$ from CRS95}
		\label{fA1_Lung_trachea_bronchus_CRS95}
	\end{subfigure}\\
	\begin{subfigure}{.3\textwidth}
		\centering
		\includegraphics[width=\textwidth]{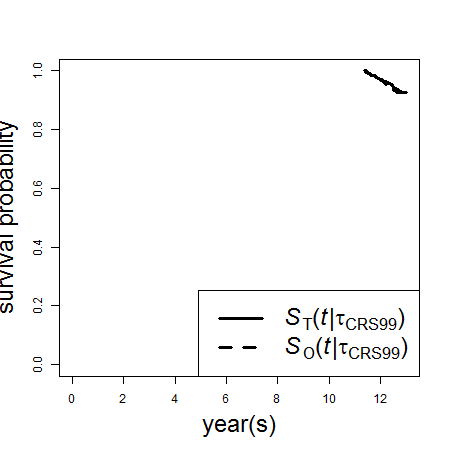}
		\caption{$\widehat{\tau}=11.38$ from CRS99}
		\label{fA1_Lung_trachea_bronchus_CRS99}
	\end{subfigure}%
	\begin{subfigure}{.3\textwidth}
		\centering
		\includegraphics[width=\textwidth]{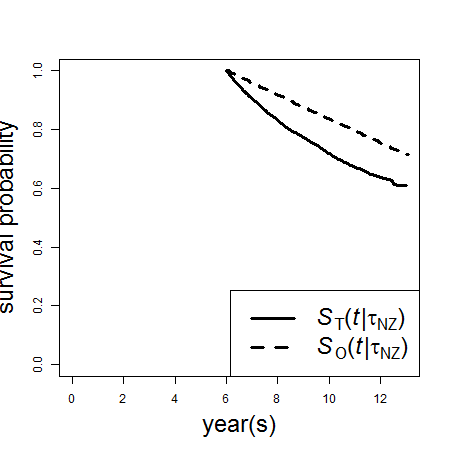}
		\caption{$\widehat{\tau}=6$ from New Zealand}
		\label{fA1_Lung_trachea_bronchus_NZ2006}
	\end{subfigure}\\
	\begin{subfigure}{.3\textwidth}
  \centering
  \includegraphics[width=\textwidth]{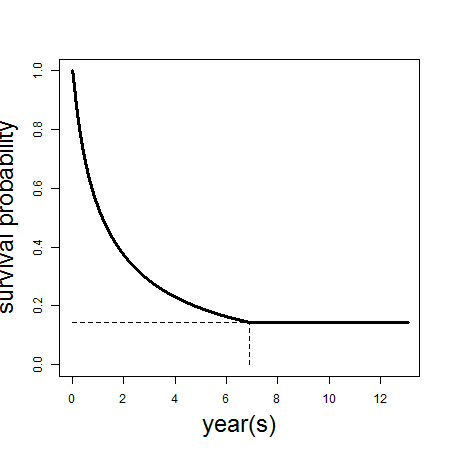}
  \caption{CTM-based net survival}
  \label{fA2_Lung_trachea_bronchus_MRS}
\end{subfigure}%
\begin{subfigure}{.3\textwidth}
  \centering
  \includegraphics[width=\textwidth]{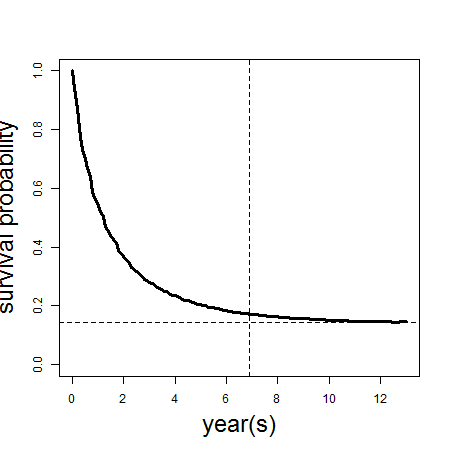}
  \caption{Relative survival}
  \label{fA2_Lung_trachea_bronchus_RS}
\end{subfigure}
	\caption{$S_T(t|k)$ and $S_O(t|k)$ of lung, trachea, and bronchus cancer data and general population in Taiwan, where $k$ is cure time estimated from CTM, CRS95, CRS99, and New Zealand (2006). (\ref{fA1_Lung_trachea_bronchus_CTM}-\ref{fA1_Lung_trachea_bronchus_NZ2006}). Model-based net survival and relative survival (\ref{fA2_Lung_trachea_bronchus_MRS}, \ref{fA2_Lung_trachea_bronchus_RS}), horizontal and vertical dashed lines represent locations of CTM-estimated cure time and cure rate, respectively.}
	\label{fA1_Lung_trachea_bronchus}
\end{figure}

\begin{figure}
	\centering
	\begin{subfigure}{.3\textwidth}
		\centering
		\includegraphics[width=\textwidth]{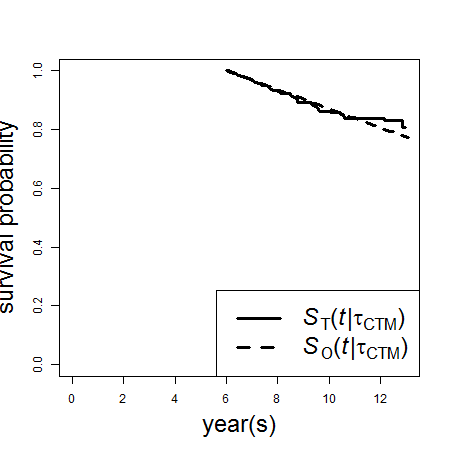}
		\caption{$\widehat{\tau}=6$ from CTM}
		\label{fA1_NHL_CTM}
	\end{subfigure}%
	\begin{subfigure}{.3\textwidth}
		\centering
		\includegraphics[width=\textwidth]{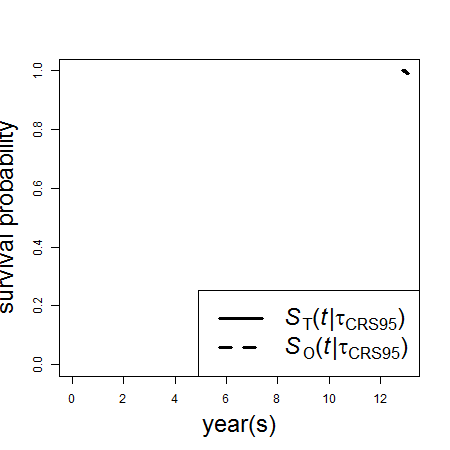}
		\caption{$\widehat{\tau}=12.86$ from CRS95}
		\label{fA1_NHL_CRS95}
	\end{subfigure}\\
	\begin{subfigure}{.3\textwidth}
		\centering
		\includegraphics[width=\textwidth]{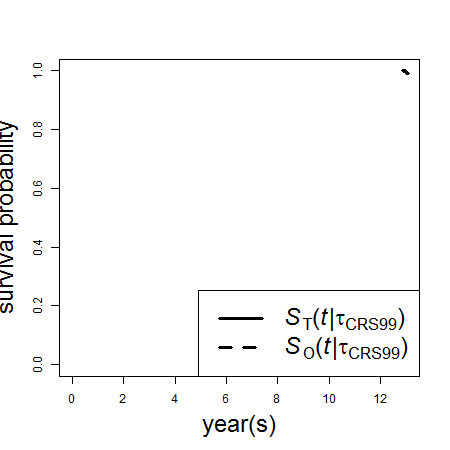}
		\caption{$\widehat{\tau}=12.86$ from CRS99}
		\label{fA1_NHL_CRS99}
	\end{subfigure}%
	\begin{subfigure}{.3\textwidth}
		\centering
		\includegraphics[width=\textwidth]{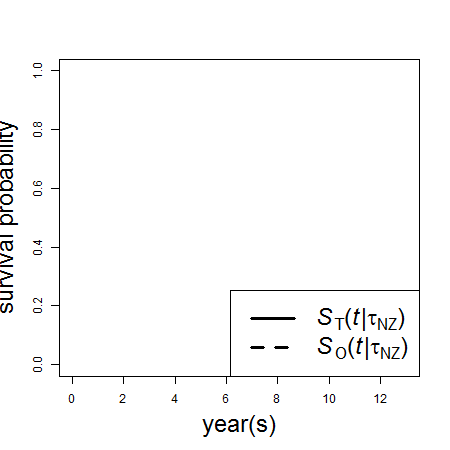}
		\caption{$\widehat{\tau}=20$ from New Zealand}
		\label{fA1_NHL_NZ2006}
	\end{subfigure}\\
	\begin{subfigure}{.3\textwidth}
  \centering
  \includegraphics[width=\textwidth]{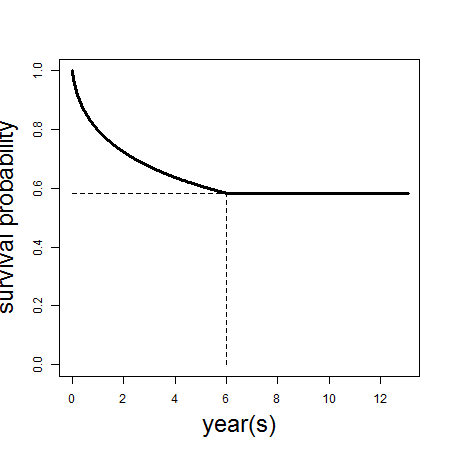}
  \caption{CTM-based net survival}
  \label{fA2_NHL_MRS}
\end{subfigure}%
\begin{subfigure}{.3\textwidth}
  \centering
  \includegraphics[width=\textwidth]{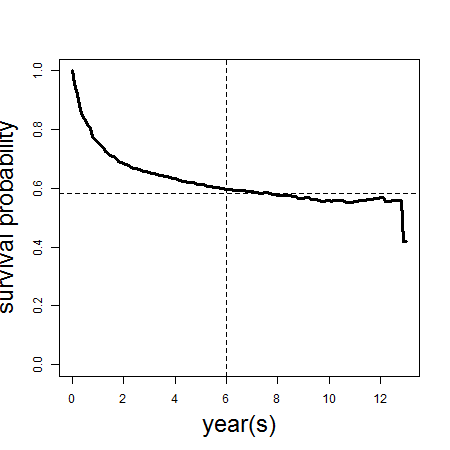}
  \caption{Relative survival}
  \label{fA2_NHL_RS}
\end{subfigure}
	\caption{$S_T(t|k)$ and $S_O(t|k)$ of non-Hodgkin lymphoma data and general population in Taiwan, where $k$ is cure time estimated from CTM, CRS95, CRS99, and New Zealand (2006). (\ref{fA1_NHL_CTM}-\ref{fA1_NHL_NZ2006}). Model-based net survival and relative survival (\ref{fA2_NHL_MRS}, \ref{fA2_NHL_RS}), horizontal and vertical dashed lines represent locations of CTM-estimated cure time and cure rate, respectively. Note that in \ref{fA1_NHL_NZ2006} there is no $S_T(t|k)$ and $S_O(t|k)$ since the follow-up time (12.96 years) is smaller than the cure time (20 years).}
	\label{fA1_NHL}
\end{figure}

\begin{figure}
	\centering
	\begin{subfigure}{.3\textwidth}
		\centering
		\includegraphics[width=\textwidth]{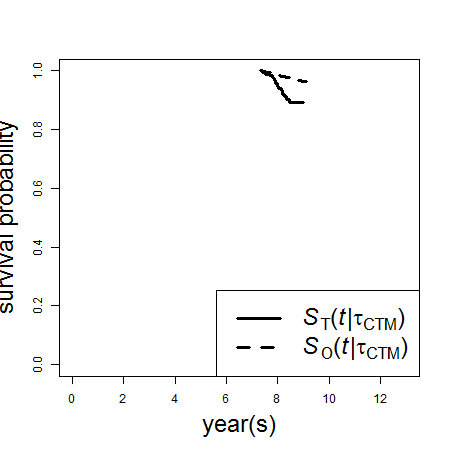}
		\caption{$\widehat{\tau}=7.33$ from CTM}
		\label{fA1_Oesophagus_CTM}
	\end{subfigure}%
	\begin{subfigure}{.3\textwidth}
		\centering
		\includegraphics[width=\textwidth]{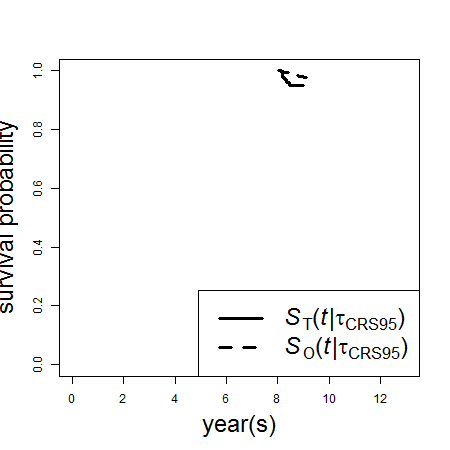}
		\caption{$\widehat{\tau}=8.02$ from CRS95}
		\label{fA1_Oesophagus_CRS95}
	\end{subfigure}\\
	\begin{subfigure}{.3\textwidth}
		\centering
		\includegraphics[width=\textwidth]{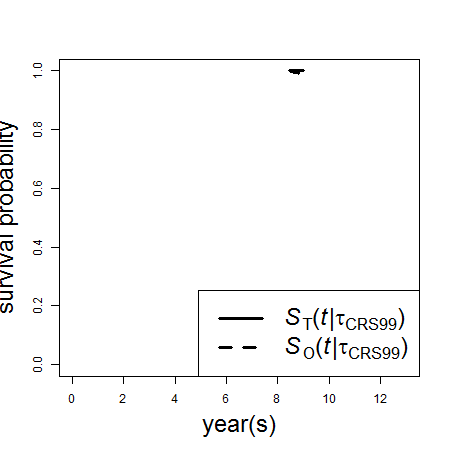}
		\caption{$\widehat{\tau}=8.45$ from CRS99}
		\label{fA1_Oesophagus_CRS99}
	\end{subfigure}%
	\begin{subfigure}{.3\textwidth}
		\centering
		\includegraphics[width=\textwidth]{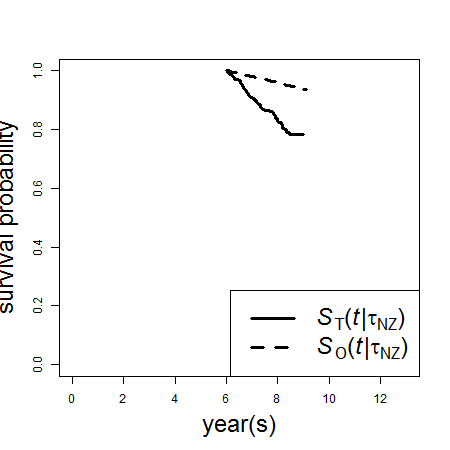}
		\caption{$\widehat{\tau}=6$ from New Zealand}
		\label{fA1_Oesophagus_NZ2006}
	\end{subfigure}\\
	\begin{subfigure}{.3\textwidth}
  \centering
  \includegraphics[width=\textwidth]{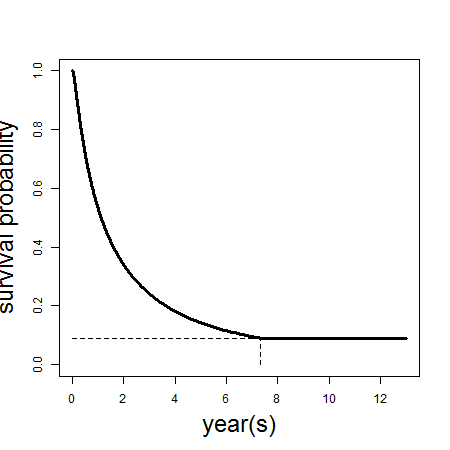}
  \caption{CTM-based net survival}
  \label{fA2_Oesophagus_MRS}
\end{subfigure}%
\begin{subfigure}{.3\textwidth}
  \centering
  \includegraphics[width=\textwidth]{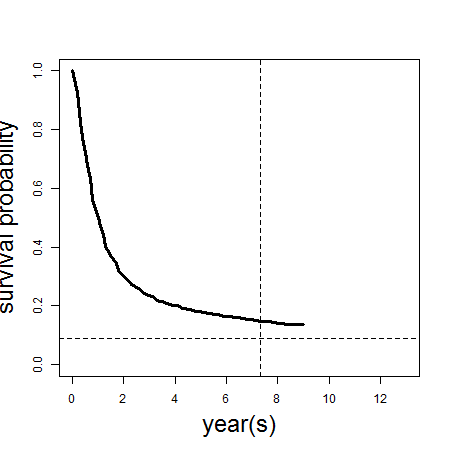}
  \caption{Relative survival}
  \label{fA2_Oesophagus_RS}
\end{subfigure}
	\caption{$S_T(t|k)$ and $S_O(t|k)$ of oesophagus cancer data and general population in Taiwan, where $k$ is cure time estimated from CTM, CRS95, CRS99, and New Zealand (2006). (\ref{fA1_Oesophagus_CTM}-\ref{fA1_Oesophagus_NZ2006}). Model-based net survival and relative survival (\ref{fA2_Oesophagus_MRS}, \ref{fA2_Oesophagus_RS}), horizontal and vertical dashed lines represent locations of CTM-estimated cure time and cure rate, respectively.}
	\label{fA1_Oesophagus}
\end{figure}

\begin{figure}
	\centering
	\begin{subfigure}{.3\textwidth}
		\centering
		\includegraphics[width=\textwidth]{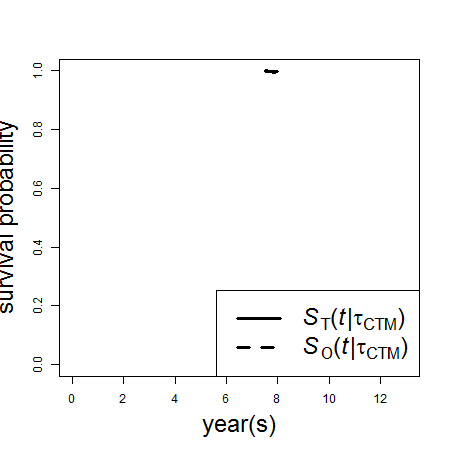}
		\caption{$\widehat{\tau}=7.51$ from CTM}
		\label{fA1_Ovary_CTM}
	\end{subfigure}%
	\begin{subfigure}{.3\textwidth}
		\centering
		\includegraphics[width=\textwidth]{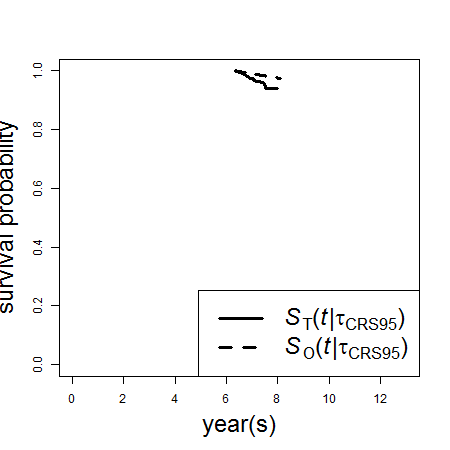}
		\caption{$\widehat{\tau}=6.34$ from CRS95}
		\label{fA1_Ovary_CRS95}
	\end{subfigure}\\
	\begin{subfigure}{.3\textwidth}
		\centering
		\includegraphics[width=\textwidth]{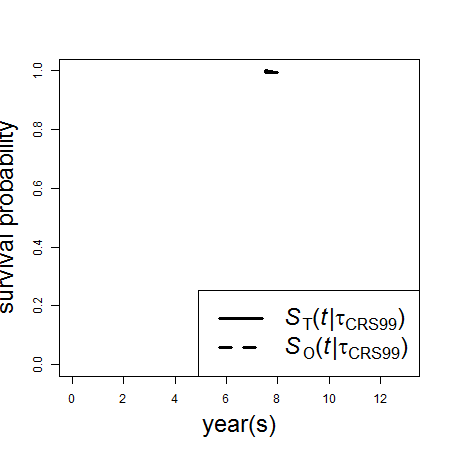}
		\caption{$\widehat{\tau}=7.50$ from CRS99}
		\label{fA1_Ovary_CRS99}
	\end{subfigure}%
	\begin{subfigure}{.3\textwidth}
		\centering
		\includegraphics[width=\textwidth]{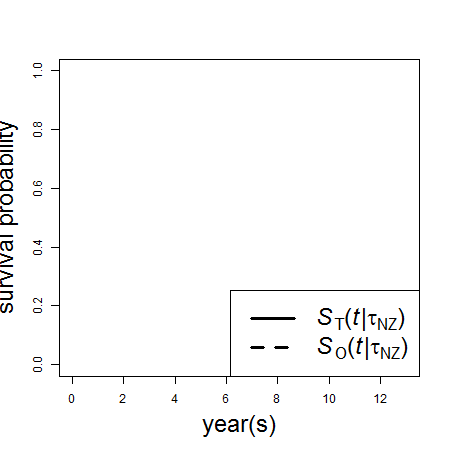}
		\caption{$\widehat{\tau}=10$ from New Zealand}
		\label{fA1_Ovary_NZ2006}
	\end{subfigure}\\
	\begin{subfigure}{.3\textwidth}
  \centering
  \includegraphics[width=\textwidth]{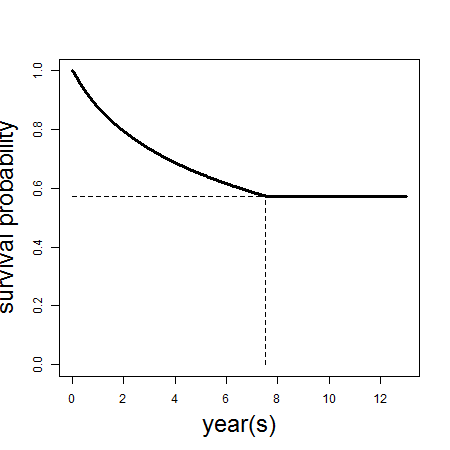}
  \caption{CTM-based net survival}
  \label{fA2_Ovary_MRS}
\end{subfigure}%
\begin{subfigure}{.3\textwidth}
  \centering
  \includegraphics[width=\textwidth]{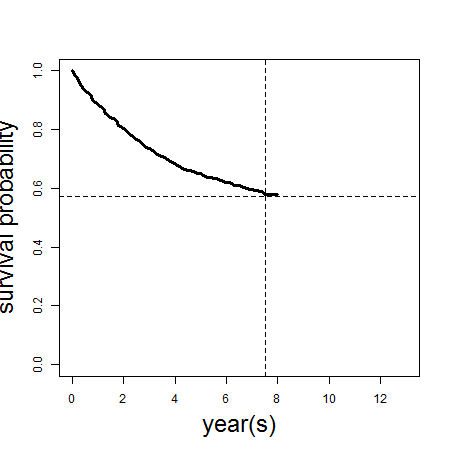}
  \caption{Relative survival}
  \label{fA2_Ovary_RS}
\end{subfigure}
	\caption{$S_T(t|k)$ and $S_O(t|k)$ of ovary cancer data and general population in Taiwan, where $k$ is cure time estimated from CTM, CRS95, CRS99, and New Zealand (2006). (\ref{fA1_Ovary_CTM}-\ref{fA1_Ovary_NZ2006}). Model-based net survival and relative survival (\ref{fA2_Ovary_MRS}, \ref{fA2_Ovary_RS}), horizontal and vertical dashed lines represent locations of CTM-estimated cure time and cure rate, respectively. Note that in \ref{fA1_Ovary_NZ2006} there is no $S_T(t|k)$ and $S_O(t|k)$ since the follow-up time (7.99 years) is smaller than the cure time (10 years).}
	\label{fA1_Ovary}
\end{figure}

\begin{figure}
	\centering
	\begin{subfigure}{.3\textwidth}
		\centering
		\includegraphics[width=\textwidth]{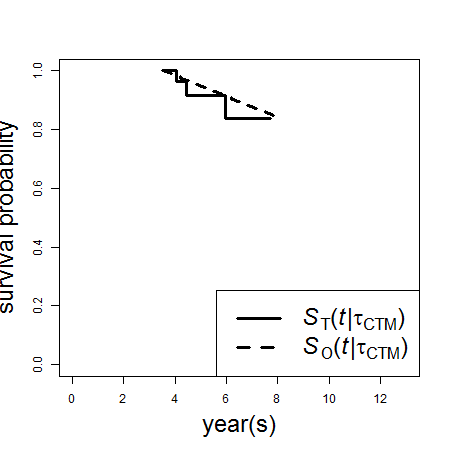}
		\caption{$\widehat{\tau}=3.52$ from CTM}
		\label{fA1_Pancreas_CTM}
	\end{subfigure}%
	\begin{subfigure}{.3\textwidth}
		\centering
		\includegraphics[width=\textwidth]{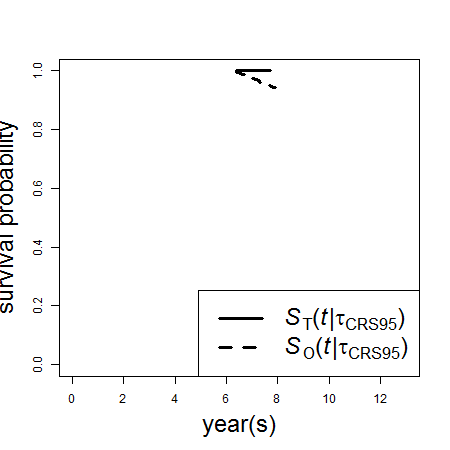}
		\caption{$\widehat{\tau}=6.39$ from CRS95}
		\label{fA1_Pancreas_CRS95}
	\end{subfigure}\\
	\begin{subfigure}{.3\textwidth}
		\centering
		\includegraphics[width=\textwidth]{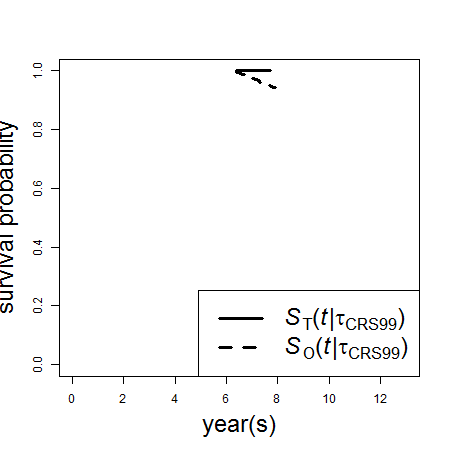}
		\caption{$\widehat{\tau}=6.39$ from CRS99}
		\label{fA1_Pancreas_CRS99}
	\end{subfigure}%
	\begin{subfigure}{.3\textwidth}
		\centering
		\includegraphics[width=\textwidth]{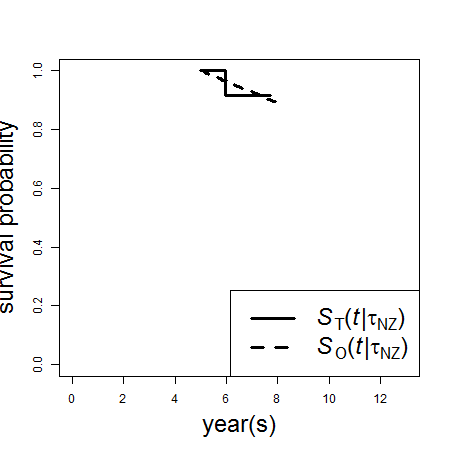}
		\caption{$\widehat{\tau}=5$ from New Zealand}
		\label{fA1_Pancreas_NZ2006}
	\end{subfigure}\\
	\begin{subfigure}{.3\textwidth}
  \centering
  \includegraphics[width=\textwidth]{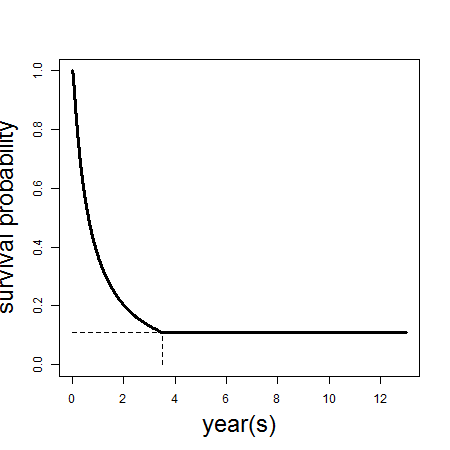}
  \caption{CTM-based net survival}
  \label{fA2_Pancreas_MRS}
\end{subfigure}%
\begin{subfigure}{.3\textwidth}
  \centering
  \includegraphics[width=\textwidth]{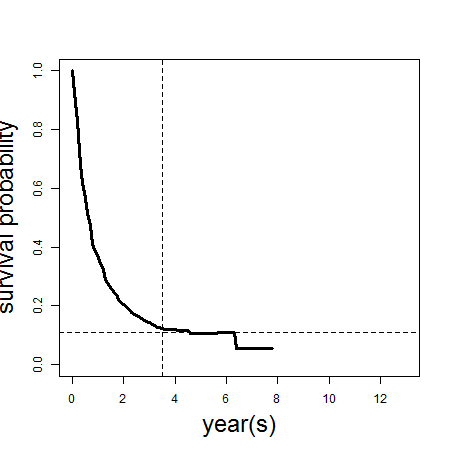}
  \caption{Relative survival}
  \label{fA2_Pancreas_RS}
\end{subfigure}
	\caption{$S_T(t|k)$ and $S_O(t|k)$ of pancreas cancer data and general population in Taiwan, where $k$ is cure time estimated from CTM, CRS95, CRS99, and New Zealand (2006). (\ref{fA1_Pancreas_CTM}-\ref{fA1_Pancreas_NZ2006}). Model-based net survival and relative survival (\ref{fA2_Pancreas_MRS}, \ref{fA2_Pancreas_RS}), horizontal and vertical dashed lines represent locations of CTM-estimated cure time and cure rate, respectively.}
	\label{fA1_Pancreas}
\end{figure}

\begin{figure}
	\centering
	\begin{subfigure}{.3\textwidth}
		\centering
		\includegraphics[width=\textwidth]{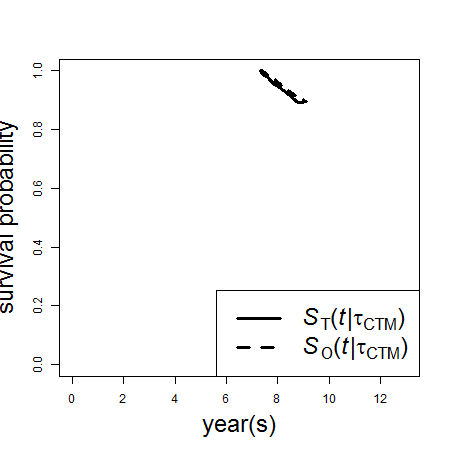}
		\caption{$\widehat{\tau}=7.31$ from CTM}
		\label{fA1_Prostate_CTM}
	\end{subfigure}%
	\begin{subfigure}{.3\textwidth}
		\centering
		\includegraphics[width=\textwidth]{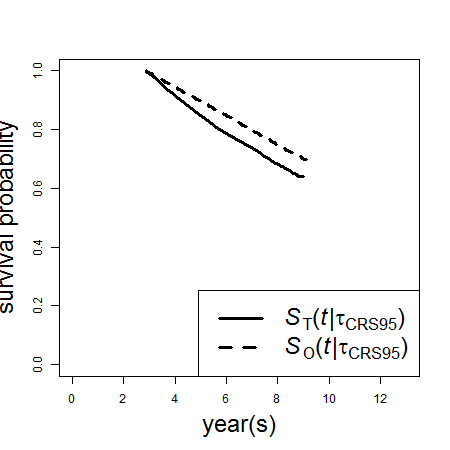}
		\caption{$\widehat{\tau}=2.87$ from CRS95}
		\label{fA1_Prostate_CRS95}
	\end{subfigure}\\
	\begin{subfigure}{.3\textwidth}
		\centering
		\includegraphics[width=\textwidth]{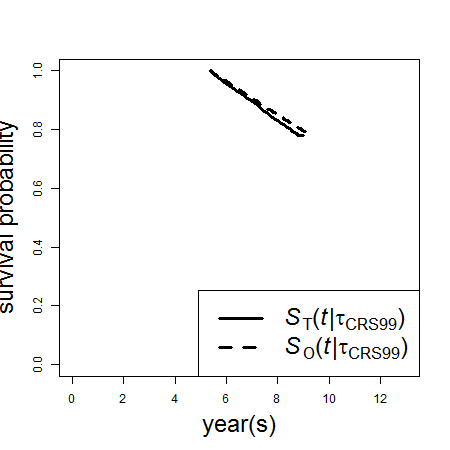}
		\caption{$\widehat{\tau}=5.38$ from CRS99}
		\label{fA1_Prostate_CRS99}
	\end{subfigure}%
	\begin{subfigure}{.3\textwidth}
		\centering
		\includegraphics[width=\textwidth]{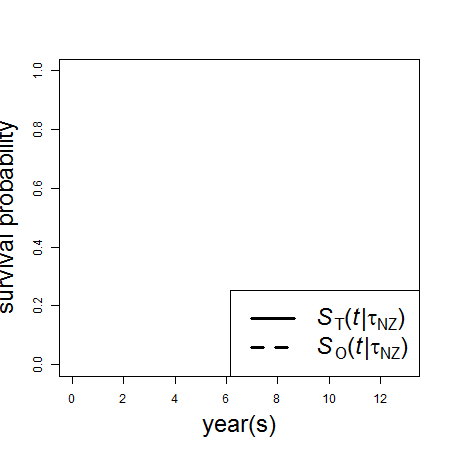}
		\caption{$\widehat{\tau}=20$ from New Zealand}
		\label{fA1_Prostate_NZ2006}
	\end{subfigure}\\
	\begin{subfigure}{.3\textwidth}
  \centering
  \includegraphics[width=\textwidth]{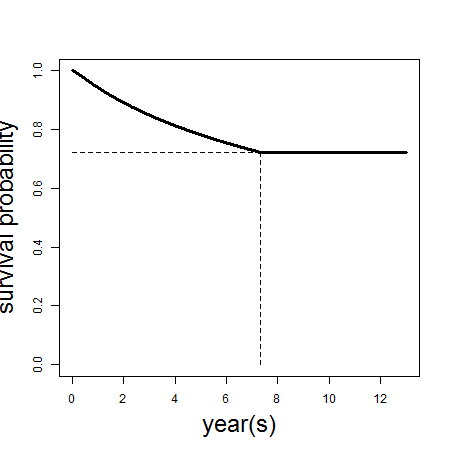}
  \caption{CTM-based net survival}
  \label{fA2_Prostate_MRS}
\end{subfigure}%
\begin{subfigure}{.3\textwidth}
  \centering
  \includegraphics[width=\textwidth]{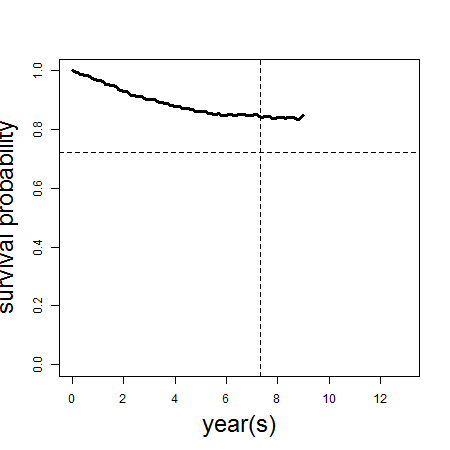}
  \caption{Relative survival}
  \label{fA2_Prostate_RS}
\end{subfigure}
	\caption{$S_T(t|k)$ and $S_O(t|k)$ of prostate cancer data and general population in Taiwan, where $k$ is cure time estimated from CTM, CRS95, CRS99, and New Zealand (2006). (\ref{fA1_Prostate_CTM}-\ref{fA1_Prostate_NZ2006}). Model-based net survival and relative survival (\ref{fA2_Prostate_MRS}, \ref{fA2_Prostate_RS}), horizontal and vertical dashed lines represent locations of CTM-estimated cure time and cure rate, respectively. Note that in \ref{fA1_Prostate_NZ2006} there is no $S_T(t|k)$ and $S_O(t|k)$ since the follow-up time (9 years) is smaller than the cure time (20 years).}
	\label{fA1_Prostate}
\end{figure}

\begin{figure}
	\centering
	\begin{subfigure}{.3\textwidth}
		\centering
		\includegraphics[width=\textwidth]{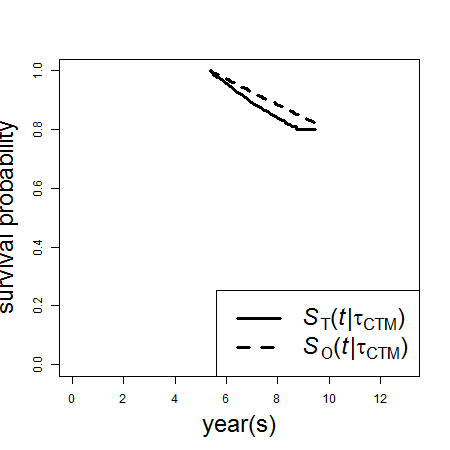}
		\caption{$\widehat{\tau}=5.36$ from CTM}
		\label{fA1_Stomach_CTM}
	\end{subfigure}%
	\begin{subfigure}{.3\textwidth}
		\centering
		\includegraphics[width=\textwidth]{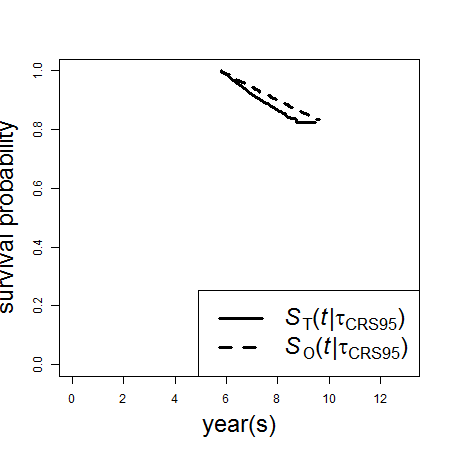}
		\caption{$\widehat{\tau}=5.79$ from CRS95}
		\label{fA1_Stomach_CRS95}
	\end{subfigure}\\
	\begin{subfigure}{.3\textwidth}
		\centering
		\includegraphics[width=\textwidth]{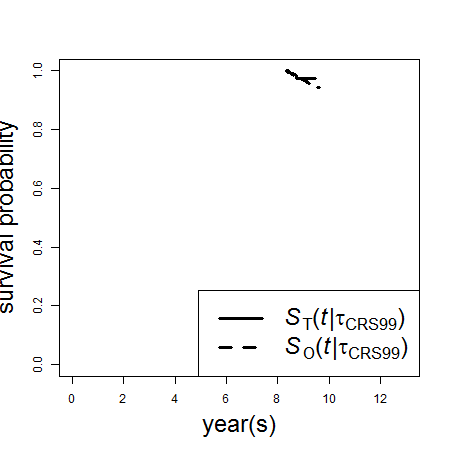}
		\caption{$\widehat{\tau}=8.33$ from CRS99}
		\label{fA1_Stomach_CRS99}
	\end{subfigure}%
	\begin{subfigure}{.3\textwidth}
		\centering
		\includegraphics[width=\textwidth]{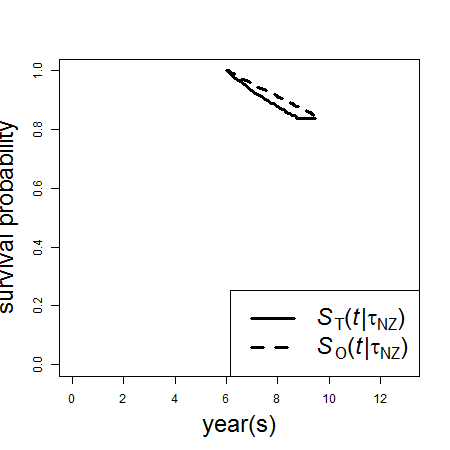}
		\caption{$\widehat{\tau}=6$ from New Zealand}
		\label{fA1_Stomach_NZ2006}
	\end{subfigure}\\
	\begin{subfigure}{.3\textwidth}
  \centering
  \includegraphics[width=\textwidth]{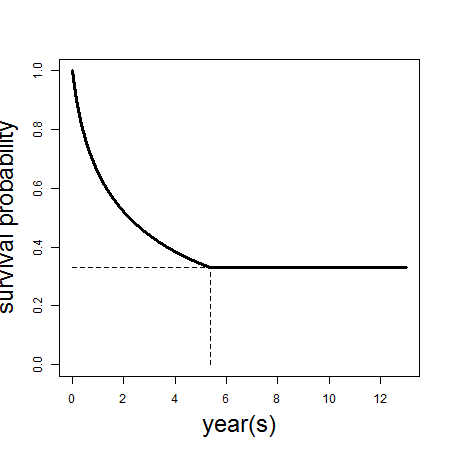}
  \caption{CTM-based net survival}
  \label{fA2_Stomach_MRS}
\end{subfigure}%
\begin{subfigure}{.3\textwidth}
  \centering
  \includegraphics[width=\textwidth]{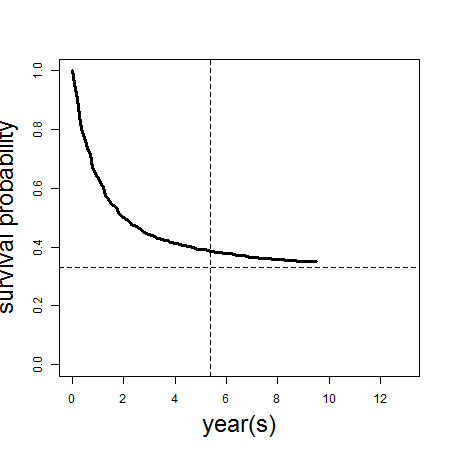}
  \caption{Relative survival}
  \label{fA2_Stomach_RS}
\end{subfigure}
	\caption{$S_T(t|k)$ and $S_O(t|k)$ of stomach cancer data and general population in Taiwan, where $k$ is cure time estimated from CTM, CRS95, CRS99, and New Zealand (2006). (\ref{fA1_Stomach_CTM}-\ref{fA1_Stomach_NZ2006}). Model-based net survival and relative survival (\ref{fA2_Stomach_MRS}, \ref{fA2_Stomach_RS}), horizontal and vertical dashed lines represent locations of CTM-estimated cure time and cure rate, respectively.}
	\label{fA1_Stomach}
\end{figure}

\begin{figure}
	\centering
	\begin{subfigure}{.3\textwidth}
		\centering
		\includegraphics[width=\textwidth]{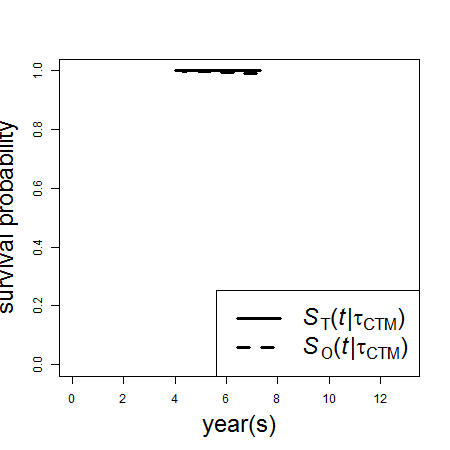}
		\caption{$\widehat{\tau}=4.02$ from CTM}
		\label{fA1_Testis_CTM}
	\end{subfigure}%
	\begin{subfigure}{.3\textwidth}
		\centering
		\includegraphics[width=\textwidth]{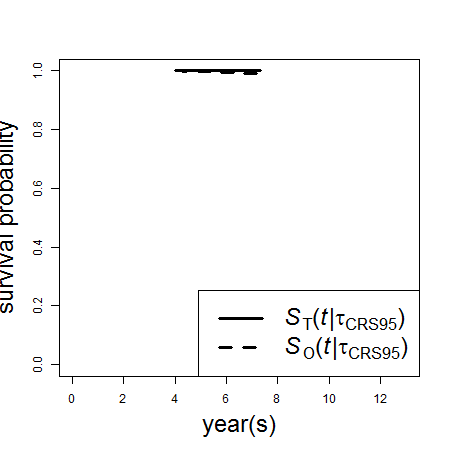}
		\caption{$\widehat{\tau}=4.02$ from CRS95}
		\label{fA1_Testis_CRS95}
	\end{subfigure}\\
	\begin{subfigure}{.3\textwidth}
		\centering
		\includegraphics[width=\textwidth]{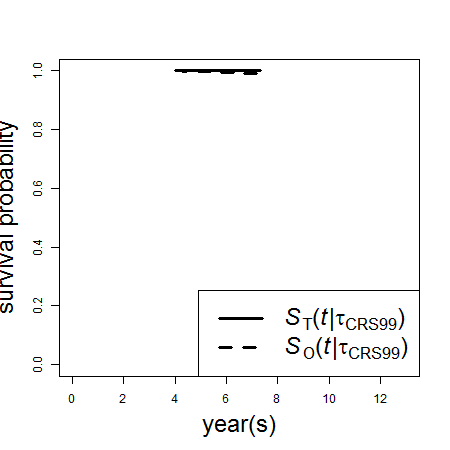}
		\caption{$\widehat{\tau}=4.02$ from CRS99}
		\label{fA1_Testis_CRS99}
	\end{subfigure}%
	\begin{subfigure}{.3\textwidth}
		\centering
		\includegraphics[width=\textwidth]{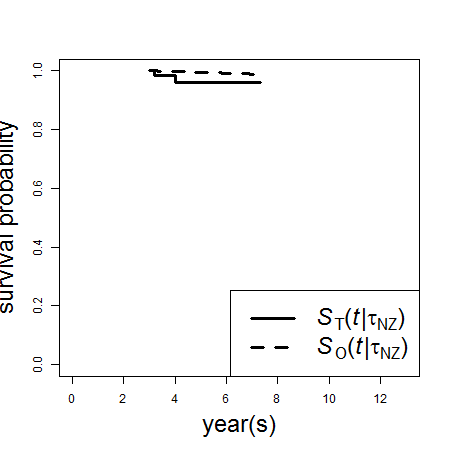}
		\caption{$\widehat{\tau}=3$ from New Zealand}
		\label{fA1_Testis_NZ2006}
	\end{subfigure}\\
	\begin{subfigure}{.3\textwidth}
  \centering
  \includegraphics[width=\textwidth]{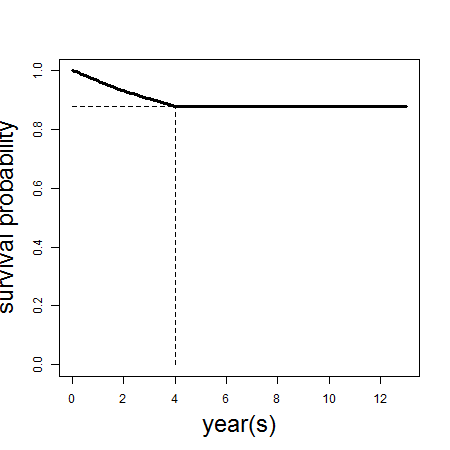}
  \caption{CTM-based net survival}
  \label{fA2_Testis_MRS}
\end{subfigure}%
\begin{subfigure}{.3\textwidth}
  \centering
  \includegraphics[width=\textwidth]{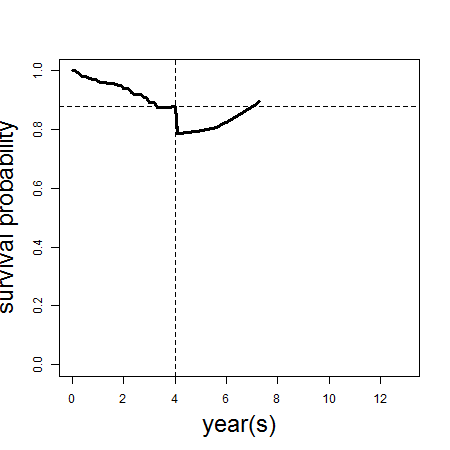}
  \caption{Relative survival}
  \label{fA2_Testis_RS}
\end{subfigure}
	\caption{$S_T(t|k)$ and $S_O(t|k)$ of testis cancer data and general population in Taiwan, where $k$ is cure time estimated from CTM, CRS95, CRS99, and New Zealand (2006). (\ref{fA1_Testis_CTM}-\ref{fA1_Testis_NZ2006}). Model-based net survival and relative survival (\ref{fA2_Testis_MRS}, \ref{fA2_Testis_RS}), horizontal and vertical dashed lines represent locations of CTM-estimated cure time and cure rate, respectively.}
	\label{fA1_Testis}
\end{figure}

\begin{figure}
	\centering
	\begin{subfigure}{.3\textwidth}
		\centering
		\includegraphics[width=\textwidth]{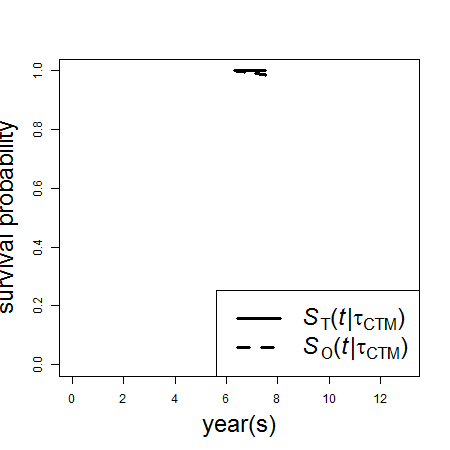}
		\caption{$\widehat{\tau}=6.31$ from CTM}
		\label{fA1_Thyroid_CTM}
	\end{subfigure}%
	\begin{subfigure}{.3\textwidth}
		\centering
		\includegraphics[width=\textwidth]{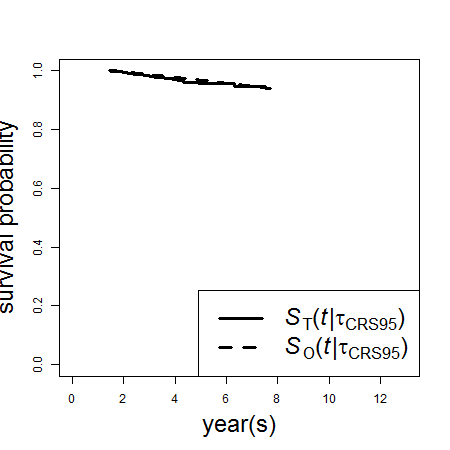}
		\caption{$\widehat{\tau}=1.46$ from CRS95}
		\label{fA1_Thyroid_CRS95}
	\end{subfigure}\\
	\begin{subfigure}{.3\textwidth}
		\centering
		\includegraphics[width=\textwidth]{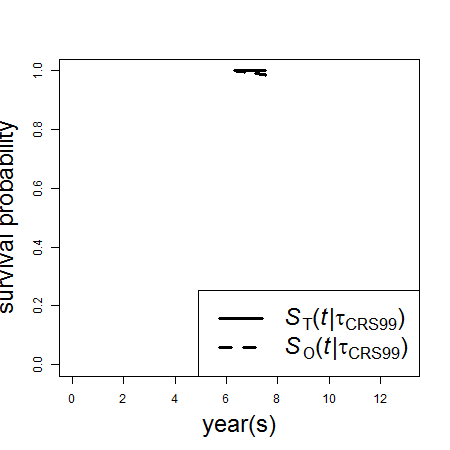}
		\caption{$\widehat{\tau}=6.31$ from CRS99}
		\label{fA1_Thyroid_CRS99}
	\end{subfigure}%
	\begin{subfigure}{.3\textwidth}
		\centering
		\includegraphics[width=\textwidth]{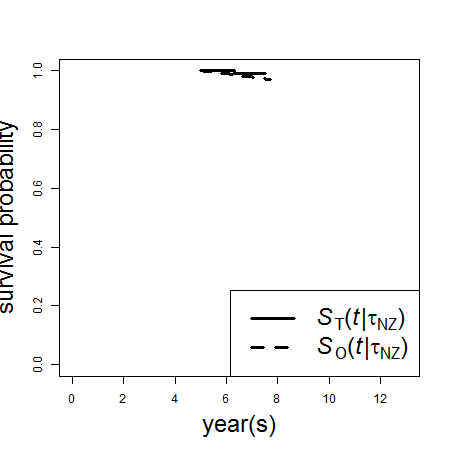}
		\caption{$\widehat{\tau}=5$ from New Zealand}
		\label{fA1_Thyroid_NZ2006}
	\end{subfigure}\\
	\begin{subfigure}{.3\textwidth}
  \centering
  \includegraphics[width=\textwidth]{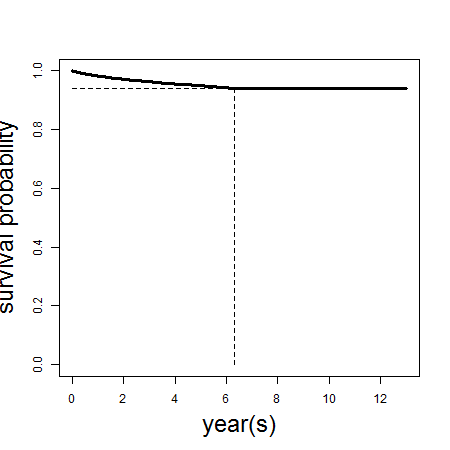}
  \caption{CTM-based net survival}
  \label{fA2_Thyroid_MRS}
\end{subfigure}%
\begin{subfigure}{.3\textwidth}
  \centering
  \includegraphics[width=\textwidth]{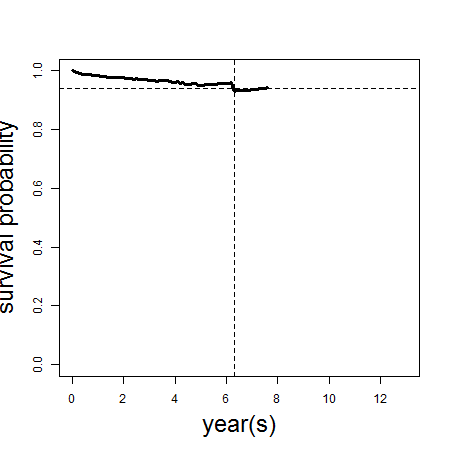}
  \caption{Relative survival}
  \label{fA2_Thyroid_RS}
\end{subfigure}
	\caption{$S_T(t|k)$ and $S_O(t|k)$ of thyroid cancer data and general population in Taiwan, where $k$ is cure time estimated from CTM, CRS95, CRS99, and New Zealand (2006). (\ref{fA1_Thyroid_CTM}-\ref{fA1_Thyroid_NZ2006}). Model-based net survival and relative survival (\ref{fA2_Thyroid_MRS}, \ref{fA2_Thyroid_RS}), horizontal and vertical dashed lines represent locations of CTM-estimated cure time and cure rate, respectively.}
	\label{fA1_Thyroid}
\end{figure}

\begin{figure}
	\centering
	\begin{subfigure}{.3\textwidth}
		\centering
		\includegraphics[width=\textwidth]{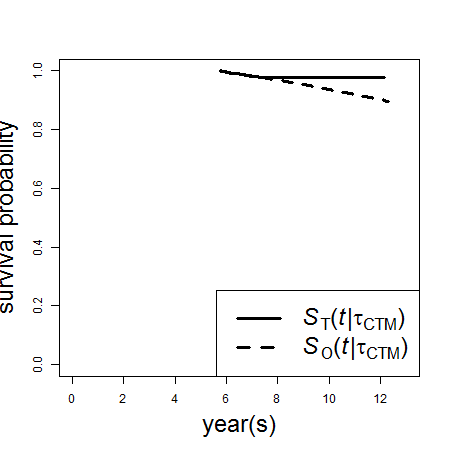}
		\caption{$\widehat{\tau}=5.76$ from CTM}
		\label{fA1_Uterus_CTM}
	\end{subfigure}%
	\begin{subfigure}{.3\textwidth}
		\centering
		\includegraphics[width=\textwidth]{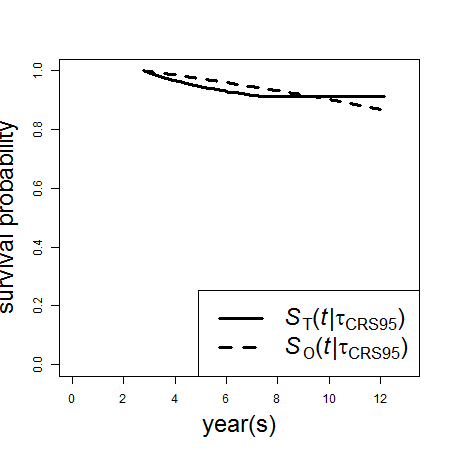}
		\caption{$\widehat{\tau}=2.77$ from CRS95}
		\label{fA1_Uterus_CRS95}
	\end{subfigure}\\
	\begin{subfigure}{.3\textwidth}
		\centering
		\includegraphics[width=\textwidth]{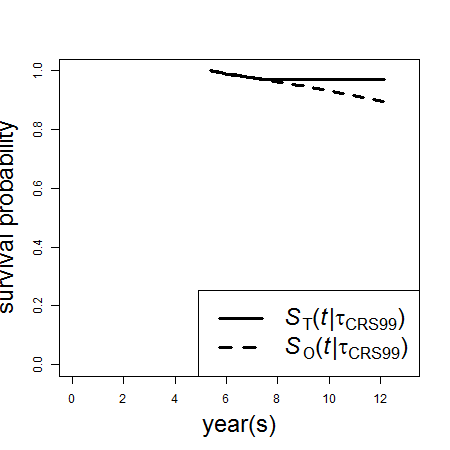}
		\caption{$\widehat{\tau}=5.37$ from CRS99}
		\label{fA1_Uterus_CRS99}
	\end{subfigure}%
	\begin{subfigure}{.3\textwidth}
		\centering
		\includegraphics[width=\textwidth]{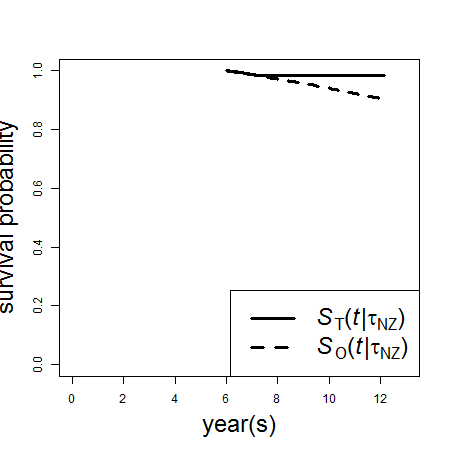}
		\caption{$\widehat{\tau}=6$ from New Zealand}
		\label{fA1_Uterus_NZ2006}
	\end{subfigure}\\
	\begin{subfigure}{.3\textwidth}
  \centering
  \includegraphics[width=\textwidth]{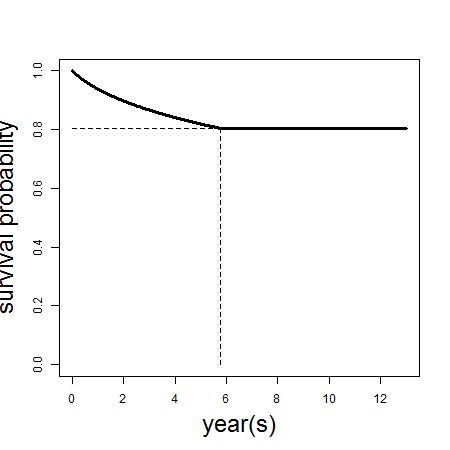}
  \caption{CTM-based net survival}
  \label{fA2_Uterus_MRS}
\end{subfigure}%
\begin{subfigure}{.3\textwidth}
  \centering
  \includegraphics[width=\textwidth]{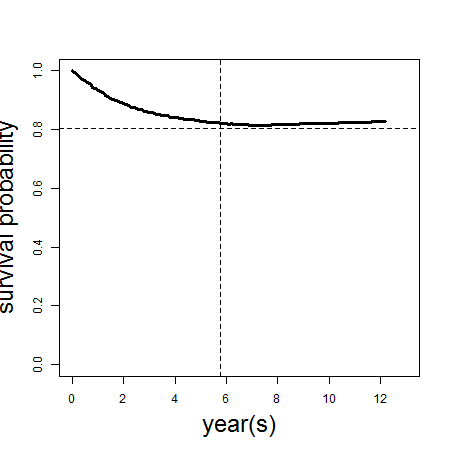}
  \caption{Relative survival}
  \label{fA2_Uterus_RS}
\end{subfigure}
	\caption{$S_T(t|k)$ and $S_O(t|k)$ of uterus cancer data and general population in Taiwan, where $k$ is cure time estimated from CTM, CRS95, CRS99, and New Zealand (2006). (\ref{fA1_Uterus_CTM}-\ref{fA1_Uterus_NZ2006}). Model-based net survival and relative survival (\ref{fA2_Uterus_MRS}, \ref{fA2_Uterus_RS}), horizontal and vertical dashed lines represent locations of CTM-estimated cure time and cure rate, respectively.}
	\label{fA1_Uterus}
\end{figure}

\end{document}